\documentclass[12pt,oneside]{article}%
\usepackage{amsmath}
\usepackage{amssymb}
\usepackage{amsfonts}
\usepackage{graphicx}
\usepackage{cite}%
\usepackage[arrow, matrix]{xy}
\allowdisplaybreaks
\newcommand{\eqa}{\begin{eqnarray}}
\newcommand{\ena}{\end{eqnarray}}

\newcommand{\nn}{\nonumber\\}

\def\one{\mbox{1 \kern-.59em {\rm l}}}
\begin{document}

\author{ Alexander Schmidt\thanks{e-mail: schmidt@theorie.physik.uni-muenchen.de},
Hartmut Wachter\thanks{e-mail: Hartmut.Wachter@physik.uni-muenchen.de}%
\vspace*{0.3in}\\Max-Planck-Institute\\for Mathematics in the Sciences \\Inselstr. 22, D-04103 Leipzig, Germany\\\hspace{0.4in}\\Arnold-Sommerfeld-Center \\Ludwig-Maximilians-Universit\"{a}t\\Theresienstr. 37, D-80333 M\"{u}nchen, Germany}
\title{Spinor calculus for q-deformed quantum spaces I}
\date{}
\maketitle

\begin{abstract}
\noindent The article is dedicated to q-deformed versions of spinor
calculus. As a kind of review, the most relevant properties of the
two-dimensional quantum plane are summarized. Additionally, the relationship
between the quantum plane and higher-dimensional quantum spaces like the
q-deformed Euclidean space in four dimensions or the q-deformed Minkowski
space is outlined. These considerations are continued by introducing
q-analogs of the Pauli matrices. Their main properties are discussed in
detail and numerous relations that could prove useful in physical applications
are presented. In this respect, q-deformed versions of the important Fierz
identities are written down.\newpage

\end{abstract}
\tableofcontents

\section{Introduction}

It is commonplace that discretizing space-time should provide an effective
method for regularizing quantum field theories \cite{Heis}. In the literature
one can find a number of serious attempts towards this goal (see for example
Refs. \cite{Fli48, Hill55, Das60, Gol63, Sny47, Yan47}). A more recent and
very promising approach, however, is based on quantum group symmetries
\cite{FLW96, CW98, MajReg, GKP96, Oec99, Blo03}.

Let us recall that quantum groups can be seen as deformations of classical
space-time symmetries, as they describe the symmetry of their comodules, which
are often referred to as quantum spaces. The most realistic quantum groups and
quantum spaces arise from q-deformation \cite{Ku83, Wor87, Dri85, Jim85,
Drin86, RFT90, Tak90, Man88, CSSW90, PW90, SWZ91, Maj91, LWW97}. (For other
deformations of space-time see Refs. \cite{Lu92, Cas93, Dob94, DFR95, ChDe95,
ChKu04, Koch04}.)

As part of the long way towards a quantum field theory on $q$-deformed spaces,
we treat a $q$-deformed version of spinor calculus in this paper. We mainly
concentrate attention on $q$-analogs of 2-spinors and Pauli matrices. Notice
that our reasonings will be continued by a second paper which is devoted to
$q$-analogs of 4-spinors and the corresponding $\gamma^{\mu}$-matrices. It
should also be mentioned that the aim of the present paper is twofold. On the
one hand it serves as review of known facts about $q$-deformed spinors
\cite{Schlieck1, Schlieck2, Schlieck3, Mey96, WSSW90}. This has the purpose to
fix\ a consistent framework of conventions and notations. On the other hand we
present numerous $q$-analogs of well-known identities found in standard
textbooks on quantum field theory (see, for instance, Refs. \cite{Peskin,
Landau, Wessbagger, bailin}).

The paper is organized as follows. In Sec.\thinspace\ref{qplanekap} we review
the foundations of the two-dimensional Manin plane and the corresponding
quantum group $SU_{q}(2)$ in great detail. Remember that the quantum group
$SU_{q}(2)$ is a Hopf algebra \cite{Hopf, Maj95, ChDe96, Klimyk}\ that gives
rise to left and right coactions on the Manin plane. For this reason we
collect all relevant formulae concerning the Hopf-structure and coactions of
$SU_{q}(2)$. Furthermore, we emphasize the existence of two different $\ast
$-structures for the algebra of $SU_{q}(2)$. If we demand for the Manin plane
to be a $\ast$-comodule algebra the right coaction of $SU_{q}(2)$ implies a
different $\ast$-structure for $SU_{q}(2)$ than the left coaction. This
observation is often neglected in the literature.

In Sec.\thinspace\ref{koppkap} we describe how to build up $q$-deformed
quantum spaces in four dimensions from two distinct quantum planes. In this
respect, we concern ourselves with the $q$-deformed Euclidean space in four
dimensions and the $q$-deformed Minkowski space. The reasonings in this
section have complete review character and were included for the sake of
completeness, since they enable a deeper understanding of the subject.
Moreover, Sec.\thinspace\ref{koppkap} shall serve as a collection of formulae
that could prove useful in what follows.

Section \ref{PaulAlg} covers all aspects concerning $q$-analogs of Pauli
matrices and spin matrices, where we restrict attention to objects of the
$q$-deformed Minkowski space as well as the three- and four-dimensional
$q$-deformed Euclidean space. These considerations form the main part of the
paper. Especially, we describe the explicit construction of the above
mentioned matrices, list their most important properties, and discuss some
peculiarities arising from the noncommutative structure. In doing so, we
obtain numerous $q$-analogs of the well-known product- and trace-relations for
Pauli matrices.

In Sec.\thinspace\ref{connectchapt} we take up the question how to switch from
spinorial objects to vectorial ones and vice versa. These results are rather
interesting from a representation theoretic point of view. In Sec.\thinspace
\ref{Fierz} we finally present formulae that can be viewed as $q$-analogs of
Fierz identities. More concretely, we list $q$-deformed versions of all the
\textit{Weyl Spinor Fierz Identities} found in App.\thinspace A of Ref.
\cite{bailin}. It is important to notice that the concrete form of these
relations depends on the choice for the braiding between the different
spinors. The influence of this choice on our results\ was carefully examined.

We would like to say a few words about the method we applied in getting most
of our results. To derive a $q$-analog of a given identity one first proposes
a reasonable ansatz. The essential idea is that normally ordered monomials of
noncommutative generators often establish a basis of the algebra under
consideration. In such cases the unknown coefficients of the ansatz can be
found by a process of normal ordering together with comparing coefficients.
This task can effectively be done by a computer algebra system like
\textit{Mathematica }\cite{Wol}.

Last but not least let us make some notational remarks\textit{.} Throughout
the article we use the shortcuts $\lambda\equiv q-q^{-1}$ and $\lambda
_{+}\equiv q+q^{-1}$. Sometimes a $n$-dimensional identity matrix is written
as $\mbox{1 \kern-.59em {\rm l}}$. The antisymmetric $q$-numbers are defined
by $[[n]]_{q^{a}}\equiv\frac{1-q^{an}}{1-q^{a}},$ where we assume $q>1$ and
$a\in\mathbb{R}$.

\section{Algebraic foundations of $q$-deformed spinor
calculus\label{qplanekap}}

In this section we give a review of the quantum group $SU_{q}(2)$ and its
two-dimensional corepresentation. This corepresentation is nothing other than
the two-dimensional quantum plane. The generators of the two-dimensional
quantum plane can be viewed as $q$-analogs of 2-spinors. The following
presentation is mainly based on the reasonings in Refs. \cite{Wess00, Bloh01,
Schlieck1, Schlieck2}. For a deeper understanding we also recommend to consult
Refs. \cite{Klimyk, ChDe96, Maj95}.

\subsection{The two-dimensional quantum plane}

Our starting point is the famous Manin plane and the coactions of the quantum
group $SL_{q}(2)$ on it. The Manin plane is generated by two noncommutative
coordinates subject to the relations%
\begin{equation}
x^{1}x^{2}=qx^{2}x^{1}. \label{qspace1}%
\end{equation}
We can write the above relation with the help of a $q$-analog of spinor metric
$\varepsilon_{ij}$, i.e.
\begin{equation}
\varepsilon_{ij}x^{i}x^{j}=0, \label{qspace1a}%
\end{equation}
where
\begin{equation}
\varepsilon_{ij}=\left(
\begin{array}
[c]{cc}%
0 & q^{-1/2}\\
-q^{1/2} & 0
\end{array}
\right)  .
\end{equation}
For the inverse of the spinor metric we have%
\begin{equation}
\varepsilon^{ij}=-\varepsilon_{ij}. \label{epsiloninverse}%
\end{equation}
It is not very difficult to verify that%
\begin{equation}
\varepsilon_{ij}\varepsilon^{jk}=\varepsilon^{kj}\varepsilon_{ji}=\delta
_{i}^{k}. \label{epsiseins}%
\end{equation}

The spinor metric enables us to raise and lower indices of quantum
plane\ coordinates:%
\begin{equation}
x_{i}=\varepsilon_{ij}x^{j},\quad x^{i}=\varepsilon^{ij}x_{j}.\label{covco}%
\end{equation}
Using this convention we transform the coordinates in Eq.\thinspace
(\ref{qspace1}) into covariant ones and get
\begin{equation}
x_{1}x_{2}=q^{-1}x_{2}x_{1}.\label{CovRelMan}%
\end{equation}
Once again, the last result can be expressed with the help of the spinor
metric. From (\ref{covco}), (\ref{qspace1a}), and (\ref{epsiseins}) we deduce%
\begin{equation}
\varepsilon^{ji}x_{i}x_{j}=0,\quad\mbox{or}\quad(\varepsilon^{T})^{ij}%
x_{i}x_{j}=0,
\end{equation}
where we introduced the transposed metric $(\varepsilon^{T})^{ij}%
\equiv\varepsilon^{ji}$. Last but not least we can write down rules for
raising the indices of\ the spinor metric itself. From (\ref{epsiseins}) we
find%
\begin{align}
\varepsilon^{lk} &  =\varepsilon^{ik}\varepsilon^{jl}\varepsilon_{ij}%
,\quad\mbox{or}\quad(\varepsilon^{T})^{kl}=\varepsilon^{ik}\varepsilon
^{jl}\varepsilon_{ij},\nonumber\\
\varepsilon_{lk} &  =\varepsilon_{ik}\varepsilon_{jl}\varepsilon^{ij}%
,\quad\mbox{or}\quad(\varepsilon^{T})_{kl}=\varepsilon_{ik}\varepsilon
_{jl}\varepsilon^{ij}.\label{epsindzieh}%
\end{align}

Next, we come to Hermitian conjugation on the quantum plane. On the
coordinates $x^{i},$ $i=1,2,$ the operation of conjugation becomes%
\begin{equation}
\overline{x^{i}}=\overline{x}_{i}=-\varepsilon_{ij}\overline{x}^{j}%
,\qquad\overline{x_{i}}=\overline{x}^{\,i}=-\varepsilon^{ij}\overline{x}_{j}.
\label{manincon1}%
\end{equation}
Notice that the conjugated coordinates $\overline{x}_{i}$, $i=1,2,$ fulfill
the same relation as the covariant coordinates $x_{i},$ $i=1,2$, as is
indicated by their lower indices. It should also be mentioned that a
conjugated spinor\ metric can be introduced via the condition
\begin{equation}
\overline{x^{i}x_{i}}\overset{!}{=}\bar{\varepsilon}^{\,ij}\bar{x_{j}}%
\bar{x_{i}}%
\end{equation}
resulting in%
\begin{equation}
\bar{\varepsilon}^{\,ij}=-\varepsilon^{ij}.
\end{equation}

There also exists an antisymmetrized version of the Manin plane characterized
by the relations \cite{MSW04, SW04}
\begin{equation}
\theta^{1}\theta^{1}=\theta^{2}\theta^{2}=0,\qquad\theta^{1}\theta^{2}%
=-q^{-1}\theta^{2}\theta^{1}. \label{qspace2}%
\end{equation}
On the antisymmetrized Manin plane covariant as well as conjugated coordinates
are introduced in very much the same way as was done for the symmetrized Manin plane.

\subsection{The quantum group $SU_{q}(2)$}

\subsubsection{Hopf algebra structure and coactions}

As next step, we analyze the symmetry of the quantum plane. This symmetry is
described by the quantum group $SL_{q}(2)$. The generators of $SL_{q}(2)$ can
be viewed as noncommuting functionals corresponding to the entries of a
$2\times2$ matrix. For this reason, $SL_{q}(2)$ is sometimes called a matrix
quantum group. In analogy to the undeformed case, we write
\begin{equation}
M^{i}{}_{j}=\left(
\begin{array}
[c]{cc}%
a & b\\
c & d
\end{array}
\right)  ,\label{MatSLq2}%
\end{equation}
with $i$ denoting the row and $j$ the column. The four entries $a,$ $b,$ $c,$
and $d$ stand for the elements generating the algebra of $SL_{q}(2)$.

It is possible to deduce the algebra relations between the entries of the
matrix in (\ref{MatSLq2}), if we demand the quantum plane to be a comodule
algebra with respect to the quantum group $SL_{q}(2)$. On coordinates of the
Manin plane the left coaction $\beta_{L}$ of $SL_{q}(2)$ should be given by
\begin{equation}
\beta_{L}(x^{i})=M^{i}{}_{j}\otimes x^{j}, \label{coact1}%
\end{equation}
whereas the right coaction $\beta_{R}$ takes the form
\begin{equation}
\beta_{R}(x^{i})=x^{j}\otimes M^{j}{}_{i}=x^{j}\otimes(M^{T})^{i}{}_{j}.
\label{coact2}%
\end{equation}
If not stated otherwise summation over all repeated indices is to be
understood. Requiring that the transformed elements again fulfill the relation
for quantum plane coordinates, i.e.
\begin{equation}
\beta_{L/R}(x^{1})\beta_{L/R}(x^{2})=q\,\beta_{L/R}(x^{2})\beta_{L/R}(x^{1}),
\end{equation}
we are able to derive the following commutation relations for the generators
of $SL_{q}(2)$:%
\begin{gather}
ab=qba,\quad ac=qca,\quad bc=cb,\quad bd=qdb,\nonumber\\
cd=qdc,\quad ad-da=(q-q^{-1})bc. \label{Mrel}%
\end{gather}

Notice that in this algebra the quantum determinant%
\begin{equation}
\det\nolimits_{q}M=ad-q\,bc,
\end{equation}
is a central element. Thus we can enlarge the algebra subjected to the
relations in (\ref{Mrel}) by\ $(\det_{q}M)^{-1}$. Furthermore, we are free to
set the quantum determinant equal to one.

Now, we are in a position to introduce a coproduct $\Delta$, an antipode $S$,
and a counit $\epsilon$ to make the algebra of $SL_{q}(2)$ into a Hopf
algebra. In our matrix notation these structures read as\footnote{Throughout
the paper $\epsilon$ denotes the counit of a Hopf structure, whereas
$\varepsilon$ stands for\ the spinor metric.}%
\begin{align}
\Delta(M^{i}{}_{j})  &  =M^{i}{}_{k}\otimes M^{k}{}_{j},\nonumber\\
\Delta(\det\nolimits_{q}M)  &  =\det\nolimits_{q}M\otimes\det\nolimits_{q}%
M,\\[0.08in]
S(M^{i}{}_{j})  &  =(\det\nolimits_{q}M)^{-1}\,\varepsilon_{lj}M^{l}{}%
_{k}\,\varepsilon^{ik},\nonumber\\
S^{-1}(M^{i}{}_{j})  &  =(\det\nolimits_{q}M)^{-1}\,\varepsilon_{jl}M^{l}%
{}_{k}\,\varepsilon^{ki},\label{AntSLq2}\\
S(\det\nolimits_{q}M)  &  =(\det\nolimits_{q}M)^{-1},\nonumber\\[0.08in]
\epsilon(M^{i}{}_{j})  &  =\delta^{i}{}_{j},\quad\epsilon(\det\nolimits_{q}%
M)=1. \label{hopfsl2}%
\end{align}
Again, this Hopf structure is consistent with the choice $\det\nolimits_{q}%
M=1,$ so we can omit the quantum determinant in the further reasonings about
the Hopf algebra $SL_{q}(2)$.

Alternatively, one can start the derivation of the relations in (\ref{Mrel})
from coactions for covariant coordinates:%
\begin{align}
\beta_{L}(x_{i})  &  =S^{-1}(M^{j}{}_{i})\otimes x_{j},\nonumber\\
\beta_{R}(x_{i})  &  =x_{j}\otimes S((M^{T})^{j}{}_{i}). \label{coact3}%
\end{align}
We deduced the explicit form of the coactions in (\ref{coact3}) from
(\ref{coact1}) and (\ref{coact2}) by lowering indices and using (\ref{AntSLq2}%
). A short look at (\ref{CovRelMan}) tells us that the coactions on covariant
coordinates now have to obey
\begin{equation}
\beta_{L/R}(x_{1})\beta_{L/R}(x_{2})=q^{-1}\,\beta_{L/R}(x_{2})\beta
_{L/R}(x_{1}).
\end{equation}
This requirement again leads us to the commutation relations in (\ref{Mrel}).

From what we have done so far, we see that the quantum plane can play the role
of a left as well as a right comodule of $SL_{q}(2).$ An essential feature of
the coactions in (\ref{coact1}), (\ref{coact2}), and (\ref{coact3}) is that
they respect the spinor metric in the sense that
\begin{align}
\varepsilon_{ij}x^{i}x^{j}  &  =\varepsilon_{kl}\beta_{L/R}(x^{k})\beta
_{L/R}(x^{l}),\nonumber\\
\varepsilon^{ji}x_{i}x_{j}  &  =\varepsilon^{lk}\beta_{L/R}(x_{k})\beta
_{L/R}(x_{l}).
\end{align}
These identities are equivalent to the relations
\begin{align}
\varepsilon_{ij}  &  =\varepsilon_{kl}M^{k}{}_{i}M^{l}{}_{j},\nonumber\\
\varepsilon_{ij}  &  =\varepsilon_{kl}(M^{T})^{k}{}_{i}(M^{T})^{l}{}%
_{j},\nonumber\\
\varepsilon^{ij}  &  =\varepsilon^{kl}S^{-1}(M^{i}{}_{k}M^{j}{}_{l}%
),\nonumber\\
\varepsilon^{ij}  &  =\varepsilon^{kl}S^{-1}((M^{T})^{i}{}_{k}(M^{T})^{j}%
{}_{l}), \label{epsMrel}%
\end{align}
where in the last two identities of (\ref{epsMrel}) we can substitute $S$ for
$S^{-1}$.

\subsubsection{Reality conditions}

Now, we want to introduce the real form of $SL_{q}(2)$, which is called
$SU_{q}(2)$. Towards this end we first have to define a conjugation on the
quantum group generators. In this manner we take attention to the objects%
\begin{equation}
\overline{M^{i}{}_{j}}=\bar{M}^{i}{}_{j}=\left(
\begin{array}
[c]{cc}%
\bar{a} & \bar{b}\\
\bar{c} & \bar{d}%
\end{array}
\right)  .
\end{equation}
It arises the question what are the relations between them and how are they
related to the generators $a,$ $b,$ $c,$ and $d$.

In what follows we assume that the quantum plane is real, i.e. $\overline
{x^{i}}=x_{i}$. If we require for the quantum plane to be a left-$\ast
$-comodule algebra, i.e.
\begin{equation}
\beta_{L}(\overline{x^{i}})=\overline{\beta_{L}(x^{i})}=\overline{M^{i}{}_{j}%
}\otimes\overline{x^{j}}, \label{ConCow}%
\end{equation}
we get as reality condition for $SU_{q}(2)$:
\begin{equation}
\overline{M^{i}{}_{j}}=S^{-1}(M^{j}{}_{i}),\quad\mbox{or}\quad(M^{\dagger
})^{i}{}_{j}=S^{-1}(M^{i}{}_{j}), \label{RelCon1}%
\end{equation}
More explicitly, we have
\begin{equation}
\bar{a}=d,\quad\bar{b}=-q\,c,\quad\bar{c}=-q^{-1}\,b,\quad\bar{d}=a.
\label{konjuexpl1}%
\end{equation}
Notice that (\ref{RelCon1}) follows from (\ref{ConCow}) by exploiting the
reality of the quantum plane together with (\ref{coact3}).

However, if we demand that the quantum plane is a right comodule algebra,
i.e.
\begin{equation}
\beta_{R}(\overline{x^{i}})=\overline{\beta_{R}(x^{i})}=\overline{x^{j}%
}\otimes\overline{M^{j}{}_{i}},
\end{equation}
we are lead to a second $\ast$-structure, namely
\begin{equation}
\overline{M^{j}{}_{i}}=S(M^{i}{}_{j}),\quad\mbox{or}\quad(M^{\dagger})^{i}%
{}_{j}=S(M^{i}{}_{j}).
\end{equation}
Notice that its explicit form is most easily obtained from (\ref{konjuexpl1})
by replacing $q$ with $q^{-1}$.

Taking the identifications in (\ref{konjuexpl1}) we are able to express the
quantum group generators in (\ref{Mrel}) by the conjugated ones. This way, we
are able to find the wanted relations for the $\overline{M^{i}{}_{j}}$ and
recognize that they follow from (\ref{Mrel}) by replacing $M^{i}{}_{j}$ with
$\overline{M^{i}{}_{j}}$ and $q$ with $q^{-1}.$ Remarkably, the second $\ast
$-structure implies the same relations for the $\overline{M^{i}{}_{j}}$.
Straightforward calculations show us that both $\ast$-structures fulfill all
axioms of a Hopf-$\ast$-algebra (for these axioms see Ref. \cite{Klimyk}).
Finally, let us also mention that we have $\overline{\det_{q}M}=\det_{q}M,$ as
one expects.

\subsubsection{The $\hat{R}$-matrix of $SU_{q}(2)$}

It is often convenient to express the defining relations of a quantum group
with the help of the so-called $\hat{R}$-matrix. The explicit form of the
$\hat{R}$-matrix of $SU_{q}(2)$ is given by \cite{Wess00}%
\begin{equation}
\hat{R}^{ij}{}_{kl}=\left(
\begin{array}
[c]{cccc}%
q & 0 & 0 & 0\\
0 & \lambda & 1 & 0\\
0 & 1 & 0 & 0\\
0 & 0 & 0 & q
\end{array}
\right)  , \label{rmanin}%
\end{equation}
where the indices $i,$ $j,$ $k,$ and $l$ run over all values from the set
${\{}1,2{\}}$. The $({ij)}$-double index labels the four rows, while the
$({kl)}$-double index refers to the columns of the $\hat{R}$-matrix. With the
$\hat{R}$-matrix of $SU_{q}(2)$ the relations in (\ref{Mrel}) can be written
as
\begin{equation}
\hat{R}^{ij}{}_{kl}M^{k}{}_{r}M^{l}{}_{s}=M^{i}{}_{k}M^{j}{}_{l}\hat{R}^{kl}%
{}_{rs}. \label{MrelR}%
\end{equation}

It is also possible to formulate the commutation relations of the quantum
plane by means of the $\hat{R}$-matrix. To reach this goal we have to take
into account that the $\hat{R}$-matrix in (\ref{rmanin}) shows the projector
decomposition
\begin{equation}
\hat{R}=qS-q^{-1}A\quad\text{with}\quad\mbox{1 \kern-.59em {\rm l}}=A+S,
\label{prodecman}%
\end{equation}
where $S$ and $A$ are $q$-analogs of a symmetrizer and an antisymmetrizer,
respectively. Using the spinor metric the projector $A$ can be written as%
\begin{equation}
A^{ij}{}_{kl}=-\lambda_{+}^{-1}\varepsilon^{ij}\varepsilon_{kl}.
\label{projmaneps}%
\end{equation}

From the projector decomposition in (\ref{prodecman}) one deduces some helpful
formulae for the contraction of the $\hat{R}$-matrix with the spinor metric:%
\begin{align}
\hat{R}^{ii^{\prime}}{}_{jk}\,\varepsilon_{i^{\prime}l}  &  =q\;\varepsilon
_{ji^{\prime}}(\hat{R}^{-1})^{i^{\prime}i}{}_{kl},\nonumber\\
\hat{R}^{ij}{}_{li^{\prime}}\,\varepsilon^{i^{\prime}k}  &  =q\;\varepsilon
^{ii^{\prime}}(\hat{R}^{-1})^{jk}{}_{i^{\prime}l}. \label{Reps2}%
\end{align}
In addition to this, we have further identities which can be derived from
those in (\ref{Reps2}) via the replacements%
\begin{equation}
\hat{R}^{ij}{}_{kl}\rightarrow(\hat{R}^{-1})^{ij}{}_{kl},\qquad q\rightarrow
q^{-1}. \label{RepMan}%
\end{equation}
Further relations concerning this stuff are
\begin{equation}
R^{ij}{}_{kl}\,\varepsilon^{kl}=-q^{-1}\,\varepsilon^{ij},\qquad\hat{R}^{ij}%
{}_{kl}\,\varepsilon_{ij}=-q^{-1}\,\varepsilon_{kl}, \label{Reps1}%
\end{equation}
and two other ones that are again obtained via the replacements in
(\ref{RepMan}).

For the moment let us have a look at some identities concerning the
\textit{undeformed} spinor metric $\varepsilon_{ij}$:%
\begin{align}
\varepsilon_{ij} &  =-\varepsilon_{ji},\label{epseig1}\\
\varepsilon^{ik}\varepsilon_{kj} &  =\delta^{i}{}_{j},\label{epseig2}\\
\varepsilon^{ij}\varepsilon_{kl} &  =\delta^{i}{}_{l}\,\delta^{j}{}_{k}%
-\delta^{i}{}_{k}\,\delta^{j}{}_{l},\label{epseig3}\\
\varepsilon_{ij}\,\varepsilon_{kl} &  +\varepsilon_{il}\,\varepsilon
_{jk}+\varepsilon_{ik}\,\varepsilon_{lj}=0.\label{epseig4}%
\end{align}
It is rather instructive to compare the above identities with their
$q$-analogs. The $q$-analog of (\ref{epseig1}) is given by (\ref{Reps1}) since
the $\hat{R}$-matrix can be recognized as a kind of twist. Relation
(\ref{epseig2}), however,\ remains unchanged under q-deformation
[cf.\thinspace Eq.\thinspace(\ref{epsiseins})]. The $q$-analog of
(\ref{epseig3}) is nothing other than
\begin{equation}
\varepsilon^{ij}\varepsilon_{kl}=\hat{R}^{ij}{}_{kl}-q\,\delta^{i}{}%
_{k}\,\delta^{j}{}_{l},\label{qepseig3}%
\end{equation}
if one realizes that $\hat{R}^{ij}{}_{kl}\overset{q\rightarrow1}{\rightarrow
}\delta^{i}{}_{l}\,\delta^{j}{}_{k}$. To derive (\ref{qepseig3}) we first
eliminate the projector $S$ from the equations in (\ref{prodecman}). In doing
so, we should arrive at
\begin{equation}
A=q\lambda_{+}^{-1}\mbox{1 \kern-.59em {\rm l}}-\lambda_{+}^{-1}\hat{R},
\end{equation}
which together with (\ref{projmaneps}) implies (\ref{qepseig3}).

Finally, the deformation of (\ref{epseig4}) reads
\begin{equation}
\varepsilon_{ij}\,\varepsilon_{kl}+q^{-2}\hat{R}^{l^{\prime}k^{\prime}}{}%
_{kl}\,\hat{R}^{l^{\prime\prime}j^{\prime}}{}_{jl^{\prime}}\,\varepsilon
_{il^{\prime\prime}}\,\varepsilon_{j^{\prime}k^{\prime}}+\hat{R}^{k^{\prime
}j^{\prime}}{}_{jk}\,\hat{R}^{l^{\prime}j^{\prime\prime}}{}_{j^{\prime}%
l}\,\varepsilon_{ik^{\prime}}\,\varepsilon_{l^{\prime}j^{\prime\prime}}=0.
\label{qepseig4}%
\end{equation}
Applying the replacements $\hat{R}\rightarrow{R}^{-1}$ and $q\rightarrow
q^{-1}$ to the expressions in (\ref{qepseig3}) and (\ref{qepseig4}) we get
further $q$-analogs of (\ref{epseig3}) and (\ref{epseig4}), respectively.
Notice that (\ref{Reps2}) becomes the trivial identity $\varepsilon
_{kl}=\varepsilon_{kl}$ if $q$ tends to $1.$ Thus, it describes a true
$q$-deformed feature.

With the help of the projectors $S$ and $A$ we are able to write down the
quantum space relations (\ref{qspace1}) and (\ref{qspace2}) in a rather
compact form:
\begin{equation}
A^{ij}{}_{kl}\,x^{k}x^{l}=0,\qquad S^{ij}{}_{kl}\,\theta^{k}\theta^{l}=0.
\end{equation}
In Ref. \cite{MSW04} it was shown that these relations can also be written in
terms of the\ $\hat{R}$-matrix (\ref{rmanin}):
\begin{equation}
x^{i}x^{j}=q\hat{R}^{ij}{}_{kl}\,x^{k}x^{l},\qquad\theta^{i}\theta^{j}%
=-q^{-1}\hat{R}^{ij}{}_{kl}\,\theta^{k}\theta^{l}.
\end{equation}

Lastly, we wish to say a few words about the commutation relations between
spinor coordinates and their conjugated versions. For these commutation
relations to be covariant under the coactions (\ref{coact1}), (\ref{coact2}),
and (\ref{coact3}) they have to be of the form%
\begin{equation}
\bar{x}^{\,\bar{\imath}}x^{j}=q\hat{R}^{\bar{\imath}j}{}_{k\bar{l}}\,x^{k}%
\bar{x}^{\bar{l}}, \label{braidqplane1}%
\end{equation}
and, likewise,
\begin{equation}
\bar{\theta}^{\,\bar{\imath}}\theta^{j}=-q^{-1}\hat{R}^{\bar{\imath}j}%
{}_{k\bar{l}}\,\theta^{k}\bar{\theta}^{\bar{l}}. \label{braidqplane2}%
\end{equation}
It should be mentioned that the relations in (\ref{braidqplane1}) and
(\ref{braidqplane2}) describe the braiding between two quantum planes. For the
details we recommend the reader to consult Ref. \cite{MSW04}.

\section{Construction of higher dimensional spaces from quantum
planes\label{koppkap}}

In the following section we review the canonical construction of higher
dimensional (symmetrized) quantum spaces from quantum planes. For physical
reasons, we restrict attention to four-dimensional $q$-deformed Euclidean
space and $q$-deformed Minkowski space. The considerations in this section are
mainly based on Refs. \cite{Schlieck1}, \cite{Schlieck2}, \cite{Schlieck3},
and \cite{Schmidke}.

To construct the four-dimensional $q$-deformed Euclidean space and the
$q$-deformed Minkowski space together with their quantum groups we need two
copies of the quantum plane and two copies of the quantum group $SL_{q}(2)$.
We denote the generators of the second quantum plane by $\tilde{x}^{i}$ and
those of the second copy of $SL_{q}(2)$ by $\tilde{M}^{i}{}_{j}$. Clearly, the
relations between\ the $\tilde{M}^{i}{}_{j}$ have to be of the same form as
those between the $M^{i}{}_{j}$:
\begin{equation}
\hat{R}^{ij}{}_{kl}\,\tilde{M}^{k}{}_{r}\,\tilde{M}^{l}{}_{s}=\tilde{M}^{i}%
{}_{k}\,\tilde{M}^{j}{}_{l}\,\hat{R}^{kl}{}_{rs}. \label{Mtrel}%
\end{equation}

Next, we build the tensor product of the two quantum planes. To achieve this
there are two possibilities: either the spinor components of distinct quantum
planes commute or they are subject to nontrivial commutation relations given
by the braiding in (\ref{braidqplane1}). The first choice leads to the
four-dimensional $q$-deformed Euclidean space, while the second one
establishes $q$-Minkowski space.

Analogously, the quantum group $SO_{q}(4)$ and the $q$-deformed Lorentz group
are build on the tensor product $SL_{q}(2)\otimes SL_{q}(2)$. The relations
between the generators of the two tensor factors depend on whether we deal
with four-dimensional $q$-deformed Euclidean space or $q$-deformed Minkowski
space. The details are explained in the following two subsections.

\subsection{Four-dimensional $q$-deformed Euclidean space}

Compared to $q$-Minkowski space the $q$-deformed Euclidean space in four
dimensions has a simpler algebraic structure, so we will start with it. In
this case, as mentioned above, the coordinates of the two distinct quantum
planes commute with each other. On these grounds\ the generators of the two
copies of $SL_{q}(2)$ shall also commute with each other, i.e.
\begin{equation}
M^{i}{}_{j}\,\tilde{M}^{\tilde{k}}{}_{\tilde{l}}=\tilde{M}^{\tilde{k}}%
{}_{\tilde{l}}\,M^{i}{}_{j}. \label{MMTreleu4}%
\end{equation}

Now, we come to the tensor product of the two commuting quantum planes. This
tensor product is recognized as\ $q$-deformed Euclidean space in four
dimensions. As coordinates on this space we define%
\begin{equation}
X^{\tilde{\imath}j}\equiv\tilde{x}^{\,\tilde{\imath}}x^{j}=\tilde{x}%
^{\,\tilde{\imath}}\otimes x^{j},\qquad\tilde{\imath},j\in\{1,2\}.
\label{Cooreu41}%
\end{equation}
We can also arrange tensor factors in reversed order. In this manner, the
coordinates of $q$-deformed Euclidean space in four dimensions alternatively
become
\begin{equation}
X^{i\tilde{j}}\equiv x^{i}\tilde{x}^{\,\tilde{j}}=x^{i}\otimes\tilde
{x}^{\,\tilde{j}},\qquad i,\tilde{j}\in\{1,2\}. \label{Cooreu42}%
\end{equation}

Next, we are seeking\ the relations of the coordinates $X^{i\tilde{j}}$. To
this end, we first make reasonable ansaetze. Then we substitute the
coordinates with their tensor representations given by (\ref{Cooreu41}).
Applying relation (\ref{qspace1}) we rearrange the expressions of each tensor
factor in a way that they become normally ordered. Proceeding this way, each
ansatz can be rewritten in terms of a tensor basis. From the coefficient of
each basis element we finally read off a system of equations. Its solutions
lead us to the commutation relations (\ref{koordeu4app}) in App.\thinspace A
if we make the identifications%
\begin{equation}
X^{1}\equiv X^{\tilde{1}1},\quad X^{2}\equiv X^{\tilde{1}2},\quad X^{3}\equiv
X^{\tilde{2}1},\quad X^{4}\equiv X^{\tilde{2}2}.
\end{equation}

We now turn to the quantum group $SO_{q}(4)$. Its left coaction on coordinates
of $q$-deformed Euclidean space in four dimensions takes the form
\begin{equation}
\beta_{L}(X^{\tilde{\imath}j})=\tilde{M}^{\tilde{\imath}}{}_{\tilde{k}}%
\,M^{j}{}_{s}\otimes X^{\tilde{k}s}=:\Lambda^{\tilde{\imath}j}{}_{\tilde{k}%
s}\otimes X^{\tilde{k}s}, \label{LeftCoac1}%
\end{equation}
or, alternatively,
\begin{equation}
\beta_{L}(X^{i\tilde{j}})={M}^{i}{}_{k}\,\tilde{M}^{\tilde{j}}{}_{\tilde{s}%
}\otimes X^{k\tilde{s}}=:\Lambda^{i\tilde{j}}{}_{k\tilde{s}}\otimes
X^{k\tilde{s}}. \label{LeftCoac2}%
\end{equation}
As one expects, each tensor factor transforms under its own copy of
$SL_{q}(2)$.

In analogy to (\ref{MrelR}) the commutation relations between the quantum
group generators $\Lambda^{i\tilde{j}}{}_{k\tilde{s}}$ should determine the
$\hat{R}$-matrix for $SO_{q}(4)$:%
\begin{equation}
\Lambda^{\tilde{\imath}j}{}_{\tilde{k}l}\,\Lambda^{\tilde{\imath}^{\prime
}j^{\prime}}{}_{\tilde{k}^{\prime}l^{\prime}}\hat{R}^{(\tilde{k}l)(\tilde
{k}^{\prime}l^{\prime})}{}_{(\tilde{r}s)(\tilde{r}^{\prime}s^{\prime})}%
=\hat{R}^{(\tilde{\imath}j)(\tilde{\imath}^{\prime}j^{\prime})}{}_{(\tilde
{k}l)(\tilde{k}^{\prime}l^{\prime})}\Lambda^{\tilde{k}l}{}_{\tilde{r}%
s}\,\Lambda^{\tilde{k}^{\prime}l^{\prime}}{}_{\tilde{r}^{\prime}s^{\prime}%
}.\label{commlambda}%
\end{equation}
Since the quantum group $SO_{q}(4)$ is build on the tensor product
$SL_{q}(2)\otimes SL_{q}(2)$ the $\hat{R}$-matrix of $SO_{q}(4)$ can be
related to the $\hat{R}$-matrix of $SL_{q}(2)$. Concretely, we have%
\begin{equation}
\hat{R}^{(\tilde{\imath}j)(\tilde{\imath}^{\prime}j^{\prime})}{}_{(\tilde
{k}l)(\tilde{k}^{\prime}l^{\prime})}=q^{-1}\hat{R}^{\tilde{\imath}%
\tilde{\imath}^{\prime}}{}_{\tilde{k}\tilde{k}^{\prime}}\hat{R}^{jj^{\prime}%
}{}_{ll^{\prime}}.\label{Reu41}%
\end{equation}
To verify this identity we insert it together with%
\begin{equation}
\Lambda^{\tilde{\imath}j}{}_{\tilde{k}l}\,\Lambda^{\tilde{\imath}^{\prime
}j^{\prime}}{}_{\tilde{k}^{\prime}l^{\prime}}=(\tilde{M}^{\tilde{\imath}}%
{}_{\tilde{k}}\otimes M^{j}{}_{l})(\tilde{M}^{\tilde{\imath}^{\prime}}%
{}_{\tilde{k}^{\prime}}\otimes M^{j^{\prime}}{}_{l^{\prime}}),
\end{equation}
into Eq.\thinspace(\ref{commlambda}). Using Eqs. (\ref{MrelR}), (\ref{Mtrel}),
and (\ref{MMTreleu4}) we are then able to rearrange the left-hand side of
Eq.\thinspace(\ref{commlambda}) in such a way that it becomes identical to the
right-hand side of (\ref{commlambda}).

We can also start our considerations from the coordinates $X^{i\tilde{j}}$.
Repeating the same steps as before we get\ instead of (\ref{Reu41}) that%
\begin{equation}
\hat{R}^{({i}\tilde{j})({i}^{\prime}\tilde{j}^{\prime})}{}_{({k}\tilde{l}%
)({k}^{\prime}\tilde{l}^{\prime})}=q^{-1}\hat{R}^{{i}{i}^{\prime}}{}_{{k}%
{k}^{\prime}}\hat{R}^{\tilde{j}\tilde{j}^{\prime}}{}_{\tilde{l}\tilde
{l}^{\prime}}. \label{Reu42}%
\end{equation}
This result differs from that in (\ref{Reu41}) insofar as indices with a tilde
are interchanged with those without a tilde. However, the entries of the
$\hat{R}$-matrix of $SL_{q}(2)$ do not depend on the style of their labels. On
these grounds, we are free to identify the expressions in (\ref{Reu41}) and
(\ref{Reu42}). As a consequence, we can restrict attention to the space
spanned by the coordinates in (\ref{Cooreu41}). The results corresponding to
(\ref{Cooreu42}) then follow from those to (\ref{Cooreu41}) by replacing
indices with a tilde by indices without a tilde and vice versa. As we will see
later on, this situation is completely different if we deal with $q$-deformed
Minkowski space. Finally, let us notice that we obtain expressions for the
inverse of the $\hat{R}$-matrix of $SO_{q}(4)$ if we replace each $\hat{R}$ in
Eqs. (\ref{Reu41}) and (\ref{Reu42}) by $\hat{R}^{-1}$ and apply the
substitution $q\rightarrow q^{-1}.$

The next natural step is to consider the projector decomposition of the
$\hat{R}$-matrix of $SO_{q}(4)$ as it was done in Ref. \cite{Schlieck1}. Here,
we only state the result taken from Ref. \cite{Schlieck1}. The projector
decomposition is given by
\begin{equation}
\hat{R}=qP_{S}-q^{-1}P_{A}+q^{-3}P_{0}. \label{ProjDec}%
\end{equation}
The projectors for the $\hat{R}$-matrix of $SO_{q}(4)$ are composed of the
projectors of the $\hat{R}$-matrix of $SL_{q}(2)$:
\begin{align}
(P_{0})^{(i\tilde{j})(i^{\prime}\tilde{j}^{\prime})}{}_{(k\tilde{l}%
)(k^{\prime}\tilde{l}^{\prime})}  &  =A^{ii^{\prime}}{}_{kk^{\prime}}%
A^{\tilde{j}\tilde{j}^{\prime}}{}_{\tilde{l}\tilde{l}^{\prime}},\nonumber\\
(P_{S})^{(i\tilde{j})(i^{\prime}\tilde{j}^{\prime})}{}_{(k\tilde{l}%
)(k^{\prime}\tilde{l}^{\prime})}  &  =S^{ii^{\prime}}{}_{kk^{\prime}}%
S^{\tilde{j}\tilde{j}^{\prime}}{}_{\tilde{l}\tilde{l}^{\prime}},\nonumber\\
(P_{A})^{(i\tilde{j})(i^{\prime}\tilde{j}^{\prime})}{}_{(k\tilde{l}%
)(k^{\prime}\tilde{l}^{\prime})}  &  =A^{ii^{\prime}}{}_{kk^{\prime}}%
S^{\tilde{j}\tilde{j}^{\prime}}{}_{\tilde{l}\tilde{l}^{\prime}}+S^{ii^{\prime
}}{}_{kk^{\prime}}A^{\tilde{j}\tilde{j}^{\prime}}{}_{\tilde{l}\tilde
{l}^{\prime}}. \label{projektkoppeu4}%
\end{align}

Let us make a few comments on the meaning of these projectors. $P_{0}$
projects onto a one-dimensional space. Via relation
\begin{equation}
(P_{0})^{(i\tilde{j})(i^{\prime}\tilde{j}^{\prime})}{}_{(k\tilde{l}%
)(k^{\prime}\tilde{l}^{\prime})}=\frac{1}{C^{(s\tilde{t})(s^{\prime}\tilde
{t}^{\prime})}C_{(s\tilde{t})(s^{\prime}\tilde{t}^{\prime})}}C^{(i\tilde
{j})(i^{\prime}\tilde{j}^{\prime})}C_{(k\tilde{l})(k^{\prime}\tilde{l}%
^{\prime})}%
\end{equation}
it determines the quantum metric of the $q$-deformed four-dimensional
Euclidean space. Exploiting (\ref{projmaneps}) we find that the\ quantum
metric and its inverse respectively take the form%
\begin{equation}
C_{(i\tilde{j})(k\tilde{l})}=-\varepsilon_{ik}\varepsilon_{\tilde{j}\tilde{l}%
},\label{meteu4spin1}%
\end{equation}
and%
\begin{equation}
C^{(k\tilde{l})(i\tilde{j})}=-\varepsilon^{ki}\varepsilon^{\tilde{l}\tilde{j}%
}.\label{meteu4spin2}%
\end{equation}
$P_{S}$ is nothing other than a $q$-analog of a symmetrizer. Likewise, $P_{A}$
can be viewed as $q$-analog of an antisymmetrizer. The first summand of its
decomposition in (\ref{projektkoppeu4}) is the antiself-dual projector $P_{-}%
$, while the second one is the self-dual projector\ $P_{+}$. The
antisymmetrizer is linked to a $q$-deformed epsilon tensor:%
\begin{equation}
\varepsilon^{i\tilde{\imath}j\tilde{j}k\tilde{k}l\tilde{l}}=-q^{2}%
(\varepsilon^{ij}\varepsilon^{\tilde{\imath}\tilde{k}^{\prime}}\hat{R}%
^{\tilde{j}\tilde{k}}{}_{\tilde{k}^{\prime}\tilde{j}^{\prime}}\varepsilon
^{\tilde{j}^{\prime}\tilde{l}}\varepsilon^{kl}-\varepsilon^{\tilde{\imath
}\tilde{j}}\varepsilon^{ik^{\prime}}\hat{R}^{jk}{}_{k^{\prime}j^{\prime}%
}\varepsilon^{j^{\prime}l}\varepsilon^{\tilde{k}\tilde{l}}%
),\label{epskoppeu4N}%
\end{equation}
The above expression for the epsilon tensor remains unchanged by substituting
$\hat{R}$ with $\hat{R}^{-1}$. The prefactors in (\ref{Reu41}), (\ref{Reu42}),
and (\ref{projektkoppeu4} - \ref{epskoppeu4N}) were chosen for the sake of
simplification of the conversion formulae between spinorial and vectorial
objects [cf.\thinspace Sec.\thinspace\ref{connectchapt}].

\subsection{$q$-deformed Minkowski space}

To proceed with $q$-deformed Minkowski space we now demand that there is a
braiding between the two quantum planes. The definition of the coordinates
on\ the tensor product of the quantum planes is exactly the same as in the
case of $q$-deformed Euclidean space, i.e. Eq.\thinspace(\ref{Cooreu41})
carries over to $q$-deformed Minkowski space. The same holds for the coactions
in (\ref{LeftCoac1}) and (\ref{LeftCoac2}). As opposed to our previous
presentation we distinguish the generators of the second quantum plane by
bars. This shall remind us of the fact that the two quantum planes are
transformed into each other by conjugation, i.e.%
\begin{equation}
\overline{X^{i\bar{j}}}=\overline{x^{k}\otimes\bar{x}^{\,\bar{j}}}%
=x_{j}\otimes\bar{x}_{\bar{k}}=X_{j\bar{k}}.
\end{equation}
On these grounds, the braiding between the two quantum planes is given by
(\ref{braidqplane1}). Proceeding in the same way as described after
Eq.\thinspace(\ref{Cooreu42}), but now taking into account the braiding of the
spinor coordinates, we can derive the commutation relations for coordinates on
$q$-deformed Minkowski space.

The braiding between the two quantum planes forces the two copies of
$SL_{q}(2)$ to be braided as well. To compute the relations between the two
copies we apply the left coaction on both sides of (\ref{braidqplane1}):%
\begin{equation}
\beta_{L}(X^{\bar{\imath}j})=q\hat{R}^{\,\bar{\imath}j}{}_{{k}\bar{l}}%
\,\beta_{L}(X^{k\bar{l}}).
\end{equation}
The above equation can be recognized as covariance of the braiding
(\ref{braidqplane1}). Exploiting this kind of consistency condition finally
yields
\begin{equation}
\hat{R}^{\,\bar{\imath}j}{}_{k\bar{l}}\,M^{k}{}_{s}\tilde{M}^{\bar{l}}{}%
_{\bar{t}}=\tilde{M}^{\bar{\imath}}{}_{\bar{m}}M^{j}{}_{n}\hat{R}^{\bar{m}n}%
{}_{s\bar{t}}.
\end{equation}

By now we are able to compute the relations between the elements
$\Lambda^{\bar{\imath}j}{}_{\bar{k}l}$ of the $q$-deformed Lorentz group. Using
similar arguments as for $q$-deformed Euclidean space we can get the $\hat{R}%
$-matrix of the $q$-deformed Lorentz group\ in terms of the $\hat{R}$-matrix of
$SL_{q}(2)$. In this manner, we have%
\begin{equation}
\hat{R}^{(\bar{\imath}j)(\bar{k}l)}{}_{(\bar{k}^{\prime}l^{\prime}%
)(\bar{\imath}^{\prime}j^{\prime})}=(\hat{R}^{-1})^{j\bar{k}}{}_{\bar{\imath
}^{\prime\prime}j^{\prime\prime}}(\hat{R}^{-1})^{\bar{\imath}\bar{\imath
}^{\prime\prime}}{}_{\bar{k}^{\prime}\bar{l}^{\prime\prime}}(\hat{R}%
^{-1})^{j^{\prime\prime}l}{}_{k^{\prime\prime}j^{\prime}}\hat{R}^{\,\bar
{l}^{\prime\prime}k^{\prime\prime}}{}_{l^{\prime}\bar{\imath}^{\prime}%
},\label{Rspinmink1}%
\end{equation}
and the corresponding inverse becomes%
\begin{equation}
(\hat{R}^{-1})^{(\bar{k}^{\prime}l^{\prime})(\bar{\imath}^{\prime}j^{\prime}%
)}{}_{(\bar{\imath}j)(\bar{k}l)}=(\hat{R}^{-1})^{l^{\prime}\bar{\imath
}^{\prime}}{}_{\bar{\imath}^{\prime\prime}l^{\prime\prime}}\hat{R}^{\,\bar
{k}^{\prime}\bar{\imath}^{\prime\prime}}{}_{\bar{\imath}\bar{k}^{\prime\prime
}}\hat{R}^{l^{\prime\prime}j^{\prime}}{}_{j^{\prime\prime}l^{\prime}}\hat
{R}^{\,\bar{k}^{\prime\prime}j^{\prime\prime}}{}_{j\bar{k}}.\label{Rspinmink2}%
\end{equation}
With the $\hat{R}$-matrix of the $q$-deformed Lorentz group at hand the
commutation relations of the generators $\Lambda^{\bar{\imath}j}{}_{\bar{k}l}$
of $q$-deformed Lorentz group can be written in the form%
\begin{equation}
\Lambda^{\bar{\imath}j}{}_{\bar{k}l}\Lambda^{\bar{\imath}^{\prime}j^{\prime}%
}{}_{\bar{k}^{\prime}l^{\prime}}\hat{R}^{(\bar{k}l)(\bar{k}^{\prime}l^{\prime
})}{}_{(\bar{r}s)(\bar{r}^{\prime}s^{\prime})}=\hat{R}^{(\bar{\imath}%
j)(\bar{\imath}^{\prime}j^{\prime})}{}_{(\bar{k}l)(\bar{k}^{\prime}l^{\prime
})}\,\Lambda^{\bar{k}l}{}_{\bar{r}s}\Lambda^{\bar{k}^{\prime}l^{\prime}}%
{}_{\bar{r}^{\prime}s^{\prime}}.
\end{equation}

Nothing prevents us from starting our considerations with the coordinates
introduced in (\ref{Cooreu42}), i.e. we reverse the order in which the two
quantum planes are arranged in their tensor product. In doing so, we end up at
a second version of the $\hat{R}$-matrix for $q$-deformed Lorentz group:%
\begin{equation}
\hat{R}^{({i}\bar{j})({k}\bar{l})}{}_{({k}^{\prime}\bar{l}^{\prime}%
)({i}^{\prime}\bar{j}^{\prime})}=\hat{R}^{\bar{j}{k}}{}_{i^{\prime\prime}%
\bar{j}^{\prime\prime}}(\hat{R}^{-1})^{{i}i^{\prime\prime}}{}_{{k}^{\prime
}l^{\prime\prime}}(\hat{R}^{-1})^{\,\bar{j}^{\prime\prime}\bar{l}}{}_{\bar
{k}^{\prime\prime}\bar{j}^{\prime}}(\hat{R}^{-1})^{l^{\prime\prime}\bar
{k}^{\prime\prime}}{}_{\bar{l}^{\prime}{i}^{\prime}}, \label{Rspinmink3}%
\end{equation}
and
\begin{equation}
(\hat{R}^{-1})^{({k}^{\prime}\bar{l}^{\prime})({i}^{\prime}\bar{j}^{\prime}%
)}{}_{({i}\bar{j})({k}\bar{l})}=\hat{R}^{\,\bar{l}^{\prime}i^{\prime}}%
{}_{i^{\prime\prime}\bar{l}^{\prime\prime}}\hat{R}^{{k}^{\prime}%
i^{\prime\prime}}{}_{{i}k^{\prime\prime}}\hat{R}^{\,\bar{l}^{\prime\prime}%
\bar{j}^{\prime}}{}_{\bar{j}^{\prime\prime}\bar{l}}(\hat{R}^{-1}%
)^{k^{\prime\prime}\bar{j}^{\prime\prime}}{}_{\bar{j}{k}}. \label{Rspinmink4}%
\end{equation}
It is important to realize that the second choice for the coordinates of
$q$-deformed Minkowski space provides us with an expression for the $q$-deformed
Lorentz $\hat{R}$-matrix that is really different from (\ref{Rspinmink1}).
This observation is a consequence of the nontrivial braiding between the two
quantum planes $q$-deformed Minkowski space is made out of.

The $\hat{R}$-matrix of $q$-deformed Lorentz group obeys the projector
decomposition%
\begin{equation}
\hat{R}=q^{-2}P_{S}+q^{2}P_{T}-(P_{+}+P_{-}).\label{ProDecLor}%
\end{equation}
The expressions of the projectors can again be reduced to the projectors of
the $\hat{R}$-matrix of\ $SL_{q}(2)$, but now we have to be aware of the
nontrivial braiding between the two quantum planes. For this reason
symmetrizers and antisymmetrizers of $SL_{q}(2)$ are accompanied by $\hat{R}%
$-matrices of\ $SL_{q}(2)$:%
\begin{align}
(P_{0})^{(\bar{\imath}{j})(\bar{k}{l})}{}_{(\bar{k}^{\prime}l^{\prime}%
)(\bar{\imath}^{\prime}j^{\prime})} &  =(\hat{R}^{-1})^{{j}\bar{k}}%
{}_{i^{\prime\prime}j^{\prime\prime}}A^{\bar{\imath}i^{\prime\prime}}{}%
_{\bar{k}^{\prime}l^{\prime\prime}}A^{j^{\prime\prime}{l}}{}_{k^{\prime\prime
}j^{\prime}}\hat{R}^{l^{\prime\prime}k^{\prime\prime}}{}_{l^{\prime}%
\bar{\imath}^{\prime}},\nonumber\\
(P_{S})^{(\bar{\imath}{j})(\bar{k}{l})}{}_{(\bar{k}^{\prime}l^{\prime}%
)(\bar{\imath}^{\prime}j^{\prime})} &  =(\hat{R}^{-1})^{{j}\bar{k}}%
{}_{i^{\prime\prime}j^{\prime\prime}}S^{\bar{\imath}i^{\prime\prime}}{}%
_{\bar{k}^{\prime}l^{\prime\prime}}S^{j^{\prime\prime}{l}}{}_{k^{\prime\prime
}j^{\prime}}\hat{R}^{l^{\prime\prime}k^{\prime\prime}}{}_{l^{\prime}%
\bar{\imath}^{\prime}},\nonumber\\
(P_{A})^{(\bar{\imath}{j})(\bar{k}{l})}{}_{(\bar{k}^{\prime}l^{\prime}%
)(\bar{\imath}^{\prime}j^{\prime})} &  =(\hat{R}^{-1})^{{j}\bar{k}}%
{}_{i^{\prime\prime}j^{\prime\prime}}(A^{\bar{\imath}i^{\prime\prime}}{}%
_{\bar{k}^{\prime}l^{\prime\prime}}S^{j^{\prime\prime}{l}}{}_{k^{\prime\prime
}j^{\prime}}+S^{\bar{\imath}i^{\prime\prime}}{}_{\bar{k}^{\prime}%
l^{\prime\prime}}A^{j^{\prime\prime}{l}}{}_{k^{\prime\prime}j^{\prime}}%
)\hat{R}^{l^{\prime\prime}k^{\prime\prime}}{}_{l^{\prime}\bar{\imath}^{\prime
}}.\label{asdf1}%
\end{align}
In complete analogy to the four-dimensional Euclidean space the first summand
in the formula for the projector $P_{A}$ is the antiself-dual projector
$P_{-}$, while the second one plays the role of the self-dual
projector\ $P_{+}$.

In spinorial form the quantum metric of $q$-deformed Minkowski space now becomes%
\begin{equation}
C_{(\bar{\imath}{j})(\bar{k}{l})}=-q^{-2}\varepsilon_{\bar{\imath}\bar{\imath
}^{\prime}}(\hat{R})^{\bar{\imath}^{\prime}l^{\prime}}{}_{{j}\bar{k}%
}\varepsilon_{{l}^{\prime}l}. \label{asdf2}%
\end{equation}
With the help of (\ref{Reps2}) one readily checks that the inverse metric is
given by
\begin{equation}
C^{(\bar{\imath}{j})(\bar{k}l)}=-q^{2}\varepsilon^{\bar{\imath}\bar{\imath
}^{\prime}}(\hat{R}^{-1})^{{j}\bar{k}}{}_{\bar{\imath}^{\prime}l^{\prime}%
}\varepsilon^{l^{\prime}{l}}. \label{asdf3}%
\end{equation}
The four-dimensional epsilon tensor in a spinorial basis reads
\begin{align}
\varepsilon^{(\bar{\imath}{i})(\bar{j}{j})(\bar{k}{k})(\bar{l}{l})}=  &
q^{3}\big((\hat{R}^{-1})^{\bar{\imath}j}{}_{j^{\prime}\bar{\imath}^{\prime}%
}\varepsilon^{ij^{\prime}}\varepsilon^{\bar{\imath}^{\prime}\bar{k}%
^{\prime\prime}}(\hat{R}^{-1})^{\bar{j}\bar{k}^{\prime}}{}_{\bar{k}%
^{\prime\prime}\bar{j}^{\prime}}\varepsilon^{k^{\prime}l}\varepsilon^{\bar
{j}^{\prime}\bar{l}}\hat{R}^{k\bar{k}}{}_{\bar{k}^{\prime}k^{\prime}%
}\nonumber\\
&  -(\hat{R}^{-1})^{\bar{k}l}{}_{l^{\prime}\bar{k}^{\prime}}\varepsilon
^{ik^{\prime}}\varepsilon^{\bar{\imath}\bar{j}^{\prime}}(\hat{R}%
^{-1})^{j^{\prime}k}{}_{k^{\prime}j^{\prime\prime}}\varepsilon^{j^{\prime
\prime}l^{\prime}}\varepsilon^{\bar{k}^{\prime}\bar{l}}\hat{R}^{j\bar{j}}%
{}_{\bar{j}^{\prime}j^{\prime}}\big). \label{epskoppmink}%
\end{align}

The above formulae for projectors, $q$-deformed Minkowski metric and
four-dimensional epsilon tensor refer to (\ref{Cooreu41}) as choice for the
coordinates on $q$-deformed Minkowski space. We saw that another choice for the
Minkowski coordinates as it was given in (\ref{Cooreu42}) leads us to
different expressions for the $q$-deformed Lorentz $\hat{R}$-matrix. The same
holds for its projectors, the $q$-deformed Minkowski metric, and the
four-dimensional epsilon tensor. Luckily, this second set of expressions can
easily be derived from the formulae in Eqs. (\ref{asdf1})-(\ref{asdf3}) by
replacing $\hat{R}$ with $\hat{R}^{-1}$ and $q$ with $q^{-1},$ if unbarred
indices are always converted into barred ones and vice versa. Such a
transition rule follows from a comparison of (\ref{Rspinmink1}) and
(\ref{Rspinmink3}) [or alternatively from comparing (\ref{Rspinmink2}) with
(\ref{Rspinmink4})]. Finally, this transition rule also applies to
(\ref{epskoppmink}) with one exception: instead of the replacement $q$
$\rightarrow$ $q^{-1}$ we have $q$ $\rightarrow$ $-q^{-1}$.

We would like to conclude with some remarks about the three-dimensional
$q$-deformed Euclidean space. Its coordinates span a module algebra of the
Hopf algebra $U_{q}(su(2))$. In this respect, three-dimensional $q$-deformed
Euclidean space is more or less a subspace of $q$-deformed Minkowski space,
since the Minkowski coordinates transform according to $1/2\otimes
1/2=0\oplus1$, i.e. they contain a vector representation of $U_{q}(su(2))$.

\section{$q$-Deformed Pauli matrices\label{PaulAlg}}

\subsection{Some general remarks\label{pauli1kap}}

In this section we treat $q$-analogs of Pauli matrices in detail. Our
considerations refer to $q$-deformed Minkowski space as well as the $q$-deformed
Euclidean spaces with three and four dimensions.

Let us recall that the Pauli matrices tell us how to combine two spinors to
form a four-vector. In principle, the method to determine the entries of the
Pauli matrices is the same for all quantum spaces under consideration. In this
respect, we make an ansatz for the decomposition of the vector coordinates
$X^{\mu}$ into symmetrized spinor coordinates,
\begin{equation}
X^{\mu}=\sum_{\alpha=1,\,\dot{\beta}=1}^{2}x^{\alpha}(\sigma^{\mu}%
)_{\alpha\dot{\beta}}\bar{x}^{\dot{\beta}},\label{anspeu31}%
\end{equation}
where $(\sigma^{\mu})_{\alpha\dot{\beta}}$ denotes the entries of the the
so-called \textit{bosonic }Pauli matrices. Exchanging the spinors on the right
hand side of Eq.\thinspace(\ref{anspeu31}), we get a second possibility for
introducing Pauli matrices:%
\begin{equation}
X^{\mu}=\sum_{\dot{\alpha}=1,\,{\beta}=1}^{2}\bar{x}^{\dot{\alpha}}%
(\bar{\sigma}^{\mu})_{\dot{\alpha}{\beta}}{x}^{{\beta}}.\label{anspeu32}%
\end{equation}
The coefficients $(\bar{\sigma}^{\mu})_{\dot{\alpha}\beta}$ of this ansatz are
the so-called \textit{conjugated} Pauli matrices.

Now, we act on both sides of Eqs. (\ref{anspeu31}) and\ (\ref{anspeu32}) with
generators of a quantum algebra describing the symmetry of the quantum space
under consideration. On the left-hand side of (\ref{anspeu31}) and
(\ref{anspeu32}) the action of the symmetry generators leads to a linear
combination of the vector coordinates $X^{\mu},$ which we replace by the
ansaetze in (\ref{anspeu31}) and (\ref{anspeu32}), respectively. Since the
monomials $x^{\alpha}\bar{x}^{\dot{\beta}}$ as well as $\bar{x}^{\dot{\alpha}%
}{x}^{{\beta}}$ set up a basis, their coefficients should vanish. This
requirement give us a system of equations for computing\ \ $(\sigma^{\mu
})_{\alpha\dot{\beta}}$ or $(\bar{\sigma}^{\mu})_{\dot{\alpha}\beta}$.

This procedure also applies\ if we use antisymmetrized spinors $\theta^{i}$
and $\bar{\theta}^{i}$ in (\ref{anspeu31}) and (\ref{anspeu32}). This way, we
are lead to the conjugated and unconjugated versions of the so-called
\textit{fermionic} Pauli matrices. To sum up, we can distinguish four
different categories of Pauli matrices. It should be noted that the Pauli
matrices of the same category are unique up to a common normalization factor.
Let us denote these factors by $k_{b},$ $\bar{k}_{b},$ $k_{f},$ and $\bar
{k}_{f},$ where the labels $b$ and $f$ stand for 'bosonic' and 'fermionic',
respectively, while the bar serves to distinguish the conjugated versions from
the unconjugated ones. The ratios $k_{b}/\bar{k}_{b}$ and $k_{f}/\bar{k}_{f}$
are fixed by the commutation relations between the spinors of different
quantum planes. The reason for this lies in the fact that commuting the
spinors in (\ref{anspeu31}) should yield the expression on the right-hand side
of (\ref{anspeu32}). Finally, there remain two degrees of freedom. They can be
removed by convenient choices for the normalization of unconjugated bosonic
and unconjugated fermionic Pauli matrices, but this is a matter of taste.

\subsection{Pauli matrices for the three-dimensional $q$-deformed Euclidean
space\label{paulieu3kap}}

First of all, it is our aim to compute the explicit form of the Pauli matrices
for the three-dimensional $q$-deformed Euclidean space. This requires to know
the following commutation relations between generators of $U_{q}(su(2))$ and
spinors of a quantum plane\ \cite{Wess00}:
\begin{align}
T^{+}x^{1} &  =qx^{1}T^{+}+q^{-1}x^{2},\nonumber\\
T^{+}x^{2} &  =q^{-1}x^{2}T^{+},\nonumber\\
T^{-}x^{1} &  =qx^{1}T^{-},\nonumber\\
T^{-}x^{2} &  =q^{-1}x^{2}T^{-}+qx^{1},\nonumber\\
\tau^{3}x^{1} &  =q^{2}x^{1}\tau^{3},\nonumber\\
\tau^{3}x^{2} &  =q^{-2}x^{2}\tau^{3}.\label{rotreleu3}%
\end{align}
For the other types of spinors, i.e. $\bar{x}^{i},\theta^{i},$ and
$\bar{\theta}^{i}$, we have the same relations. (From the point of view
provided by $U_{q}(su(2))$ there is no difference between conjugated and
unconjugated coordinates. Thus, the only purpose of the bar on top of the
coordinates is to distinguish the two quantum planes from one another.) With
(\ref{rotreleu3}) at hand we are able to find the action of $U_{q}%
(su(2))$-generators on a tensor product of quantum planes. Acting on both
sides of Eq.\thinspace(\ref{anspeu31}) with the generators of $U_{q}(su(2))$
(for the vectorial and spinorial representation of $U_{q}(su(2))$ see, for
example, Ref. \cite{qliealg}) and comparing coefficients we find the explicit
form of the Pauli matrices up to normalization factors $k$ and $\bar{k},$
i.e.
\begin{gather}
(\sigma^{+})_{\alpha\dot{\beta}}=k\,q^{1/2}\lambda_{+}^{1/2}\left(
\begin{array}
[c]{cc}%
0 & 0\\
0 & 1
\end{array}
\right)  ,\quad(\sigma^{-})_{\alpha\dot{\beta}}=k\,q^{1/2}\lambda_{+}%
^{1/2}\left(
\begin{array}
[c]{cc}%
1 & 0\\
0 & 0
\end{array}
\right)  ,\nonumber\\
(\sigma^{3})_{\alpha\dot{\beta}}=k\left(
\begin{array}
[c]{cc}%
0 & q\\
1 & 0
\end{array}
\right)  ,\label{paulieu3}%
\end{gather}
and
\begin{gather}
(\bar{\sigma}^{+})_{\dot{\alpha}{\beta}}=\bar{k}\,q^{1/2}\lambda_{+}%
^{1/2}\left(
\begin{array}
[c]{cc}%
0 & 0\\
0 & 1
\end{array}
\right)  ,\quad(\bar{\sigma}^{-})_{\dot{\alpha}{\beta}}=\bar{k}\,q^{1/2}%
\lambda_{+}^{1/2}\left(
\begin{array}
[c]{cc}%
1 & 0\\
0 & 0
\end{array}
\right)  ,\nonumber\\
(\bar{\sigma}^{3})_{\dot{\alpha}{\beta}}=\bar{k}\left(
\begin{array}
[c]{cc}%
0 & q\\
1 & 0
\end{array}
\right)  .\label{paulieu3k}%
\end{gather}

Apart from a global normalization there is no difference between the
conjugated and the unconjugated matrices. This is a consequence of the fact
that all types of spinor coordinates behave under transformations of
$U_{q}(su(2))$ in the same way. The braiding for symmetrized spinors in
(\ref{braidqplane1}) requires for the Pauli matrices to fulfill%
\begin{equation}
(\bar{\sigma}^{A})_{\dot{\gamma}\delta}=q^{-1}(\hat{R}^{-1})^{\alpha\dot
{\beta}}{}_{\dot{\gamma}\delta}(\sigma^{A})_{\alpha\dot{\beta}},\qquad
A\in\{+,3-\}.
\end{equation}
As one can prove by inserting this is indeed the case if $k_{b}=q^{2}\bar
{k}_{b}.$ Likewise, the braiding for antisymmetrized spinors in
(\ref{braidqplane2}) implies $k_{f}=-\bar{k}_{f}$.

We do not discuss the three-dimensional Pauli matrices any further here,
because they are a subset of the Pauli matrices of $q$-deformed Minkowski
space. In other words, the results for\ $q$-deformed Minkowski space include
those for three-dimensional $q$-deformed Euclidean space.

\subsection{Pauli matrices for the four-dimensional $q$-deformed Euclidean
space\label{paulieu4kap}}

\subsubsection{Computation of Pauli matrices \label{sigmaeu4kap}}

The treatment of the four-dimensional $q$-deformed Euclidean space differs from
the three-dimensional one insofar as the generators of $U_{q}(so(4))$ have to
act on the two quantum planes differently. As we know, $U_{q}(so(4))$ is
algebraically isomorphic to $SL_{q}(2)\otimes SL_{q}(2)$ and the first quantum
plane carries a $(1/2,0)$-representation of $U_{q}(so(4))$. Thus, the
commutation relations between the generators of $U_{q}(so(4))$ and spinor
components of the first quantum plane read as \cite{qliealg}
\begin{align}
L_{1}^{+}x^{1}  &  =-q^{-1}x^{2}+qx^{1}L_{1}^{+},\nonumber\\
L_{1}^{+}x^{2}  &  =x^{2}L_{1}^{+},\nonumber\\[0.06in]
L_{1}^{-}x^{1}  &  =qx^{1}L_{1}^{-},\nonumber\\
L_{1}^{-}x^{2}  &  =q^{-1}x^{2}L_{1}^{-}-qx^{1},\nonumber\\[0.06in]
K_{1}x^{1}  &  =q^{-1}x^{1}K_{1},\nonumber\\
K_{1}x^{2}  &  =qx^{2}K_{1},\nonumber\\[0.06in]
L_{2}^{\pm}x^{i}  &  =x^{i}L_{2}^{\pm},\nonumber\\
K_{2}x^{i}  &  =x^{i}K_{2},\quad i\in{\{1,2\}}. \label{spinreleu4}%
\end{align}
The second quantum plane transforms under the action of $U_{q}(so(4))$ as a
$(0,1/2)$-representation. The corresponding commutation relations are obtained
from (\ref{spinreleu4}) via the interchanges
\begin{equation}
K_{1}\leftrightarrow K_{2},\qquad x^{i}\leftrightarrow\tilde{x}^{i},\qquad
L_{1}^{\pm}\rightarrow L_{2}^{\pm}.
\end{equation}
The $x^{i}$ as well as the $\tilde{x}^{i}$ are coordinates of symmetrized
quantum planes subject to (\ref{qspace1}). In complete analogy to the three
dimensional case the generators of $U_{q}(so(4))$ commute with antisymmetrized
spinors in the same way as with symmetrized ones.

Notice that the above commutation relations are compatible with the relations
of the symmetrized and the antisymmetrized quantum plane in the sense that
\begin{align}
T(x^{1}x^{2}-qx^{2}x^{1})  &  =(x^{1}x^{2}-qx^{2}x^{1})T,\\[0.1in]
T(\theta^{1}\theta^{2}+q^{-1}\theta^{2}\theta^{1})  &  =(\theta^{1}\theta
^{2}+q^{-1}\theta^{2}\theta^{1})T,\nonumber\\
T\theta^{k}\theta^{k}  &  =0,\quad k\in\{1,2\},
\end{align}
where $T$ stands for a generator of $U_{q}(so(4))$, i.e. $L_{1}^{+},$
$L_{1}^{-},$ $L_{2}^{+},$ $L_{2}^{-},$ $K_{1},$ or $K_{2}$. Moreover, the
commutation relations between generators of $U_{q}(so(4))$ and spinor
coordinates are consistent with trivial braidings between different copies of
quantum planes, i.e. we have%
\begin{equation}
x^{i}\tilde{x}^{j}=\tilde{x}^{j}x^{i},\qquad\theta^{i}\tilde{\theta}%
^{j}=-\tilde{\theta}^{j}\theta^{i}, \label{sdlfkj}%
\end{equation}
where $x^{i},$ $\tilde{x}^{i}$ denote symmetrized spinors and $\theta^{i},$
$\tilde{\theta}^{i}$ antisymmetrized ones.

Now we want to describe how to compute the Pauli matrices for the
four-dimensional $q$-deformed Euclidean space with its special features. First
we compute the Pauli matrices for symmetrized spinors. As ansatz we use
(\ref{anspeu31}). $X^{\mu}$ stands for one of the vector coordinates of
four-dimensional $q$-deformed Euclidean space, i.e. $X^{1},$ $X^{2},$ $X^{3},$
or $X^{4}.$ Their commutation relations with the generators of $U_{q}(so(4))$
are listed in the appendix [cf.\thinspace(\ref{koordeu4app})]. We apply the
generators of $U_{q}(so(4))$ to both sides of (\ref{anspeu31}) and substitute
the vector coordinates by\ (\ref{anspeu31}). Comparing coefficients we finally
get for the unconjugated bosonic Pauli matrices
\begin{align}
(\sigma^{1})_{\alpha\dot{\beta}} &  =\left(
\begin{array}
[c]{cc}%
k_{b} & 0\\
0 & 0
\end{array}
\right)  ,\quad(\sigma^{2})_{\alpha\dot{\beta}}=\left(
\begin{array}
[c]{cc}%
0 & 0\\
k_{b} & 0
\end{array}
\right)  ,\nonumber\\
(\sigma^{3})_{\alpha\dot{\beta}} &  =\left(
\begin{array}
[c]{cc}%
0 & k_{b}\\
0 & 0
\end{array}
\right)  ,\quad(\sigma^{4})_{\alpha\dot{\beta}}=\left(
\begin{array}
[c]{cc}%
0 & 0\\
0 & -k_{b}%
\end{array}
\right)  ,\label{paulieu4}%
\end{align}
with $k_{b}$ being undetermined at the moment.

Starting from the ansatz (\ref{anspeu32}) instead, we obtain the 'conjugated'
bosonic Pauli matrices:
\begin{align}
(\tilde{\sigma}^{1})_{\dot{\alpha}{\beta}} &  =\left(
\begin{array}
[c]{cc}%
\tilde{k}_{b} & 0\\
0 & 0
\end{array}
\right)  ,\quad(\tilde{\sigma}^{2})_{\dot{\alpha}{\beta}}=\left(
\begin{array}
[c]{cc}%
0 & \tilde{k}_{b}\\
0 & 0
\end{array}
\right)  ,\nonumber\\
(\tilde{\sigma}^{3})_{\dot{\alpha}{\beta}} &  =\left(
\begin{array}
[c]{cc}%
0 & 0\\
\tilde{k}_{b} & 0
\end{array}
\right)  ,\quad(\tilde{\sigma}^{4})_{\dot{\alpha}{\beta}}=\left(
\begin{array}
[c]{cc}%
0 & 0\\
0 & -\tilde{k}_{b}%
\end{array}
\right)  .\label{paulieu4k}%
\end{align}
However,\ (\ref{anspeu31}) and (\ref{anspeu32}) are not really different since
the two spinors $x^{i}$ and\ $\tilde{x}^{i}$ commute with each other
[cf.\thinspace the first identity in (\ref{sdlfkj})]. On these grounds, it
should hold
\begin{equation}
(\sigma^{\mu})_{\alpha\dot{\beta}}=(\tilde{\sigma}^{\mu})_{\dot{\beta}\alpha
}=(\tilde{\sigma}^{\mu})_{\alpha\dot{\beta}}^{T},\label{braidtriveu4}%
\end{equation}
if we set $k_{b}=\tilde{k}_{b}$.

A short look at (\ref{paulieu4}) and (\ref{paulieu4k}) tells us that the Pauli
matrices of four-dimensional $q$-deformed Euclidean space take on a rather
simple form. The reason for this lies in the fact that the vector coordinates
of four-dimensional $q$-deformed Euclidean space are given by (\ref{Cooreu41})
or (\ref{Cooreu42}) (up to additional factors). If we choose $k_{b}=1$ we
have
\begin{gather}
X^{1}=X^{1\dot{1}}=X^{\dot{1}1},\qquad X^{2}=X^{1\dot{2}}=X^{\dot{2}%
1},\nonumber\\
X^{3}=X^{2\dot{1}}=X^{\dot{1}2},\qquad X^{4}=-X^{2\dot{2}}=-X^{\dot{2}2}.
\end{gather}

We can repeat the same steps as above for antisymmetrized spinors. The only
thing we have to do is to replace symmetrized spinors in (\ref{anspeu31}) and
(\ref{anspeu32}) with antisymmetrized ones. As already mentioned both types of
spinors behave in the same way under transformations of $U_{q}(so(4)).$ Hence,
we should get the same matrices as for symmetrized spinors, apart from
different normalizations $k_{f}$ and $\tilde{k}_{f}$. Since $\theta^{i}$
anticommutes with $\tilde{\theta}^{j}$ [cf.\thinspace the second identity in
(\ref{sdlfkj})] we should have $k_{f}=-\tilde{k}_{f}$. As we will see later
on, it is not convenient to incorporate the minus sign into the 'conjugated'
Pauli matrices. In the case of antisymmetrized spinors it would be better to
redefine the 'conjugated' Pauli matrices:
\begin{equation}
X^{\mu}=-\sum_{\dot{\alpha}=1,\,\beta=1}^{2}\tilde{\theta}^{\dot{\alpha}%
}(\tilde{\sigma}^{\mu})_{\dot{\alpha}{\beta}}\theta^{{\beta}}%
.\label{fermpaulieu4}%
\end{equation}
In this manner we have $k_{f}=\tilde{k}_{f}.$ The advantage of the new
definition becomes apparent, if we realize that we are free to choose the
values for\ $k_{b}$ and $k_{f}$, so we take the convention $k_{b}=k_{f}=1.$
Now, there is no difference between the bosonic and the fermionic Pauli
matrices, thus we can restrict ourselves to the bosonic ones.

Let us recall that our Pauli matrices combine two spinors to give a vector. We
can also ask for matrices that allow to switch from vector coordinates
$X^{\mu}$ to a tensor product of two spinors. This problem leads us to the
so-called '\textit{inverse}' Pauli matrices $(\sigma_{\mu}^{-1})^{\alpha
\dot{\beta}}$. It is important to notice that these matrices are \textit{not}
inverse to the matrices $\sigma^{\mu}$ in the sense of matrix multiplication.
The terminus '\textit{inverse}' shall remind us of the fact that the matrices
$\sigma_{\mu}^{-1}$ give rise to mappings from vectorial to spinorial objects.
This way, we have%
\begin{align}
X^{\alpha\dot{\beta}} &  \equiv x^{\alpha}\tilde{x}^{\dot{\beta}}=\sum_{\mu
=1}^{4}X^{\mu}(\sigma_{\mu}^{-1})^{\alpha\dot{\beta}},\nonumber\\
X^{\dot{\alpha}{\beta}} &  \equiv\tilde{x}^{\dot{\alpha}}x^{{\beta}}=\sum
_{\mu=1}^{4}X^{\mu}(\tilde{\sigma}_{\mu}^{-1})^{\dot{\alpha}{\beta}}.
\end{align}
Remarkably, the 'inverse' Pauli matrices are identical with the Pauli matrices
in (\ref{paulieu4}) and (\ref{paulieu4k}):
\begin{equation}
(\sigma_{\mu}^{-1})^{\alpha\dot{\beta}}=(\sigma^{\mu})_{\alpha\dot{\beta}%
},\qquad(\tilde{\sigma}_{\mu}^{-1})^{\dot{\alpha}\beta}=(\tilde{\sigma}^{\mu
})_{\dot{\alpha}\beta}.\label{wertsigmainveu4}%
\end{equation}
These identifications are a characteristic feature of the $q$-deformed
Euclidean space in four dimensions, since for $q$-deformed Minkowski space the
situation is more involved, as we will see later on.

\subsubsection{Elementary properties of Pauli matrices\label{EleProp}}

The matrices $\sigma^{\mu}$ and $\sigma_{\mu}^{-1}$ as well as their
'conjugated' counterparts fulfill certain orthogonality and completeness
relations, which are a direct consequence of the definitions of the Pauli
matrices. The orthogonality relations read%
\begin{equation}
(\sigma^{\mu})_{\alpha\dot{\beta}}(\sigma_{\nu}^{-1})^{\alpha\dot{\beta}%
}=\delta_{\nu}^{\mu},\qquad(\tilde{\sigma}^{\mu})_{\dot{\alpha}\beta}%
(\tilde{\sigma}_{\nu}^{-1})^{\dot{\alpha}\beta}=\delta_{\nu}^{\mu
},\label{OrtRelPau}%
\end{equation}
and the completeness relations take the form
\begin{equation}
(\sigma^{\mu})_{\alpha\dot{\beta}}(\sigma_{\mu}^{-1})^{\alpha^{\prime}%
\dot{\beta}^{\prime}}=\delta_{\alpha}^{\alpha^{\prime}}\delta_{\dot{\beta}%
}^{\dot{\beta}^{\prime}},\qquad(\tilde{\sigma}^{\mu})_{\dot{\alpha}\beta
}(\tilde{\sigma}_{\mu}^{-1})^{\dot{\alpha}^{\prime}\beta^{\prime}}%
=\delta_{\dot{\alpha}}^{\dot{\alpha}^{\prime}}\delta_{\beta}^{\beta^{\prime}%
}.\label{ComRelPau}%
\end{equation}
In addition to this we have
\begin{equation}
(\tilde{\sigma}_{\mu}^{-1})^{\dot{\alpha}\beta}(\sigma^{\mu})_{\gamma
\dot{\delta}}=\delta_{\gamma}^{\dot{\alpha}}\,\delta_{\dot{\delta}}^{\beta
},\qquad({\sigma}_{\mu}^{-1})^{\alpha\dot{\beta}}(\tilde{\sigma}^{\mu}%
)_{\dot{\gamma}\delta}=\delta_{\dot{\gamma}}^{\alpha}\,\delta_{\delta}%
^{\dot{\beta}},
\end{equation}
as can be proven by inserting the explicit form of the Pauli matrices.

There are $q$-analogs of trace formulae for Pauli matrices\ \cite{Schlieck1}:%
\begin{align}
\mbox{Tr}_{q}\;(\sigma^{\mu}\tilde{\sigma}^{\nu}) &  \equiv\varepsilon
^{{\alpha}{\alpha}^{\prime}}({\sigma}^{\mu})_{{\alpha}\dot{\beta}}%
\varepsilon^{\dot{\beta}\dot{\beta}^{\prime}}(\tilde{\sigma}^{\nu}%
)_{\dot{\beta}^{\prime}{\alpha}^{\prime}}=-g^{\mu\nu},\nonumber\\
\mbox{Tr}_{q}\,(\tilde{\sigma}^{\mu}{\sigma}^{\nu}) &  \equiv\varepsilon
^{\dot{\alpha}\dot{\alpha}^{\prime}}(\tilde{\sigma}^{\mu})_{\dot{\alpha}%
{\beta}}\varepsilon^{{\beta}{\beta}^{\prime}}({\sigma}^{\nu})_{{\beta^{\prime
}}\dot{\alpha}^{\prime}}=-g^{\mu\nu},\label{qtraceeu40}\\[0.1in]
\mbox{Tr}_{q}\;(\sigma_{\mu}^{-1}\tilde{\sigma}_{\nu}^{-1}) &  \equiv
\varepsilon_{{\alpha}{\alpha}^{\prime}}({\sigma}_{\mu}^{-1})^{{\alpha}%
^{\prime}\dot{\beta}}\varepsilon_{\dot{\beta}\dot{\beta}^{\prime}}%
(\tilde{\sigma}_{\nu}^{-1})^{\dot{\beta}^{\prime}{\alpha}}=-g_{\mu\nu
},\nonumber\\
\mbox{Tr}_{q}\,(\tilde{\sigma}_{\mu}^{-1}{\sigma}_{\nu}^{-1}) &
\equiv\varepsilon_{\dot{\alpha}\dot{\alpha}^{\prime}}(\tilde{\sigma}_{\mu
}^{-1})^{\dot{\alpha}^{\prime}{\beta}}\varepsilon_{{\beta}{\beta}^{\prime}%
}({\sigma}_{\nu}^{-1})^{{\beta^{\prime}}\dot{\alpha}}=-g_{\mu\nu
},\label{qtraceeu41}%
\end{align}
where $g^{\mu\nu}$ denotes the metric of the four-dimensional $q$-deformed
Euclidean space (see also App.\thinspace\ref{AppA}). Notice that the relations
in (\ref{qtraceeu41}) follow from those in (\ref{qtraceeu40}) by means
of\ Eqs. (\ref{epsiloninverse}) and (\ref{wertsigmainveu4}) together with the
identity $g^{\mu\nu}=g_{\mu\nu}$.

Next, let us say some words about raising and lowering indices of Pauli
matrices. We assume that a four-vector can be build up from spinors in the
following ways:
\begin{align}
X^{\mu} &  =x^{\alpha}(\sigma^{\mu})_{\alpha\dot{\beta}}{\tilde{x}}%
^{\dot{\beta}}=x_{\alpha}(\sigma^{\mu})^{\alpha\dot{\beta}}{\tilde{x}}%
_{\dot{\beta}}\nonumber\\
&  =x_{\alpha}(\sigma^{\mu})_{\hspace{0.07in}\dot{\beta}}^{\alpha}{\tilde{x}%
}^{\dot{\beta}}=x^{\alpha}(\sigma^{\mu})_{\alpha}^{\hspace{0.07in}\dot{\beta}%
}{\tilde{x}}_{\dot{\beta}}.\label{indrai1eu4N}%
\end{align}
Since we can apply the spinor metric to raise and lower indices of spinors
[cf.\thinspace Eq.\thinspace(\ref{covco})] it follows
\begin{align}
X^{\mu} &  =\varepsilon^{\alpha\alpha^{\prime}}x_{\alpha^{\prime}}(\sigma
^{\mu})_{\alpha\dot{\beta}}\varepsilon^{\dot{\beta}\dot{\beta}^{\prime}%
}{\tilde{x}}_{\dot{\beta}^{\prime}}\nonumber\\
&  =\varepsilon^{\alpha\alpha^{\prime}}x_{\alpha^{\prime}}(\sigma^{\mu
})_{\alpha\dot{\beta}}{\tilde{x}}^{\dot{\beta}}=x^{\alpha}(\sigma^{\mu
})_{\alpha\dot{\beta}}\varepsilon^{\dot{\beta}\dot{\beta}^{\prime}}{\tilde{x}%
}_{\dot{\beta}^{\prime}}.\label{indrai2eu4}%
\end{align}
Comparing (\ref{indrai1eu4N}) with (\ref{indrai2eu4}) yields%
\begin{gather}
(\sigma^{\mu})^{\alpha\dot{\beta}}=(\sigma^{\mu})_{\alpha^{\prime}\dot{\beta
}^{\prime}}\varepsilon^{\alpha^{\prime}\alpha}\varepsilon^{\dot{\beta}%
^{\prime}\dot{\beta}},\nonumber\\
(\sigma^{\mu})_{\hspace{0.07in}\dot{\beta}}^{\alpha}=(\sigma^{\mu}%
)_{\alpha^{\prime}\dot{\beta}}\varepsilon^{\alpha^{\prime}\alpha}%
,\qquad(\sigma^{\mu})_{\alpha}^{\hspace{0.07in}\dot{\beta}}=(\sigma^{\mu
})_{\alpha^{\prime}\dot{\beta}^{\prime}}\varepsilon^{\dot{\beta}^{\prime}%
\dot{\beta}},\label{indcon1}%
\end{gather}
i.e. raising and lowering of spinor indices of the Pauli matrices $\sigma
^{\mu}$ is different to that of spinor coordinates. Similar arguments lead us
to
\begin{gather}
(\tilde{\sigma}^{\mu})^{\dot{\alpha}\beta}=(\tilde{\sigma}^{\mu})_{\dot
{\alpha}^{\prime}\beta^{\prime}}\varepsilon^{\dot{\alpha}^{\prime}\dot{\alpha
}}\varepsilon^{\beta^{\prime}\beta},\nonumber\\
(\tilde{\sigma}^{\mu})_{\hspace{0.07in}\dot{\alpha}}^{\dot{\alpha}}%
=(\tilde{\sigma}^{\mu})_{\dot{\alpha}^{\prime}\beta}\varepsilon^{\dot{\alpha
}^{\prime}\dot{\alpha}},\qquad(\tilde{\sigma}^{\mu})_{\dot{\alpha}}%
^{\hspace{0.07in}\beta}=(\tilde{\sigma}^{\mu})_{\dot{\alpha}^{\prime}%
\beta^{\prime}}\varepsilon^{\beta^{\prime}\beta}.
\end{gather}

Next, we come to the behavior of the vector index of the matrices $\sigma
^{\mu}$ and $\tilde{\sigma}_{\mu}$. It should hold%
\begin{equation}
X_{\mu}=x^{\alpha}(\sigma_{\mu})_{\alpha\dot{\beta}}{\tilde{x}}^{\dot{\beta}%
}={\tilde{x}}^{\dot{\alpha}}(\tilde{\sigma}_{\mu})_{\dot{\alpha}\beta}%
x^{\beta}.
\end{equation}
On the other hand we have
\begin{equation}
X_{\mu}(=g_{\mu\nu}X^{\nu})=g_{\mu\nu}x^{\alpha}(\sigma^{\nu})_{\alpha
\dot{\beta}}{\tilde{x}}^{\dot{\beta}}=g_{\mu\nu}{\tilde{x}}^{\dot{\alpha}%
}(\tilde{\sigma}^{\nu})_{\dot{\alpha}\beta}x^{\beta}.
\end{equation}
Thus, we can conclude that $\sigma_{\mu}=g_{\mu\nu}\sigma^{\nu}$ and
$\tilde{\sigma}_{\mu}=g_{\mu\nu}\tilde{\sigma}^{\nu}$, so the vector indices
of the Pauli matrices $\sigma^{\mu}$ and $\tilde{\sigma}^{\mu}$ are raised and
lowered in the same way as indices of vector coordinates.

The above considerations about raising and lowering of indices of the Pauli
matrices $\sigma^{\mu}$ and $\tilde{\sigma}^{\mu}$ can be adapted to 'inverse'
Pauli matrices. In doing so, we find that the spinor indices of $\sigma_{\mu
}^{-1}$ and $\tilde{\sigma}_{\mu}^{-1}$ follow the convention given in
Eq.\thinspace(\ref{covco}), but the vector indices of these matrices are now
raised according to
\begin{equation}
(\sigma^{-1})^{\mu}=\sigma_{\nu}^{-1}g^{\nu\mu},\qquad(\tilde{\sigma}%
^{-1})^{\mu}=\tilde{\sigma}_{\nu}^{-1}g^{\nu\mu}.
\end{equation}

\subsubsection{Relations for products of Pauli matrices\label{RelPro}}

Using the index conventions of the previous subsection we define matrix
multiplication of Paul matrices\textit{ }as
\begin{align}
\sigma^{\mu}\tilde{\sigma}^{\nu}  &  \equiv(\sigma^{\mu}\tilde{\sigma}^{\nu
})_{\alpha}{}^{\beta}=(\sigma^{\mu})_{\alpha}{}^{\dot{\gamma}}(\tilde{\sigma
})_{\dot{\gamma}}{}^{\beta},\nonumber\\
\tilde{\sigma}^{\mu}\sigma^{\nu}  &  \equiv(\tilde{\sigma}^{\mu}\sigma^{\nu
})_{\dot{\alpha}}{}^{\dot{\beta}}=(\tilde{\sigma})_{\dot{\alpha}}{}^{\gamma
}(\sigma^{\mu})_{\gamma}{}^{\dot{\beta}}. \label{matmulteu4}%
\end{align}
In the remainder of section \ref{paulieu4kap} it is understood that the Pauli
matrices $\sigma^{\mu}$ have the index structure $(\sigma^{\mu})_{\alpha}%
{}^{\dot{\beta}}$, if not stated otherwise. Products of $\sigma$-matrices are
then multiplied according to (\ref{matmulteu4}).

With this matrix multiplication at hand, we are now able to write down
$q$-analogs of some standard relations in a rather compact form. First,
symmetrizing the product of a Pauli matrix with a 'conjugated' one is equal to
zero, i.e.
\begin{equation}
(P_{S})^{\mu\nu}{}_{\mu^{\prime}\nu^{\prime}}\tilde{\sigma}^{\mu^{\prime}%
}\sigma^{\nu^{\prime}}=(P_{S})^{\mu\nu}{}_{\mu^{\prime}\nu^{\prime}}%
\sigma^{\mu^{\prime}}\tilde{\sigma}^{\nu^{\prime}}=0,\label{SymSigm}%
\end{equation}
where $P_{S}$ denotes the symmetric projector of the four-dimensional
Euclidean space. Its explicit form can be obtained from (\ref{projektkoppeu4})
and the general conversion formulae (see also the discussion in Sec.\thinspace
\ref{connectchapt})%
\begin{align}
T^{\mu\mu^{\prime}\ldots}{}_{\nu\nu^{\prime}\ldots} &  =(\sigma^{\mu
})_{i\tilde{j}}(\sigma^{\nu})_{i^{\prime}\tilde{j}^{\prime}}\ldots
T^{(i\tilde{j})(i^{\prime}\tilde{j}^{\prime})\ldots}{}_{(k\tilde{l}%
)(k^{\prime}\tilde{l}^{\prime})\ldots}(\sigma_{\nu}^{-1})^{k\tilde{l}}%
(\sigma_{\nu^{\prime}}^{-1})^{k^{\prime}\tilde{l}^{\prime}}\nonumber\\
&  =(\tilde{\sigma}^{\mu})_{\tilde{\imath}j}(\tilde{\sigma}^{\nu}%
)_{\tilde{\imath}^{\prime}j^{\prime}}\ldots T^{(\tilde{\imath}j)(\tilde
{\imath}^{\prime}j^{\prime})\ldots}{}_{(\tilde{k}l)(\tilde{k}^{\prime
}l^{\prime})\ldots}(\tilde{\sigma}_{\nu}^{-1})^{\tilde{k}l}(\tilde{\sigma
}_{\nu^{\prime}}^{-1})^{\tilde{k}^{\prime}l^{\prime}}.
\end{align}
The identities in (\ref{SymSigm}) become clear from the following
representation theoretic argument. The index structure of the product
$(\sigma^{\mu}\tilde{\sigma}^{\nu})_{\alpha}{}^{\beta}$ should tell us that it
behaves like a spinorial object living in the representation space
\begin{equation}
(1/2,0)\otimes(1/2,0)=(1,0)\oplus(0,0).\label{IrrBes1}%
\end{equation}
The point now is that $P_{S}$ projects onto the representation $(1,1)$ which
is not contained in (\ref{IrrBes1}).

By means of the projector decomposition (\ref{ProjDec}) and the trace formulae
of (\ref{qtraceeu40}) the identities in (\ref{SymSigm}) are equivalent to%
\begin{align}
(\sigma^{\mu}\tilde{\sigma}^{\nu})_{\alpha}{}^{\beta}  &  =-q\,\hat{R}^{\mu
\nu}{}_{\mu^{\prime}\nu^{\prime}}({\sigma}^{\mu^{\prime}}\tilde{\sigma}%
^{\nu^{\prime}})_{\alpha}{}^{\beta}+q^{-1}g^{\mu\nu}\delta_{\alpha}{}^{\beta
},\nonumber\\
(\tilde{\sigma}^{\mu}{\sigma}^{\nu})_{\dot{\alpha}}{}^{\dot{\beta}}  &
=-q\,\hat{R}^{\mu\nu}{}_{\mu^{\prime}\nu^{\prime}}(\tilde{\sigma}^{\mu
^{\prime}}{\sigma}^{\nu^{\prime}})_{\dot{\alpha}}{}^{\dot{\beta}}+q^{-1}%
g^{\mu\nu}\delta_{\dot{\alpha}}{}^{\dot{\beta}}, \label{Cliffeu4klein1}%
\end{align}
where $\hat{R}^{\mu\nu}{}_{\rho\sigma}$ is the $\hat{R}$-matrix of the
four-dimensional $q$-deformed Euclidean space. In part two of this paper the
above relations will lead us to the Clifford algebra for $q$-deformed $\gamma
$-matrices. Notice that applying the substitutions $\hat{R}\rightarrow\hat
{R}^{-1}$ and $q\rightarrow q^{-1}$ to (\ref{Cliffeu4klein1}) gives equally
good relations.

In analogy to the undeformed case, the product of a Pauli matrix with a
'conjugated' one can be decomposed into a symmetric and an antisymmetric part,
since we have%
\begin{align}
\sigma^{\mu}\tilde{\sigma}^{\nu} &  =\lambda_{+}^{-1}g^{\mu\nu}-\;q^{-2}%
\lambda_{+}^{-1}\varepsilon^{\mu\nu\rho\sigma}g_{\sigma\gamma}g_{\rho\delta
}\sigma^{\gamma}\tilde{\sigma}^{\delta},\nonumber\\
\tilde{\sigma}^{\mu}{\sigma}^{\nu} &  =\lambda_{+}^{-1}g^{\mu\nu}%
+\;q^{-2}\lambda_{+}^{-1}\varepsilon^{\mu\nu\rho\sigma}g_{\sigma\gamma}%
g_{\rho\delta}\tilde{\sigma}^{\gamma}{\sigma}^{\delta},\label{ZerAntSym}%
\end{align}
with $\varepsilon^{\mu\nu\rho\sigma}$ being the epsilon tensor for the
four-dimensional $q$-deformed Euclidean space [cf.\thinspace(\ref{epseu4app}) in
App.\thinspace\ref{AppA}]. In this way the equalities of (\ref{ZerAntSym})
reflect the decomposition in (\ref{IrrBes1}). Notice that in the first summand
on the right-hand side we omitted the Kronecker delta in spinor space.
Contracting the two vector indices by the quantum metric enables us to extract
the symmetric part contained in the products of (\ref{ZerAntSym}):
\begin{equation}
g_{\mu\nu}\sigma^{\mu}\tilde{\sigma}^{\nu}=g_{\mu\nu}\tilde{\sigma}^{\mu
}\sigma^{\nu}=\lambda_{+}\mbox{1 \kern-.59em {\rm l}}.
\end{equation}

The product of three Pauli matrices reduces to linear combinations of Pauli
matrices. The classical formulae inspired us to make the ansaetze
\begin{align}
2\,\sigma^{\mu}\tilde{\sigma}^{\nu}\sigma^{\rho}=\, &  \;k_{1}\;g^{\mu\nu
}\sigma^{\rho}+k_{2}\;g^{\nu\rho}\sigma^{\mu}\nonumber\\
&  -\;\hat{R}^{\mu\nu}{}_{\mu^{\prime}\nu^{\prime}}g^{\nu^{\prime}\rho}%
\sigma^{\mu^{\prime}}-q^{-2}\;\varepsilon^{\mu\nu\rho\sigma}g_{\sigma\lambda
}\sigma^{\lambda},\\[0.08in]
2\,\tilde{\sigma}^{\mu}{\sigma}^{\nu}\tilde{\sigma}^{\rho}=\, &
\;k_{1}\;g^{\mu\nu}\tilde{\sigma}^{\rho}+k_{2}\;g^{\nu\rho}\tilde{\sigma}%
^{\mu}\nonumber\\
&  -\;\hat{R}^{\mu\nu}{}_{\mu^{\prime}\nu^{\prime}}g^{\nu^{\prime}\rho}%
\tilde{\sigma}^{\mu^{\prime}}+q^{-2}\;\varepsilon^{\mu\nu\rho\sigma}%
g_{\sigma\lambda}\tilde{\sigma}^{\lambda}.
\end{align}
For both formulae the coefficients $k_{1},$ $k_{2}$ take as values
$k_{1}=q^{-1},$ $k_{2}=q$. Further relations are obtained from the above ones
by replacing\ $\hat{R}$ with $\hat{R}^{-1}$ and performing the interchange
$k_{1}\leftrightarrow k_{2}$.

Next, we turn to trace formulae for products of four Pauli matrices. They
should be of the form
\begin{align}
2\,\mbox{Tr}_{q}\;(\tilde{\sigma}^{\mu}\sigma^{\nu}\tilde{\sigma}^{\rho}%
\sigma^{\lambda})=  &  \;2(\tilde{\sigma}^{\mu}\sigma^{\nu}\tilde{\sigma
}^{\rho}\sigma^{\lambda})^{\dot{\alpha}}{}_{\dot{\alpha}}\nonumber\\
=  &  \;k_{1}\;g^{\mu\nu}g^{\rho\lambda}+k_{2}\;g^{\mu\lambda}g^{\nu\rho
}\nonumber\\
&  +\;g^{\mu\rho^{\prime}}\hat{R}^{\nu\rho}{}_{\rho^{\prime}\nu^{\prime}%
}g^{\nu^{\prime}\lambda}+q^{-2}\varepsilon^{\mu\nu\rho\lambda},\\[0.08in]
2\,\mbox{Tr}_{q}\;({\sigma}^{\mu}\tilde{\sigma}^{\nu}{\sigma}^{\rho}%
\tilde{\sigma}^{\lambda})=  &  \;2({\sigma}^{\mu}\tilde{\sigma}^{\nu}{\sigma
}^{\rho}\tilde{\sigma}^{\lambda})^{\alpha}{}_{\alpha}\nonumber\\
=  &  \;k_{1}\;g^{\mu\nu}g^{\rho\lambda}+k_{2}\;g^{\mu\lambda}g^{\nu\rho
}\nonumber\\
&  +\;g^{\mu\rho^{\prime}}\hat{R}^{\nu\rho}{}_{\rho^{\prime}\nu^{\prime}%
}g^{\nu^{\prime}\lambda}-q^{-2}\varepsilon^{\mu\nu\rho\lambda}.
\end{align}
Here, we find that $k_{1}=-q$, $k_{2}=-q^{-1}$ is a consistent choice. Again
the interchanges\ $\hat{R}$ $\leftrightarrow$ $\hat{R}^{-1}$ together with
$k_{1}\leftrightarrow k_{2}$ give a second set of relations.

Contraction of the four-dimensional epsilon tensor with a product of
four\ Pauli matrices finally yields the scalars%
\begin{align}
\varepsilon_{\mu\nu\rho\lambda}\sigma^{\lambda}\tilde{\sigma}^{\rho}%
\sigma^{\nu}\tilde{\sigma}^{\mu}  &  =-q^{-1}[[2]]_{q^{2}}[[3]]_{q^{2}%
}\mbox{1 \kern-.59em {\rm l}},\nonumber\\
\varepsilon_{\mu\nu\rho\lambda}\tilde{\sigma}^{\lambda}\sigma^{\rho}%
\tilde{\sigma}^{\nu}\sigma^{\mu}  &  =q^{-1}[[2]]_{q^{2}}[[3]]_{q^{2}%
}\mbox{1 \kern-.59em {\rm l}},
\end{align}
as can be proven by insertion.

At last, we wish to write down a trace expansion for an arbitrary $2\times2$
matrix $A_{\alpha}{}^{\beta}$:%
\begin{align}
A_{\dot{\alpha}}{}^{\beta}= &  \,\;\mbox{Tr}(\sigma_{1}^{-1}A)\,(\sigma
^{1})^{\beta}{}_{\dot{\alpha}}+q^{-1}\,\mbox{Tr}(\sigma_{2}^{-1}%
A)\,(\sigma^{2})^{\beta}{}_{\dot{\alpha}}\nonumber\\
&  \,+\mbox{Tr}(\sigma_{3}^{-1}A)\,(\sigma^{3})^{\beta}{}_{\dot{\alpha}%
}+\mbox{Tr}(\sigma_{4}^{-1}A)\,(\sigma^{4})^{\beta}{}_{\dot{\alpha}}.
\end{align}
Notice that we have to take the standard trace and not the quantum trace. The
above identity is a direct consequence of completeness relation
(\ref{ComRelPau}).

\subsubsection{The spin matrices $\sigma^{\mu\nu}$ and $\tilde{\sigma}^{\mu
\nu}$}

\textbf{Definition:} In the undeformed case the spin matrices are defined as
antisymmetrized products of two Pauli matrices. This definition pertains in
the $q$-deformed case. With $P_{A}$ being the antisymmetric projector of the
four-dimensional $q$-deformed Euclidean space the $q$-deformed spin matrices
take the form
\begin{align}
(\sigma^{\mu\nu})_{\alpha}{}^{\beta}  &  =(P_{A})^{\mu\nu}{}_{\kappa\lambda
}(\sigma^{\kappa}\tilde{\sigma}^{\lambda})_{\alpha}{}^{\beta},\nonumber\\
(\tilde{\sigma}^{\mu\nu})_{\dot{\alpha}}{}^{\dot{\beta}}  &  =(P_{A})^{\mu\nu
}{}_{\kappa\lambda}(\tilde{\sigma}^{\kappa}{\sigma}^{\lambda})_{\dot{\alpha}%
}{}^{\dot{\beta}}, \label{spinmeu4}%
\end{align}
and
\begin{align}
(\sigma_{\mu\nu}^{-1})_{\alpha}{}^{\beta}  &  =(P_{A})^{\kappa\lambda}{}%
_{\mu\nu}(\sigma_{\kappa}^{-1}\tilde{\sigma}_{\lambda}^{-1})_{\alpha}{}%
^{\beta},\nonumber\\
(\tilde{\sigma}_{\mu\nu}^{-1})_{\dot{\alpha}}{}^{\dot{\beta}}  &
=(P_{A})^{\kappa\lambda}{}_{\mu\nu}(\tilde{\sigma}_{\kappa}^{-1}%
\sigma_{\lambda}^{-1})_{\dot{\alpha}}{}^{\dot{\beta}}. \label{spinmeu41}%
\end{align}
Raising of indices and matrix multiplication works in exactly the same manner
as for the Pauli matrices.

We would like to list the non-vanishing matrices $\sigma^{\mu\nu}$ and
$\tilde{\sigma}^{\mu\nu},$ explicitly:%
\begin{align}
\sigma^{13}  &  =\tilde{\sigma}^{12}=\left(
\begin{array}
[c]{cc}%
0 & q^{-1}\\
0 & 0
\end{array}
\right)  , & \sigma^{14}  &  =\tilde{\sigma}^{14}=\lambda_{+}^{-1}\left(
\begin{array}
[c]{cc}%
q & 0\\
0 & -q^{-1}%
\end{array}
\right)  ,\nonumber\\[0.03in]
\sigma^{23}  &  =\tilde{\sigma}^{32}=q^{-1}\lambda_{+}^{-1}\left(
\begin{array}
[c]{cc}%
-q & 0\\
0 & q^{-1}%
\end{array}
\right)  , & \sigma^{24}  &  =\tilde{\sigma}^{34}=\left(
\begin{array}
[c]{cc}%
0 & 0\\
1 & 0
\end{array}
\right)  ,\nonumber\\[0.03in]
\sigma^{31}  &  =\tilde{\sigma}^{21}=\left(
\begin{array}
[c]{cc}%
0 & -1\\
0 & 0
\end{array}
\right)  , & \sigma^{32}  &  =\tilde{\sigma}^{23}=q\lambda_{+}^{-1}\left(
\begin{array}
[c]{cc}%
q & 0\\
0 & -q^{-1}%
\end{array}
\right)  ,\nonumber\\[0.03in]
\sigma^{41}  &  =\tilde{\sigma}^{41}=\lambda_{+}^{-1}\left(
\begin{array}
[c]{cc}%
-q & 0\\
0 & q^{-1}%
\end{array}
\right)  , & \sigma^{42}  &  =\tilde{\sigma}^{43}=\left(
\begin{array}
[c]{cc}%
0 & 0\\
-q & 0
\end{array}
\right)  .
\end{align}
The reader may realize that the two types of spin matrices differ from one
another by an interchange of the vector indices $2$ and $3$.

\textbf{Fundamental properties: }Next, we come to some fundamental properties
of the spin matrices. First of all, the spin matrices are antisymmetric in
their vector indices, i.e.
\begin{align}
(P_{A})^{\mu\nu}{}_{\mu^{\prime}\nu^{\prime}}\,\sigma^{\mu^{\prime}\nu
^{\prime}} &  =\sigma^{\mu\nu},\qquad(P_{A})^{\mu\nu}{}_{\mu^{\prime}%
\nu^{\prime}}\,\tilde{\sigma}^{\mu^{\prime}\nu^{\prime}}=\tilde{\sigma}%
^{\mu\nu},\nonumber\\
(P_{A})^{\mu^{\prime}\nu^{\prime}}{}_{\mu\nu}\,\sigma_{\mu^{\prime}\nu
^{\prime}}^{-1} &  =\sigma_{\mu\nu}^{-1},\qquad(P_{A})^{\mu^{\prime}%
\nu^{\prime}}{}_{\mu\nu}\,\tilde{\sigma}_{\mu^{\prime}\nu^{\prime}}%
^{-1}=\tilde{\sigma}_{\mu\nu}^{-1}.\label{AntSpinMat}%
\end{align}
This property is a direct consequence of the definitions in (\ref{spinmeu4})
and (\ref{spinmeu41}). We know that the projector $P_{A}$ of the
four-dimensional $q$-deformed Euclidean space splits into a self-dual and an
antiself-dual part, denoted by $P_{+}$ and $P_{-}$, respectively. $P_{-}$
projects onto the representation $(1,0).$ In the same manner, $P_{+}$
corresponds to the representation $(0,1).$ Essentially for us is the fact that
the spin matrices $\sigma^{\mu\nu}$ and $\tilde{\sigma}^{\mu\nu}$ respectively
organize mappings of the representations $(1,0)$ and $(0,1)$ onto themselves.
In this manner, we have
\begin{align}
(P_{+})^{\mu\nu}{}_{\rho\lambda}\sigma^{\rho\lambda} &  =\sigma^{\mu\nu}%
,\quad(P_{-})^{\mu\nu}{}_{\rho\lambda}\sigma^{\rho\lambda}=0,\nonumber\\
(P_{-})^{\mu\nu}{}_{\rho\lambda}\tilde{\sigma}^{\rho\lambda} &  =\tilde
{\sigma}^{\mu\nu},\quad(P_{+})^{\mu\nu}{}_{\rho\lambda}\tilde{\sigma}%
^{\rho\lambda}=0.\label{ProSig}%
\end{align}
With the remarkable identity%
\begin{equation}
\varepsilon^{\mu\nu}{}_{\rho\lambda}=g_{\rho^{\prime}\rho}g_{\lambda^{\prime
}\lambda}\varepsilon^{\mu\nu\rho^{\prime}\lambda^{\prime}}=q^{2}\lambda
_{+}((P_{+})^{\mu\nu}{}_{\rho\lambda}-(P_{-})^{\mu\nu}{}_{\rho\lambda}),
\end{equation}
we get from (\ref{ProSig}) that%
\begin{align}
\varepsilon^{\mu\nu}{}_{\rho\lambda}\sigma^{\rho\lambda} &  =q^{2}\lambda
_{+}((P_{+})^{\mu\nu}{}_{\rho\lambda}-(P_{-})^{\mu\nu}{}_{\rho\lambda}%
)\sigma^{\rho\lambda}=q^{2}\lambda_{+}\sigma^{\rho\lambda},\nonumber\\
\varepsilon^{\mu\nu}{}_{\rho\lambda}\tilde{\sigma}^{\rho\lambda} &
=q^{2}\lambda_{+}((P_{+})^{\mu\nu}{}_{\rho\lambda}-(P_{-})^{\mu\nu}{}%
_{\rho\lambda})\tilde{\sigma}^{\rho\lambda}=-q^{2}\lambda_{+}\tilde{\sigma
}^{\rho\lambda}.
\end{align}

The conjunction between spin matrices and the representations $(1,0)$ and
$(0,1)$ also shows up in the formulae%
\begin{align}
(\sigma_{\mu\nu}^{-1})^{\gamma\delta}(\sigma^{\mu\nu})_{\alpha\beta} &
=\lambda_{+}S^{\gamma\delta}{}_{\alpha\beta},\quad(\tilde{\sigma}_{\mu\nu
}^{-1})^{\dot{\gamma}\dot{\delta}}(\tilde{\sigma}^{\mu\nu})_{\dot{\alpha}%
\dot{\beta}}=\lambda_{+}S^{\dot{\gamma}\dot{\delta}}{}_{\dot{\alpha}\dot
{\beta}},\nonumber\\
(\sigma^{\mu\nu})_{\alpha\beta}(\sigma_{\mu\nu}^{-1})^{\gamma\delta} &
=\lambda_{+}S^{\gamma\delta}{}_{\alpha\beta},\quad(\tilde{\sigma}^{\mu\nu
})_{\dot{\alpha}\dot{\beta}}(\tilde{\sigma}_{\mu\nu}^{-1})^{\dot{\gamma}%
\dot{\delta}}=\lambda_{+}S^{\dot{\gamma}\dot{\delta}}{}_{\dot{\alpha}%
\dot{\beta}},\label{ContVec}%
\end{align}
and%
\begin{align}
(\sigma_{\mu\nu}^{-1})^{\gamma\delta}(\tilde{\sigma}^{\mu\nu})_{\dot{\alpha
}\dot{\beta}} &  =0,\quad(\tilde{\sigma}_{\mu\nu}^{-1})^{\dot{\gamma}%
\dot{\delta}}({\sigma}^{\mu\nu})_{{\alpha}{\beta}}=0,\nonumber\\
(\sigma^{\mu\nu})_{\alpha\beta}(\tilde{\sigma}_{\mu\nu}^{-1})^{\dot{\gamma
}\dot{\delta}} &  =0,\quad(\tilde{\sigma}^{\mu\nu})_{\dot{\alpha}\dot{\beta}%
}({\sigma}_{\mu\nu}^{-1})^{{\gamma}{\delta}}=0,
\end{align}
where $S$ denotes the symmetric projector for the quantum plane [cf.\thinspace
(\ref{prodecman})]. Notice that $S$ corresponds to the spin-$1$ representation
of $U_{q}(su(2))$, while the spinor metric corresponds to the spin-$0$
representation of $U_{q}(su(2)).$ Thus, we additionally have%
\begin{align}
(\sigma^{\mu\nu})_{\alpha\beta}\varepsilon^{\alpha\beta} &  =0,\quad
(\sigma_{\mu\nu}^{-1})^{\alpha\beta}\varepsilon_{\alpha\beta}=0,\nonumber\\
(\tilde{\sigma}^{\mu\nu})_{\dot{\alpha}\dot{\beta}}\varepsilon^{\dot{\alpha
}\dot{\beta}} &  =0,\quad(\tilde{\sigma}_{\mu\nu}^{-1})^{\dot{\alpha}%
\dot{\beta}}\varepsilon_{\dot{\alpha}\dot{\beta}}=0,\label{EpsSpiMat}%
\end{align}
which can also be written as%
\begin{equation}
\mbox{Tr}_{q}\,\sigma^{\mu\nu}=0,\quad\mbox{Tr}_{q}\,\sigma_{\mu\nu}%
^{-1}=0,\quad\mbox{Tr}_{q}\,\tilde{\sigma}^{\mu\nu}=0,\quad\mbox{Tr}_{q}%
\,\tilde{\sigma}_{\mu\nu}^{-1}=0.
\end{equation}
In other words, spin matrices are traceless and symmetric in their spinor indices.

The identities in (\ref{ContVec}) describe a kind of completeness on the space
onto which $S$ projects. In a similar fashion, but now with spinor indices
being contracted, we have%
\begin{equation}
(\sigma^{\mu\nu})_{\alpha\beta}(\sigma_{\rho\lambda}^{-1})^{\alpha\beta
}=\lambda_{+}(P_{+})^{\mu\nu}{}_{\rho\lambda},\quad(\tilde{\sigma}^{\mu\nu
})_{\dot{\alpha}\dot{\beta}}(\tilde{\sigma}_{{\rho}{\lambda}}^{-1}%
)^{\dot{\alpha}\dot{\beta}}=\lambda_{+}(P_{-})^{\mu\nu}{}_{\rho\lambda}.
\end{equation}
Once again, these relations confirm the observation that the spin matrices are
linked to the representations $(1,0)$ and\ $(0,1).$

\textbf{Relations concerning spin matrices: }There are numerous rearrangement
formulae including spin matrices that could prove useful in physical
applications. It is our aim to write down q-analogs of such formulae. (For
their undeformed counterparts we refer the reader to Refs. \cite{Wessbagger},
\cite{Core}, and \cite{Grimm}, for example.) In this manner, we found%
\begin{align}
\sigma^{\mu\nu}= &  \;-\lambda_{+}^{-1}g^{\mu\nu}+\sigma^{\mu}\tilde{\sigma
}^{\nu},\nonumber\\
\tilde{\sigma}^{\mu\nu}= &  \;-\lambda_{+}^{-1}g^{\mu\nu}+\tilde{\sigma}^{\mu
}{\sigma}^{\nu},\nonumber\\[0.08in]
\sigma^{\mu\nu}\sigma^{\lambda}= &  \;\frac{\lambda_{+}}{2}(P_{A})^{\mu\nu}%
{}_{\mu^{\prime}\nu^{\prime}}g^{\nu^{\prime}\lambda}\sigma^{\mu^{\prime}%
}-\frac{1}{2}q^{-2}\varepsilon^{\mu\nu\lambda\rho}\sigma_{\rho},\nonumber\\
\tilde{\sigma}^{\mu\nu}\tilde{\sigma}^{\lambda}= &  \;\frac{\lambda_{+}}%
{2}(P_{A})^{\mu\nu}{}_{\mu^{\prime}\nu^{\prime}}g^{\nu^{\prime}\lambda}%
\tilde{\sigma}^{\mu^{\prime}}+\frac{1}{2}q^{-2}\varepsilon^{\mu\nu\lambda\rho
}\tilde{\sigma}_{\rho},\nonumber\\[0.08in]
\tilde{\sigma}^{\mu}\sigma^{\nu\lambda}= &  \;\frac{\lambda_{+}}{2}%
(P_{A})^{\nu\lambda}{}_{\nu^{\prime}\lambda^{\prime}}g^{\mu\nu^{\prime}}%
\tilde{\sigma}^{\lambda^{\prime}}+\;\frac{1}{2}q^{-2}\varepsilon^{\mu
\nu\lambda\rho}\tilde{\sigma}_{\rho},\nonumber\\
{\sigma}^{\mu}\tilde{\sigma}^{\nu\lambda}= &  \;\frac{\;\lambda_{+}}{2}%
(P_{A})^{\nu\lambda}{}_{\nu^{\prime}\lambda^{\prime}}g^{\mu\nu^{\prime}}%
\tilde{\sigma}^{\lambda^{\prime}}-\;\frac{1}{2}q^{-2}\varepsilon^{\mu
\nu\lambda\rho}\tilde{\sigma}_{\rho},\nonumber\\[0.08in]
\sigma^{\mu\nu}\sigma^{\kappa\lambda}= &  \;\frac{1}{2}(P_{A})^{\mu\nu}{}%
_{\mu^{\prime}\nu^{\prime}}(P_{A})^{\kappa\lambda}{}_{\kappa^{\prime}%
\lambda^{\prime}}g^{\mu^{\prime}\lambda^{\prime}}g^{\nu^{\prime}\kappa
^{\prime}}-\frac{1}{2}q^{-2}\lambda_{+}^{-1}\varepsilon^{\mu\nu\kappa\lambda
}\nonumber\\
&  +\;\lambda_{+}(P_{A})^{\mu\nu}{}_{\mu^{\prime}\nu^{\prime}}(P_{A}%
)^{\kappa\lambda}{}_{\kappa^{\prime}\lambda^{\prime}}g^{\nu^{\prime}%
\kappa^{\prime}}\sigma^{\mu^{\prime}\lambda^{\prime}},\nonumber\\
\tilde{\sigma}^{\mu\nu}\tilde{\sigma}^{\kappa\lambda}= &  \;\frac{1}{2}%
(P_{A})^{\mu\nu}{}_{\mu^{\prime}\nu^{\prime}}(P_{A})^{\kappa\lambda}{}%
_{\kappa^{\prime}\lambda^{\prime}}g^{\mu^{\prime}\lambda^{\prime}}%
g^{\nu^{\prime}\kappa^{\prime}}+\frac{1}{2}q^{-2}\lambda_{+}^{-1}%
\varepsilon^{\mu\nu\kappa\lambda}\nonumber\\
&  +\;\lambda_{+}(P_{A})^{\mu\nu}{}_{\mu^{\prime}\nu^{\prime}}(P_{A}%
)^{\kappa\lambda}{}_{\kappa^{\prime}\lambda^{\prime}}g^{\nu^{\prime}%
\kappa^{\prime}}\tilde{\sigma}^{\mu^{\prime}\lambda^{\prime}}.
\end{align}

In addition to this, there are some formulae concerning contraction of vector
indices:
\begin{align}
(\sigma^{\mu\nu})_{\alpha\alpha^{\prime}}(\sigma^{\rho})_{\beta\dot{\beta}%
}\,g_{\nu\rho} &  =-\varepsilon_{\alpha^{\prime}\beta}(\sigma^{\mu}%
)_{\alpha\dot{\beta}}-\lambda_{+}^{-1}\varepsilon_{\alpha\alpha^{\prime}%
}(\sigma^{\mu})_{\beta\dot{\beta}},\nonumber\\
(\sigma^{\nu})_{\alpha\dot{\alpha}}(\tilde{\sigma}^{\rho\mu})_{\beta
\beta^{\prime}}\,g_{\nu\rho} &  =-\varepsilon_{\beta^{\prime}\alpha}%
(\sigma^{\mu})_{\beta\dot{\alpha}}-\lambda_{+}^{-1}\varepsilon_{\beta
\beta^{\prime}}(\sigma^{\mu})_{\alpha\dot{\alpha}}.\label{ProSigMet}%
\end{align}
Through the interchanges $\sigma^{\mu}\leftrightarrow\tilde{\sigma}^{\mu}$ and
$\sigma^{\mu\nu}\leftrightarrow\tilde{\sigma}^{\mu\nu}$ one gets further
relations from (\ref{ProSigMet}).
Moreover, we could verify the trace formulae
\begin{align}
\mbox{Tr}_{q}(\sigma^{\mu\nu}\sigma^{\kappa\lambda}) &  =\frac{\lambda_{+}}%
{2}(P_{A})^{\mu\nu}{}_{\mu^{\prime}\nu^{\prime}}(P_{A})^{\kappa\lambda}%
{}_{\kappa^{\prime}\lambda^{\prime}}\,g^{\mu^{\prime}\lambda^{\prime}}%
g^{\nu^{\prime}\kappa^{\prime}}-\,\frac{1}{2q^{2}}\varepsilon^{\mu\nu
\kappa\lambda},\nonumber\\
\mbox{Tr}_{q}(\tilde{\sigma}^{\mu\nu}\tilde{\sigma}^{\kappa\lambda}) &
=\frac{\lambda_{+}}{2}(P_{A})^{\mu\nu}{}_{\mu^{\prime}\nu^{\prime}}%
(P_{A})^{\kappa\lambda}{}_{\kappa^{\prime}\lambda^{\prime}}\,g^{\mu^{\prime
}\lambda^{\prime}}g^{\nu^{\prime}\kappa^{\prime}}+\,\frac{1}{2q^{2}%
}\varepsilon^{\mu\nu\kappa\lambda}.
\end{align}

Last but not least, we would like to give the following very useful formula:
\begin{align}
(\sigma^{\mu})_{\alpha\dot{\alpha}}(\sigma^{\nu})_{\beta\dot{\beta}}= &
\;k_{1}\varepsilon_{\alpha\beta}\varepsilon_{\dot{\alpha}\dot{\beta}}%
\,g^{\mu\nu}+k_{2}\,\varepsilon_{\alpha\beta}(\tilde{\sigma}^{\mu\nu}%
)_{\dot{\alpha}\dot{\beta}}+k_{3}\,\varepsilon_{\dot{\alpha}\dot{\beta}%
}(\sigma^{\mu\nu})_{\alpha\beta}\nonumber\\
&  +\,k_{4}\,(\sigma^{\mu\rho})_{\alpha\beta}\,g_{\rho\kappa}(\tilde{\sigma
}^{\kappa\nu})_{\dot{\alpha}\dot{\beta}},\label{fierzsigmaeu4neu}%
\end{align}
where the constants take on the values $k_{1}=-\lambda_{+}^{-2},$ $k_{2}%
=k_{3}=-\lambda_{+}^{-1},$ $k_{4}=-1.$ Applying the interchanges $\sigma^{\mu
}\leftrightarrow\tilde{\sigma}^{\mu}$ and $\sigma^{\mu\nu}\leftrightarrow
\tilde{\sigma}^{\mu\nu}$ yields another $q$-deformed Fierz identity. Notice that
the formula in (\ref{fierzsigmaeu4neu}) describes how a tensor product of two
four-vectors decomposes into its irreducible constituents.

\subsection{Pauli matrices for the $q$-deformed Minkowski
space\label{paulimink1}}

\subsubsection{Computation of Pauli matrices}

From a physical point of view $q$-deformed Minkowski space is one of the most
important quantum spaces. In this section, we will consider its Pauli and spin
matrices in some detail. Once again, we begin with explicit calculation of
Pauli matrices. This requires to know\ how the generators of the $q$-deformed
Lorentz algebra commute with the spinors $x^{\alpha}$ and $\bar{x}%
^{\dot{\alpha}}.$ In the first case, we have \cite{SWZ91}%
\begin{align}
\tau^{1}x^{1}  &  =x^{1}\tau^{1},\nonumber\\
\tau^{1}x^{2}  &  =x^{2}\tau^{1}-q\lambda^{2}x^{1}T^{2},\nonumber\\[0.06in]
T^{2}x^{1}  &  =q^{-1}x^{1}T^{2},\nonumber\\
T^{2}x^{2}  &  =qx^{2}T^{2},\nonumber\\[0.06in]
S^{1}x^{1}  &  =qx^{1}S^{1},\nonumber\\
S^{1}x^{2}  &  =q^{-1}x^{2}S^{1}-x^{1}\sigma^{2},\nonumber\\[0.06in]
\sigma^{2}x^{1}  &  =x^{1}\sigma^{2},\nonumber\\
\sigma^{2}x^{2}  &  =x^{2}\sigma^{2},
\end{align}
where the elements $\tau^{1},$ $\sigma^{2},$ $T^{2},$ $S^{1}$ together with
the generators of $U_{q}(su(2))$ span the $q$-deformed Lorentz algebra. For
conjugated spinors the corresponding commutation relations read as
\begin{align}
\tau^{1}\bar{x}^{1}  &  =q^{-1}\bar{x}^{1}\tau^{1},\nonumber\\
\tau^{1}\bar{x}^{2}  &  =q\bar{x}^{2}\tau^{1},\nonumber\\[0.06in]
T^{2}\bar{x}^{1}  &  =\bar{x}^{1}T^{2}+q^{-1}\bar{x}^{2}\tau^{1},\nonumber\\
T^{2}\bar{x}^{2}  &  =\bar{x}^{2}T^{2},\nonumber\\[0.06in]
S^{1}\bar{x}^{1}  &  =\bar{x}^{1}S^{1},\nonumber\\
S^{1}\bar{x}^{2}  &  =\bar{x}^{2}S^{1},\nonumber\\[0.06in]
\sigma^{2}\bar{x}^{1}  &  =q\bar{x}^{1}\sigma^{2}+\lambda^{2}\bar{x}^{2}%
S^{1},\nonumber\\
\sigma^{2}\bar{x}^{2}  &  =q^{-1}\bar{x}^{2}\sigma^{2}.
\end{align}
We have not written down the relations of the generators of $U_{q}(su(2))$
with the spinors $x^{\alpha}$ and $\bar{x}^{\dot{\alpha}},$ since they take
the same form as in (\ref{rotreleu3}). These commutation relations make the
quantum planes to the spinors $x^{\alpha}$ and $\bar{x}^{\dot{\alpha}}$ into
module algebras of the $q$-deformed Lorentz algebra. The same assertion is valid
if we consider the quantum planes to the antisymmetrized spinors
$\theta^{\alpha}$ and $\bar{\theta}^{\dot{\alpha}}.$

The commutation relations between Lorentz generators and spinors are
completely determined by the following two requirements. First, the Lorentz
generators have to commute with the relations of the symmetrized and the
antisymmetrized quantum plane, i.e.
\begin{gather}
T(x^{1}x^{2}-qx^{2}x^{1})=(x^{1}x^{2}-qx^{2}x^{1})T,\nonumber\\
T(\bar{x}^{1}\bar{x}^{2}-q\bar{x}^{2}\bar{x}^{1})=(\bar{x}^{1}\bar{x}%
^{2}-q\bar{x}^{2}\bar{x}^{1})T,\nonumber\\[0.06in]
T(\theta^{1}\theta^{2}+q^{-1}\theta^{2}\theta^{1})=(\theta^{1}\theta
^{2}+q^{-1}\theta^{2}\theta^{1})T,\nonumber\\
T(\bar{\theta}^{1}\bar{\theta}^{2}+q^{-1}\bar{\theta}^{2}\bar{\theta}%
^{1})=(\bar{\theta}^{1}\bar{\theta}^{2}+q^{-1}\bar{\theta}^{2}\bar{\theta}%
^{1})T,\nonumber\\
T^{i}\theta^{\alpha}\theta^{\alpha}=T\bar{\theta}^{\dot{\alpha}}\bar{\theta
}^{\dot{\alpha}}=0,\quad\alpha,\dot{\alpha}\in\{1,2\},
\end{gather}
where $T$ stands for a generator from the set $\{T^{+},$ $T^{-},$ $\tau^{3},$
$\tau^{1},$ $\sigma^{2},$ $S^{1},$ $T^{2}\}.$ Second, the commutation
relations between Lorentz generators and spinors should respect the braiding
between spinors and their conjugates [cf.\thinspace Eqs. (\ref{braidqplane1})
and (\ref{braidqplane2})]:%
\begin{align}
T(\bar{x}^{\dot{\alpha}}{x}^{\beta}-q\hat{R}^{\dot{\alpha}\beta}{}%
_{\beta^{\prime}\dot{\alpha}^{\prime}}x^{\beta^{\prime}}\bar{x}^{\dot{\alpha
}^{\prime}}) &  =(\bar{x}^{\dot{\alpha}}{x}^{\beta}-q\hat{R}^{\dot{\alpha
}\beta}{}_{\beta^{\prime}\dot{\alpha}^{\prime}}x^{\beta^{\prime}}\bar{x}%
^{\dot{\alpha}^{\prime}})T,\nonumber\\
T(\bar{\theta}^{\dot{\alpha}}\theta^{\beta}+q^{-1}\hat{R}^{\dot{\alpha}\beta
}{}_{\beta^{\prime}\dot{\alpha}^{\prime}}\theta^{\beta^{\prime}}\bar{\theta
}^{\dot{\alpha}^{\prime}}) &  =(\bar{\theta}^{\dot{\alpha}}\theta^{\beta
}+q^{-1}\hat{R}^{\dot{\alpha}\beta}{}_{\beta^{\prime}\dot{\alpha}^{\prime}%
}\theta^{\beta^{\prime}}\bar{\theta}^{\dot{\alpha}^{\prime}})T.
\end{align}

Following the line of reasonings described in sections \ref{pauli1kap} and
\ref{sigmaeu4kap} we calculate the explicit form of the Pauli matrices for
$q$-deformed Minkowski space. In doing so, we find the matrices%
\begin{align}
(\sigma^{+})_{\alpha\dot{\beta}} &  =k\left(
\begin{array}
[c]{cc}%
0 & 0\\
0 & q^{1/2}\lambda_{+}^{1/2}%
\end{array}
\right)  , & (\sigma^{3})_{\alpha\dot{\beta}} &  =k\left(
\begin{array}
[c]{cc}%
0 & q\\
1 & 0
\end{array}
\right)  ,\nonumber\\[0.03in]
(\sigma^{-})_{\alpha\dot{\beta}} &  =k\left(
\begin{array}
[c]{cc}%
q^{1/2}\lambda_{+}^{1/2} & 0\\
0 & 0
\end{array}
\right)  , & (\sigma^{0})_{\alpha\dot{\beta}} &  =k\left(
\begin{array}
[c]{cc}%
0 & -q^{-1}\\
1 & 0
\end{array}
\right)  ,
\end{align}
and their conjugated counterparts%
\begin{align}
(\bar{\sigma}^{+})_{\alpha\dot{\beta}} &  =\bar{k}\left(
\begin{array}
[c]{cc}%
0 & 0\\
0 & q^{-1/2}\lambda_{+}^{1/2}%
\end{array}
\right)  , & (\bar{\sigma}^{3})_{\alpha\dot{\beta}} &  =\bar{k}\left(
\begin{array}
[c]{cc}%
0 & 1\\
q^{-1} & 0
\end{array}
\right)  ,\nonumber\\[0.03in]
(\bar{\sigma}^{-})_{\alpha\dot{\beta}} &  =\bar{k}\left(
\begin{array}
[c]{cc}%
q^{-1/2}\lambda_{+}^{1/2} & 0\\
0 & 0
\end{array}
\right)  , & (\bar{\sigma}^{0})_{\alpha\dot{\beta}} &  =\bar{k}\left(
\begin{array}
[c]{cc}%
0 & 1\\
-q & 0
\end{array}
\right)  .
\end{align}
Again, our results depend on the parameters $k\in{\{}k_{b},k_{f}\}$ and
$\bar{k}=\{\bar{k}_{b},\bar{k}_{f}{\}.}$ Their values are restricted by the
braidings between spinors and their conjugates. For symmetrized and
antisymmetrized spinors these braidings are respectively given by
(\ref{braidqplane1}) and (\ref{braidqplane2}). As was already mentioned, the
relations describing the braiding between spinors and their conjugates enable
us to transform the two sets of $\sigma$-matrices into each other. In this
manner, we obtain by means of (\ref{braidqplane1}) and (\ref{braidqplane2})
that
\begin{equation}
(\sigma_{b}^{\mu})_{\alpha\dot{\beta}}=q\,(\sigma_{f}^{\mu})_{\alpha\dot
{\beta}},\qquad(\bar{\sigma}_{b}^{\mu})_{\dot{\alpha}{\beta}}=q^{-1}%
\,(\bar{\sigma}_{f}^{\mu})_{\dot{\alpha}{\beta}},\label{bosferm}%
\end{equation}
which, in turn, implies
\begin{equation}
\bar{k}_{b}=q^{-1}k_{b},\qquad\bar{k}_{f}=qk_{f}.
\end{equation}
Notice that for the fermionic Pauli matrices we used definition
(\ref{fermpaulieu4}). As we will see later on, the choices
\begin{equation}
k_{b}=q^{1/2}\lambda_{+}^{-1/2},\qquad k_{f}=q^{-1/2}\lambda_{+}%
^{-1/2}\label{ConvBosFer}%
\end{equation}
lead to simplifications in what follows.

Now, we write down the bosonic Pauli matrices in their final form:%
\begin{align}
(\sigma^{+})_{\alpha\dot{\beta}}  &  =\left(
\begin{array}
[c]{cc}%
0 & 0\\
0 & q
\end{array}
\right)  , & (\sigma^{3})_{\alpha\dot{\beta}}  &  =q\lambda_{+}^{-1/2}\left(
\begin{array}
[c]{cc}%
0 & q^{1/2}\\
q^{-1/2} & 0
\end{array}
\right)  ,\nonumber\\[0.03in]
(\sigma^{-})_{\alpha\dot{\beta}}  &  =\left(
\begin{array}
[c]{cc}%
q & 0\\
0 & 0
\end{array}
\right)  , & (\sigma^{0})_{\alpha\dot{\beta}}  &  =\lambda_{+}^{-1/2}\left(
\begin{array}
[c]{cc}%
0 & -q^{-1/2}\\
q^{1/2} & 0
\end{array}
\right)  . \label{PauExpBos1}%
\end{align}
Likewise, the conjugated Pauli matrices are:%
\begin{align}
(\bar{\sigma}^{+})_{\dot{\alpha}\beta}  &  =\left(
\begin{array}
[c]{cc}%
0 & 0\\
0 & q^{-1}%
\end{array}
\right)  , & (\bar{\sigma}^{3})_{\dot{\alpha}\beta}  &  =q^{-1}\lambda
_{+}^{-1/2}\left(
\begin{array}
[c]{cc}%
0 & q^{1/2}\\
q^{-1/2} & 0
\end{array}
\right)  ,\nonumber\\[0.03in]
(\bar{\sigma}^{-})_{\dot{\alpha}\beta}  &  =\left(
\begin{array}
[c]{cc}%
q^{-1} & 0\\
0 & 0
\end{array}
\right)  , & (\bar{\sigma}^{0})_{\dot{\alpha}\beta}  &  =\lambda_{+}%
^{-1/2}\left(
\begin{array}
[c]{cc}%
0 & q^{-1/2}\\
-q^{1/2} & 0
\end{array}
\right)  . \label{PauExpBos2}%
\end{align}
In the remainder of this section will concentrate our attention on the bosonic
Pauli matrices.

A short look at (\ref{anspeu31}) and\ (\ref{anspeu32}) tells us that the Pauli
matrices give us the decomposition of a four-vector into a tensor product of
spinors. With (\ref{PauExpBos1}) and (\ref{PauExpBos2}) we find
\begin{align}
X^{+}  &  =qx^{2}\bar{x}^{\dot{2}},\nonumber\\
X^{3}  &  =\lambda_{+}^{-1/2}(q^{3/2}x^{1}\bar{x}^{\dot{2}}+q^{1/2}x^{2}%
\bar{x}^{\dot{1}}),\nonumber\\
X^{0}  &  =\lambda_{+}^{-1/2}(-q^{-1/2}x^{1}\bar{x}^{\dot{2}}+q^{1/2}x^{2}%
\bar{x}^{\dot{1}}),\nonumber\\
X^{-}  &  =qx^{1}\bar{x}^{\dot{1}}, \label{umkopp1}%
\end{align}
and
\begin{align}
X^{+}  &  =q^{-1}\bar{x}^{\dot{1}}{x}^{1},\nonumber\\
X^{3}  &  =\lambda_{+}^{-1/2}(q^{-1/2}\bar{x}^{\dot{1}}{x}^{2}+q^{-3/2}\bar
{x}^{\dot{2}}{x}^{{1}}),\nonumber\\
X^{0}  &  =\lambda_{+}^{-1/2}(q^{-1/2}\bar{x}^{\dot{1}}{x}^{{2}}-q^{1/2}%
\bar{x}^{\dot{2}}{x}^{{1}}),\nonumber\\
X^{-}  &  =q^{-1}\bar{x}^{\dot{1}}{x}^{{1}}. \label{umkopp2}%
\end{align}

It is our next aim to invert the equations in (\ref{umkopp1}) and
(\ref{umkopp2}) in order to express tensor products of two spinor components
by\ vector components. In doing so, we obtain\ relations from which we can
read off the explicit form of the 'inverse' Pauli matrices:%
\begin{equation}
X^{\alpha\dot{\beta}}=X^{\mu}(\sigma_{\mu}^{-1})^{\alpha\dot{\beta}},\qquad
X^{\dot{\alpha}\beta}=X^{\mu}(\bar{\sigma}_{\mu}^{-1})^{\dot{\alpha}\beta},
\end{equation}
where
\begin{align}
(\sigma_{+}^{-1})^{\alpha\dot{\beta}} &  =\left(
\begin{array}
[c]{cc}%
0 & 0\\
0 & q^{-1}%
\end{array}
\right)  , & (\sigma_{3}^{-1})^{\alpha\dot{\beta}} &  =q^{-1}\lambda
_{+}^{-1/2}\left(
\begin{array}
[c]{cc}%
0 & q^{1/2}\\
q^{-1/2} & 0
\end{array}
\right)  ,\nonumber\\[0.03in]
(\sigma_{-}^{-1})^{\alpha\dot{\beta}} &  =\left(
\begin{array}
[c]{cc}%
q^{-1} & 0\\
0 & 0
\end{array}
\right)  , & (\sigma_{0}^{-1})^{\alpha\dot{\beta}} &  =\lambda_{+}%
^{-1/2}\left(
\begin{array}
[c]{cc}%
0 & -q^{-1/2}\\
q^{1/2} & 0
\end{array}
\right)  ,
\end{align}
and
\begin{align}
(\bar{\sigma}_{+}^{-1})^{\dot{\alpha}\beta} &  =\left(
\begin{array}
[c]{cc}%
0 & 0\\
0 & q
\end{array}
\right)  , & (\bar{\sigma}_{3}^{-1})^{\dot{\alpha}\beta} &  =q\lambda
_{+}^{-1/2}\left(
\begin{array}
[c]{cc}%
0 & q^{1/2}\\
q^{-1/2} & 0
\end{array}
\right)  ,\nonumber\\[0.03in]
(\bar{\sigma}_{-}^{-1})^{\dot{\alpha}\beta} &  =\left(
\begin{array}
[c]{cc}%
q & 0\\
0 & 0
\end{array}
\right)  , & (\bar{\sigma}_{0}^{-1})^{\dot{\alpha}\beta} &  =\lambda
_{+}^{-1/2}\left(
\begin{array}
[c]{cc}%
0 & q^{-1/2}\\
-q^{1/2} & 0
\end{array}
\right)  .
\end{align}
As in Sec.\thinspace\ref{sigmaeu4kap}, the reader should not confuse the
terminus 'inverse' with being inverse in the sense of matrix multiplication.
If we consider antisymmetrized spinors we have to deal with inverse Pauli
matrices that are connected with the bosonic ones via
\begin{equation}
\sigma_{f,\mu}^{-1}=q\,\sigma_{b,\mu}^{-1},\qquad(\bar{\sigma}_{f,\mu}%
^{-1})=q^{-1}\,\bar{\sigma}_{b,\mu}^{-1}.\label{bosferm2}%
\end{equation}

\subsubsection{Elementary properties of Pauli matrices}

From a direct inspection of the expressions for Pauli matrices one can read
off the following rules between the entries of Pauli matrices and their
conjugated counterparts:%
\begin{equation}
(\sigma^{\mu})_{\alpha\dot{\beta}}=(-1)^{\delta_{0\mu}}(\bar{\sigma}_{\mu
})^{\dot{\alpha}\beta},\qquad(\bar{\sigma}^{\mu})_{\dot{\alpha}\beta}%
^{-1}=(-1)^{\delta_{0\mu}}(\sigma_{\mu}^{-1})^{\alpha\dot{\beta}},
\end{equation}
where $\delta_{0\mu}$ denotes a Kronecker delta. These rules correspond to the
conjugation properties of four-vectors given by%
\begin{equation}
\overline{X^{\mu}}=(-1)^{\delta_{0\mu}}X_{\mu}.
\end{equation}

Now we come to some orthogonality and completeness relations for our matrices.
Orthogonality becomes%
\begin{equation}
(\sigma^{\mu})_{\alpha\dot{\beta}}(\sigma_{\nu}^{-1})^{\alpha\dot{\beta}%
}=\delta^{\mu}{}_{\nu},\qquad(\bar{\sigma}^{\mu})_{\dot{\alpha}\beta}%
(\bar{\sigma}_{\nu}^{-1})^{\dot{\alpha}\beta}=\delta^{\mu}{}_{\nu}.
\end{equation}
For completeness we have
\begin{equation}
(\sigma^{\mu})_{\alpha\dot{\beta}}(\sigma_{\mu}^{-1})^{\alpha^{\prime}%
\dot{\beta}^{\prime}}=\delta_{\alpha}^{\alpha^{\prime}}\delta_{\dot{\beta}%
}^{\dot{\beta}^{\prime}},\qquad(\bar{\sigma}^{\mu})_{\dot{\alpha}\beta}%
(\bar{\sigma}_{\mu}^{-1})^{\dot{\alpha}^{\prime}\beta^{\prime}}=\delta
_{\dot{\alpha}}^{\dot{\alpha}^{\prime}}\delta_{\beta}^{\beta^{\prime}}.
\end{equation}

Applying relation (\ref{braidqplane1}) to formula (\ref{anspeu31}) we are able
to commute spinors with their conjugates and should obtain formula
(\ref{anspeu32}). For this to be true we must have%
\begin{align}
(\bar{\sigma}^{\mu})_{\dot{\gamma}\delta}  &  =q^{-1}(\hat{R}^{-1}%
)^{\alpha\dot{\beta}}{}_{\dot{\gamma}\delta}(\sigma^{\mu})_{\alpha\dot{\beta}%
},\nonumber\\
({\sigma^{\mu}})_{\gamma\dot{\delta}}  &  =q\hat{R}^{\dot{\alpha}\beta}%
{}_{\gamma\dot{\delta}}(\bar{\sigma}^{\mu})_{\dot{\alpha}\beta},
\label{braidmin}%
\end{align}
if $\hat{R}$ stands for the $\hat{R}$-matrix of $U_{q}(su(2))$. From the
orthonormality and completeness relations we then conclude that for the
inverse Pauli matrices it holds
\begin{align}
(\bar{\sigma}_{\mu}^{-1})_{\dot{\gamma}\delta}  &  =q\hat{R}^{\alpha\dot
{\beta}}{}_{\dot{\gamma}\delta}(\sigma_{\mu}^{-1})_{\alpha\dot{\beta}%
},\nonumber\\
({\sigma_{\mu}^{-1}})_{\gamma\dot{\delta}}  &  =q^{-1}(\hat{R}^{-1}%
)^{\dot{\alpha}\beta}{}_{\gamma\dot{\delta}}(\bar{\sigma}_{\mu}^{-1}%
)_{\dot{\alpha}\beta},
\end{align}
and
\begin{align}
(\bar{\sigma}_{\mu}^{-1})^{\dot{\alpha}\beta}(\sigma^{\mu})_{\gamma\dot
{\delta}}  &  =q(\hat{R})^{\dot{\alpha}\beta}{}_{\gamma\dot{\delta}%
},\nonumber\\
({\sigma}_{\mu}^{-1})^{\alpha\dot{\beta}}(\bar{\sigma}^{\mu})_{\dot{\gamma
}\delta}  &  =q^{-1}(\hat{R}^{-1})^{\alpha\dot{\beta}}{}_{\dot{\gamma}\delta}.
\end{align}
The corresponding formulae for fermionic Pauli matrices follow from exploiting
relations (\ref{bosferm}) and (\ref{bosferm2}).

In very much the same way as was done for the four-dimensional $q$-deformed
Euclidean space we introduce a quantum trace for $q$-deformed Minkowski space.
This quantum trace taken for certain products of two $q$-deformed Pauli
matrices can be expressed in terms of the quantum metric of $q$-deformed
Minkowski space:%
\begin{align}
\mbox{Tr}_{q}\,(\sigma^{\mu}\bar{\sigma}^{\nu})  &  \equiv\varepsilon
^{\alpha\alpha^{\prime}}(\sigma^{\mu})_{\alpha\dot{\beta}}\varepsilon
^{\dot{\beta}\dot{\beta}^{\prime}}(\bar{\sigma}^{\nu})_{\dot{\beta}^{\prime
}\alpha^{\prime}}=-\eta_{\mu\nu},\nonumber\\
\mbox{Tr}_{q}\,(\bar{\sigma}^{\mu}{\sigma}^{\nu})  &  \equiv\varepsilon
^{\dot{\alpha}\dot{\alpha}^{\prime}}(\bar{\sigma}^{\mu})_{\dot{\alpha}{\beta}%
}\varepsilon^{{\beta}{\beta}^{\prime}}({\sigma}^{\nu})_{{\beta}^{\prime}%
\dot{\alpha}^{\prime}}=-\eta_{\mu\nu},\label{trsigma}\\
\mbox{Tr}_{q}\,(\sigma_{\mu}^{-1}\bar{\sigma}_{\nu}^{-1})  &  \equiv
\varepsilon_{\alpha\alpha^{\prime}}(\sigma_{\mu}^{-1})^{\alpha\dot{\beta}%
}\varepsilon_{\dot{\beta}\dot{\beta}^{\prime}}(\bar{\sigma}_{\nu}^{-1}%
)^{\dot{\beta}^{\prime}\alpha^{\prime}}=-\eta_{\mu\nu},\nonumber\\
\mbox{Tr}_{q}\,(\bar{\sigma}_{\mu}^{-1}{\sigma}_{\nu}^{-1})  &  \equiv
\varepsilon_{\dot{\alpha}\dot{\alpha}^{\prime}}(\bar{\sigma}_{\mu}^{-1}%
)^{\dot{\alpha}{\beta}}\varepsilon_{{\beta}{\beta}^{\prime}}({\sigma}_{\nu
}^{-1})^{{\beta}^{\prime}\dot{\alpha}^{\prime}}=-\eta_{\mu\nu}.
\label{trsigmainv}%
\end{align}

The Pauli matrices for $q$-deformed Minkowski space were introduced in very
much the same way as was done for the four dimensional $q$-deformed Euclidean
space. Thus, the considerations about rasing and lowering of indices in
Sec.\thinspace\ref{EleProp} carry over to $q$-deformed Minkowski space with
some minor modifications. Once again, we have%
\begin{align}
X^{\mu} &  =x^{\alpha}(\sigma^{\mu})_{\alpha\dot{\beta}}{\bar{x}}^{\dot{\beta
}}=x_{\alpha}(\sigma^{\mu})^{\alpha\dot{\beta}}{\bar{x}}_{\dot{\beta}%
}\nonumber\\
&  =x_{\alpha}(\sigma^{\mu})_{\hspace{0.07in}\dot{\beta}}^{\alpha}{\bar{x}%
}^{\dot{\beta}}=x^{\alpha}(\sigma^{\mu})_{\alpha}^{\hspace{0.07in}\dot{\beta}%
}{\bar{x}}_{\dot{\beta}}.\label{IndPauMin1}%
\end{align}
Lowering indices of spinor coordinates according to Eqs. (\ref{covco}) and
(\ref{manincon1}) it follows
\begin{align}
X^{\mu} &  =\varepsilon^{\alpha\alpha^{\prime}}x_{\alpha^{\prime}}(\sigma
^{\mu})_{\alpha\dot{\beta}}(-\varepsilon^{\dot{\beta}\dot{\beta}^{\prime}%
}{\bar{x}}_{\dot{\beta}^{\prime}})\nonumber\\
&  =\varepsilon^{\alpha\alpha^{\prime}}x_{\alpha^{\prime}}(\sigma^{\mu
})_{\alpha\dot{\beta}}\,{\bar{x}}^{\dot{\beta}}=x^{\alpha}(\sigma^{\mu
})_{\alpha\dot{\beta}}(-\varepsilon^{\dot{\beta}\dot{\beta}^{\prime}}{\bar{x}%
}_{\dot{\beta}^{\prime}}).\label{IndPauMin2}%
\end{align}
Comparing (\ref{IndPauMin1}) with (\ref{IndPauMin2}) we obtain the rules for
raising the spinorial indices of Pauli matrices of $q$-deformed Minkowski
space. Similar considerations hold for the conjugated Pauli matrices. In this
manner, we should arrive at%
\begin{gather}
(\sigma^{\mu})^{\alpha\dot{\beta}}=-(\sigma^{\mu})_{\alpha^{\prime}\dot{\beta
}^{\prime}}\varepsilon^{\alpha^{\prime}\alpha}\varepsilon^{\dot{\beta}%
^{\prime}\dot{\beta}},\nonumber\\
(\sigma^{\mu})_{\hspace{0.07in}\dot{\beta}}^{\alpha}=(\sigma^{\mu}%
)_{\alpha^{\prime}\dot{\beta}}\varepsilon^{\alpha^{\prime}\alpha}%
,\qquad(\sigma^{\mu})_{\alpha}^{\hspace{0.07in}\dot{\beta}}=-(\sigma^{\mu
})_{\alpha^{\prime}\dot{\beta}^{\prime}}\varepsilon^{\dot{\beta}^{\prime}%
\dot{\beta}},\label{indcon1N}%
\end{gather}
and
\begin{gather}
(\bar{\sigma}^{\mu})^{\dot{\alpha}\beta}=-(\bar{\sigma}^{\mu})_{\dot{\alpha
}^{\prime}\beta^{\prime}}\varepsilon^{\dot{\alpha}^{\prime}\dot{\alpha}%
}\varepsilon^{\beta^{\prime}\beta},\nonumber\\
(\bar{\sigma}^{\mu})_{\hspace{0.07in}\dot{\alpha}}^{\dot{\alpha}}%
=-(\bar{\sigma}^{\mu})_{\dot{\alpha}^{\prime}\beta}\varepsilon^{\dot{\alpha
}^{\prime}\dot{\alpha}},\qquad(\bar{\sigma}^{\mu})_{\dot{\alpha}}%
^{\hspace{0.07in}\beta}=(\bar{\sigma}^{\mu})_{\dot{\alpha}^{\prime}%
\beta^{\prime}}\varepsilon^{\beta^{\prime}\beta}.\label{indcon2N}%
\end{gather}
Contrary to the Euclidean case we have to take care of a minus sign if a
dotted index is lowered. The reason for this lies in the fact that spinors
with dotted indices are now subject to (\ref{manincon1}). Finally, the Lorentz
index is raised and lowered by applying the $q$-deformed Minkowski metric
$\eta$ [cf.\thinspace(\ref{metricminkapp})], i.e. we have
\begin{align}
\sigma^{\mu} &  =\eta^{\mu\nu}\sigma_{\nu},\qquad\sigma_{\mu}=\eta_{\mu\nu
}\sigma^{\nu},\nonumber\\
\bar{\sigma}^{\mu} &  =\eta^{\mu\nu}\bar{\sigma}_{\nu},\qquad\bar{\sigma}%
_{\mu}=\eta_{\mu\nu}\bar{\sigma}^{\nu}.
\end{align}

Now, we come to the 'inverse' Pauli matrices. Their spinorial indices behave
in the same way as the spinors do in Eqs. (\ref{covco}) and (\ref{manincon1}),
but their Lorentz indices obey the rules%
\begin{align}
\sigma_{\mu}^{-1}  &  =(\sigma^{-1})^{\nu}\eta_{\nu\mu},\qquad(\sigma
^{-1})^{\mu}=\sigma_{\nu}^{-1}\eta^{\nu\mu},\nonumber\\
\bar{\sigma}_{\mu}^{-1}  &  =(\bar{\sigma}^{-1})^{\nu}\eta_{\nu\mu}%
,\qquad(\bar{\sigma}^{-1})^{\mu}=\bar{\sigma}_{\nu}^{-1}\eta^{\nu\mu}.
\end{align}

\subsubsection{Relations for products of Pauli matrices}

Now we will present $q$-analogs to some well-known relations concerning
products of bosonic Pauli matrices. If not stated otherwise it is always
assumed that Pauli matrices have the following index structure:%
\begin{align}
(\sigma^{\mu})_{\alpha}{}^{\dot{\beta}}  &  =-(\sigma^{\mu})_{\alpha
\dot{\gamma}}\varepsilon^{\dot{\gamma}\dot{\beta}},\qquad(\bar{\sigma}^{\mu
})_{\dot{\alpha}}{}^{{\beta}}=(\bar{\sigma}^{\mu})_{\dot{\alpha}{\gamma}%
}\varepsilon^{{\gamma}{\beta}},\label{defsigmamink1}\\
(\sigma_{\mu}^{-1})_{\alpha}{}^{\dot{\beta}}  &  =\varepsilon_{\alpha\gamma
}(\sigma_{\mu}^{-1})^{\gamma\dot{\beta}},\qquad(\bar{\sigma}_{\mu}^{-1}%
)_{\dot{\alpha}}{}^{\beta}=-\varepsilon_{\dot{\alpha}\dot{\gamma}}(\bar
{\sigma}_{\mu}^{-1})^{\dot{\gamma}\beta}. \label{defsigmamink2}%
\end{align}
Moreover, multiplication of Pauli matrices is understood as
\begin{align}
\sigma^{\mu}\bar{\sigma}^{\nu}  &  \equiv(\sigma^{\mu}\bar{\sigma}^{\nu
})_{\alpha}{}^{\beta}=(\sigma^{\mu})_{\alpha}{}^{\dot{\gamma}}(\bar{\sigma
}^{\nu})_{\dot{\gamma}}{}^{{\beta}},\nonumber\\
\bar{\sigma}^{\mu}{\sigma}^{\nu}  &  \equiv(\bar{\sigma}^{\mu}\sigma^{\nu
})_{\dot{\alpha}}{}^{\dot{\beta}}=(\bar{\sigma}^{\mu})_{\dot{\alpha}}%
{}^{{\gamma}}({\sigma}^{\nu})_{{\gamma}}{}^{\dot{\beta}},\\[0.08in]
\sigma_{\mu}^{-1}\bar{\sigma}_{\nu}^{-1}  &  \equiv(\sigma_{\mu}^{-1}%
\bar{\sigma}_{\nu}^{-1})_{\alpha}{}^{\beta}=(\sigma_{\mu}^{-1})_{\alpha}%
{}^{\dot{\gamma}}(\bar{\sigma}_{\nu}^{-1})_{\dot{\gamma}}{}^{\beta
},\nonumber\\
\bar{\sigma}_{\mu}^{-1}{\sigma}_{\nu}^{-1}  &  \equiv(\bar{\sigma}_{\mu}%
^{-1}{\sigma}_{\nu}^{-1})_{\dot{\alpha}}{}^{\dot{\beta}}=(\bar{\sigma}_{\mu
}^{-1})_{\dot{\alpha}}{}^{{\gamma}}({\sigma}_{\nu}^{-1})_{{\gamma}}{}%
^{\dot{\beta}}. \label{matrixmultmink1}%
\end{align}

The reasonings leading to the following relations are in complete analogy to
those in Sec.\thinspace\ref{RelPro}. Thus, we can restrict ourselves to
stating the results, only. Symmetrizing a product of Pauli matrices in its
Lorentz indices gives zero:
\begin{equation}
(P_{S})^{\mu\nu}{}_{\mu^{\prime}\nu^{\prime}}\bar{\sigma}^{\mu^{\prime}}%
\sigma^{\nu^{\prime}}=(P_{S})^{\mu\nu}{}_{\mu^{\prime}\nu^{\prime}}\sigma
^{\mu^{\prime}}\bar{\sigma}^{\nu^{\prime}}=0.
\end{equation}
Together with the projector decomposition (\ref{ProDecLor}) and the trace
formulae in (\ref{trsigma}) the last identities imply%
\begin{align}
\sigma^{\mu}\bar{\sigma}^{\nu} &  =-\,\hat{R}^{\mu\nu}{}_{\mu^{\prime}%
\nu^{\prime}}{\sigma}^{\mu^{\prime}}\bar{\sigma}^{\nu^{\prime}}-q\,\eta
^{\mu\nu},\nonumber\\
\bar{\sigma}^{\mu}{\sigma}^{\nu} &  =-\,\hat{R}^{\mu\nu}{}_{\mu^{\prime}%
\nu^{\prime}}\bar{\sigma}^{\mu^{\prime}}{\sigma}^{\nu^{\prime}}-q\,\eta
^{\mu\nu}.\label{Cliffmink1}%
\end{align}
Notice that in the above formulae we are allowed to replace $\hat{R}$ with
$\hat{R}^{-1}$ and at the same time $q$ with $q^{-1}$.

The product of two Pauli matrices splits into an antisymmetric and a symmetric
part. In this sense we found%
\begin{align}
\sigma^{\mu}\bar{\sigma}^{\nu} &  =-\;\lambda_{+}^{-1}\eta^{\mu\nu
}\mbox{1 \kern-.59em {\rm l}}+\;q^{2}\lambda_{+}^{-1}\varepsilon^{\mu\nu
\rho\sigma}\eta_{\sigma\gamma}\eta_{\rho\delta}\sigma^{\gamma}\bar{\sigma
}^{\delta},\nonumber\\
\bar{\sigma}^{\mu}{\sigma}^{\nu} &  =-\;\lambda_{+}^{-1}\eta^{\mu\nu
}\mbox{1 \kern-.59em {\rm l}}-\;q^{2}\lambda_{+}^{-1}\varepsilon^{\mu\nu
\rho\sigma}\eta_{\sigma\gamma}\eta_{\rho\delta}\bar{\sigma}^{\gamma}{\sigma
}^{\delta},\label{minksigmazerleg}%
\end{align}
where $\varepsilon^{\mu\nu\rho\sigma}$ denotes the epsilon tensor of
$q$-deformed Minkowski space (for its non-vanishing components see
App.\thinspace\ref{AppA}). Similar decompositions exist for products of
'inverse' Pauli matrices. They are obtained from (\ref{minksigmazerleg}) by
applying\ the replacements
\begin{equation}
\sigma^{\mu}\rightarrow\bar{\sigma}_{\mu}^{-1},\qquad\eta^{\mu\nu}%
\rightarrow\eta_{\mu\nu},\qquad\varepsilon^{\mu\nu\rho\sigma}\rightarrow
\varepsilon_{\sigma\rho\nu\mu}.
\end{equation}
If we contract the Lorentz indices in (\ref{minksigmazerleg}) with a
$q$-deformed Minkowski metric and realize that%
\begin{equation}
\eta^{\mu\nu}\eta_{\mu\nu}=\lambda_{+}^{2},\qquad\eta_{\mu\nu}\varepsilon
^{\mu\nu\rho\sigma}=0,
\end{equation}
we arrive at the identities
\begin{equation}
\eta_{\mu\nu}\sigma^{\mu}\bar{\sigma}^{\nu}=\eta_{\mu\nu}\bar{\sigma}^{\mu
}{\sigma^{\nu}}=-\lambda_{+}\mbox{1 \kern-.59em {\rm l}}.\label{minkcon2s}%
\end{equation}

Products of three Pauli matrices can be decomposed as follows:
\begin{align}
2\,\sigma^{\mu}\bar{\sigma}^{\nu}\sigma^{\rho}=  &  \;k_{1}\,\eta^{\mu\nu
}\sigma^{\rho}+k_{2}\,\eta^{\nu\rho}\sigma^{\mu}\nonumber\\
&  +\;k_{3}\,\hat{R}^{\mu\nu}{}_{\mu^{\prime}\nu^{\prime}}\eta^{\nu^{\prime
}\rho}\sigma^{\mu^{\prime}}-{q^{2}}\varepsilon^{\mu\nu\rho\sigma}\eta
_{\sigma\lambda}\sigma^{\lambda}\nonumber\\
2\,\bar{\sigma}^{\mu}{\sigma}^{\nu}\bar{\sigma}^{\rho}=  &  \;k_{1}\,\eta
^{\mu\nu}\bar{\sigma}^{\rho}+k_{2}\,\eta^{\nu\rho}\bar{\sigma}^{\mu
},\nonumber\\
&  +\;k_{3}\,\hat{R}^{\mu\nu}{}_{\mu^{\prime}\nu^{\prime}}\eta^{\nu^{\prime
}\rho}\bar{\sigma}^{\mu^{\prime}}+{q^{2}}\varepsilon^{\mu\nu\rho\sigma}%
\eta_{\sigma\lambda}\bar{\sigma}^{\lambda}. \label{sigma3mink}%
\end{align}
In both formulae the coefficients $k_{i}$ take on the values $k_{1}=-q,$
$k_{2}=-q^{-1},$ $k_{3}=q$. The decompositions in (\ref{sigma3mink}) can also
be written with $\hat{R}^{-1}$ instead of $\hat{R}$. If we do so, the
coefficients $k_{i}$ become $k_{1}=-q^{-1},$ $k_{2}=-q,$ $k_{3}=q^{-1}$.

Next, we come to formulae for the quantum trace of products with four Pauli
matrices:
\begin{align}
2\,\mbox{Tr}_{q}\;{\sigma}^{\mu}\bar{\sigma}^{\nu}{\sigma}^{\rho}\bar{\sigma
}^{\lambda}=  &  \;k_{1}\,\eta^{\mu\nu}\eta^{\rho\lambda}-k_{2}\,\eta
^{\mu\lambda}\eta^{\nu\rho}\nonumber\\
&  +\;k_{3}\,\eta^{\mu\rho^{\prime}}\hat{R}^{\nu\rho}{}_{\rho^{\prime}%
\nu^{\prime}}\eta^{\nu^{\prime}\lambda}+{q^{2}}\varepsilon^{\mu\nu\rho\lambda
},\nonumber\\
2\,\mbox{Tr}_{q}\;\bar{\sigma}^{\mu}\sigma^{\nu}\bar{\sigma}^{\rho}%
\sigma^{\lambda}=  &  \;k_{1}\,\eta^{\mu\nu}\eta^{\rho\lambda}+k_{2}%
\,\eta^{\mu\lambda}\eta^{\nu\rho}\nonumber\\
&  +\;k_{3}\,\eta^{\mu\rho^{\prime}}\hat{R}^{\nu\rho}{}_{\rho^{\prime}%
\nu^{\prime}}\eta^{\nu^{\prime}\lambda}-{q^{2}}\varepsilon^{\mu\nu\rho\lambda
}. \label{spurenminksigma}%
\end{align}
For this to be true we have to insert $k_{1}=q^{-1},$ $k_{2}=q,$ $k_{3}=-q$.
Substituting $\hat{R}^{-1}$ for $\hat{R}$ the values of the coefficients
$k_{i}$ instead become $k_{1}=q,$ $k_{2}=q^{-1},$ $k_{3}=-q^{-1}$.

Last but not least, we would like to give formulae for contracting four Pauli
matrices with the epsilon tensor of $q$-deformed Minkowski space:
\begin{align}
\varepsilon_{\mu\nu\rho\lambda}\sigma^{\lambda}\bar{\sigma}^{\rho}\sigma^{\nu
}\bar{\sigma}^{\mu}  &  =-q^{-5}[[2]]_{q^{2}}[[3]]_{q^{3}}%
\mbox{1 \kern-.59em {\rm l}},\nonumber\\
\varepsilon_{\mu\nu\rho\lambda}\bar{\sigma}^{\lambda}{\sigma}^{\rho}%
\bar{\sigma}^{\nu}{\sigma}^{\mu}  &  =q^{-5}[[2]]_{q^{2}}[[3]]_{q^{3}%
}\mbox{1 \kern-.59em {\rm l}}. \label{4sigmacontractmink}%
\end{align}

There exist variants of the\ formulae in (\ref{minkcon2s}%
)-(\ref{4sigmacontractmink}) that use the 'inverse' Pauli matrices
$\sigma_{\mu}^{-1}$ and $\bar{\sigma}_{\mu}^{-1}$ instead of $\sigma^{\mu}$
and $\bar{\sigma}^{\mu}$. They are easily derived from (\ref{minkcon2s}%
)-(\ref{4sigmacontractmink}) by performing the replacements
\begin{gather}
\sigma^{\mu}\leftrightarrow\sigma_{\mu}^{-1},\qquad\bar{\sigma}^{\mu
}\leftrightarrow\bar{\sigma}_{\mu}^{-1},\nonumber\\
\eta^{\mu\nu}\leftrightarrow\eta_{\mu\nu},\qquad\varepsilon^{\mu\nu\rho\sigma
}\leftrightarrow-\varepsilon_{\sigma\rho\nu\mu},\nonumber\\
\hat{R}^{\mu\nu}{}_{\mu^{\prime}\nu^{\prime}}\leftrightarrow\hat{R}%
^{\mu^{\prime}\nu^{\prime}}{}_{\mu\nu},\qquad(\hat{R}^{-1})^{\mu\nu}{}%
_{\mu^{\prime}\nu^{\prime}}\leftrightarrow(\hat{R}^{-1})^{\mu^{\prime}%
\nu^{\prime}}{}_{\mu\nu}.\label{sigmainvrules}%
\end{gather}

It should be mentioned that each\ $2\times2$ matrix $A_{\alpha}{}^{\beta}$ can
be expanded in terms of Pauli\ matrices:
\begin{align}
A_{\dot{\alpha}}{}^{\beta}= &  \;\lambda_{+}^{-1/2}\mbox{Tr}(A)\,(\sigma
^{0})^{\beta}{}_{\dot{\alpha}}+\mbox{Tr}\,(\sigma_{+}^{-1}A)\,(\sigma
^{+})^{\beta}{}_{\dot{\alpha}}\nonumber\\
&  +\mbox{Tr}\,(\sigma_{3}^{-1}A)\,(\sigma^{3})^{\beta}{}_{\dot{\alpha}%
}+\mbox{Tr}\,(\sigma_{-}^{-1}A)\,(\sigma^{-})^{\beta}{}_{\dot{\alpha}}.
\end{align}
There is also an expansion in terms of the conjugated Pauli matrices. It
follows from the above formulae via the substitutions $\sigma\rightarrow
\bar{\sigma}$ and $\lambda_{+}^{-1/2}\rightarrow-\lambda_{+}^{-1/2}$.

Let us end with the following remark. Now, the advantage of the convention
(\ref{ConvBosFer}) for bosonic and fermionic Pauli matrices should become
obvious. It is responsible for the fact that a product of a Pauli matrix with
a conjugated one gives the same result for\ the bosonic and the fermionic
case. Thus, all of the above relations also hold for fermionic Pauli matrices.

\subsubsection{Additional relations for the matrices $\sigma^{+},$ $\sigma
^{3},$ $\sigma^{-}$ \label{eu3reladd}}

As already remarked in Sec.\thinspace\ref{paulieu3kap} the Pauli matrices for
the three-dimensional $q$-deformed Euclidean space constitute a subset of the
Pauli matrices of $q$-deformed Minkowski space. Thus, there is a rather simple
method to obtain relations for Pauli matrices of the three-dimensional
$q$-deformed Euclidean space: in the results of $q$-deformed Minkowski space we
omit all relations with indices corresponding to the time element. In doing
so, it suffices to deal with the matrices $\sigma^{A},$ $A\in\{+,3,-\}$, since
the 'conjugated' Pauli matrices $\bar{\sigma}^{A},$ $A\in\{+,3,-\}$ differ
from them by a global factor, only:%
\begin{equation}
\bar{\sigma}_{b}^{A}=q^{-1}\sigma_{b}^{A},\qquad\bar{\sigma}_{f}^{A}%
=\sigma_{f}^{A}.
\end{equation}
Similar reasonings hold for 'inverse' Pauli matrices.

The completeness and orthogonality relations are
\eqa
(\sigma_A^{-1})^{\alpha\beta}(\sigma^A)_{\gamma\delta}&=&S^{\alpha\beta}{}_{\gamma\delta},\nn
(\sigma^A)_{\alpha\beta}(\sigma_B^{-1})^{\alpha\beta}&=&\delta^A{}_B.
\ena

For the undeformed counterparts of the following relations we refer the reader
to Ref. \cite{Core} (see pp. 4 and 5 therein). First some simple identities:
\begin{align}
\sigma^{+}\sigma^{3}  &  =-q^{-2}\sigma^{3}\sigma^{+}=\lambda_{+}^{-1/2}%
\sigma^{+},\nonumber\\
\sigma^{-}\sigma^{3}  &  =-q^{-2}\sigma^{3}\sigma^{-}=\lambda_{+}^{-1/2}%
\sigma^{-},\nonumber\\
\sigma^{+}\sigma^{-}  &  =-q^{3}\,\lambda_{+}^{-1}%
\mbox{1 \kern-.59em {\rm l}}-q\,\lambda_{+}^{-1/2}\sigma^{3},\nonumber\\
\sigma^{-}\sigma^{+}  &  =-q\,\lambda_{+}^{-1}%
\mbox{1 \kern-.59em {\rm l}}+q\,\lambda_{+}^{-1/2}\sigma^{3},\nonumber\\
(\sigma^{+}\sigma^{-})^{n}  &  =(-1)^{n-1}q^{2(n-1)}\sigma^{+}\sigma
^{-},\nonumber\\
(\sigma^{-}\sigma^{+})^{n}  &  =(-1)^{n-1}q^{2(n-1)}\sigma^{-}\sigma^{+}.
\end{align}

The Pauli matrices now obey the Clifford algebra
\begin{align}
\sigma^{A}\sigma^{B} &  =-q^{4}\hat{R}^{AB}{}_{A^{\prime}B^{\prime}}%
\sigma^{A^{\prime}}\sigma^{B^{\prime}}+q\,g^{AB},\nonumber\\
\sigma^{A}\sigma^{B} &  =-q^{-4}(\hat{R}^{-1})^{AB}{}_{A^{\prime}B^{\prime}%
}\sigma^{A^{\prime}}\sigma^{B^{\prime}}+q^{3}\,g^{AB},
\end{align}
where $\hat{R}$ and $g$ respectively denote the $\hat{R}$-matrix and the
quantum metric of the three-dimensional $q$-deformed Euclidean space
\cite{LWW97}.

The formulae for decomposing products of Pauli matrices reduce to
\begin{align}
\sigma^{A}\sigma^{B}= &  \;q^{3}\lambda_{+}^{-1/2}\varepsilon^{ABC}%
g_{CC^{\prime}}\sigma^{C^{\prime}}+q^{2}\lambda_{+}^{-1}g^{AB}%
\mbox{1 \kern-.59em {\rm l}},\nonumber\\
\sigma^{A}\sigma^{B}\sigma^{C}= &  \;k_{1}\,g^{AB}\sigma^{C}+k_{2}%
\,g^{BC}\sigma^{A}\nonumber\\
&  -\,k_{2}\,\hat{R}^{AB}{}_{A^{\prime}B^{\prime}}g^{B^{\prime}C}%
\sigma^{A^{\prime}}+q^{5}\,\lambda_{+}^{-3/2}\varepsilon^{ABC}%
\mbox{1 \kern-.59em {\rm l}},
\end{align}
where the coefficients $k_{i}$ are given by $k_{1}=\lambda_{+}^{-1}%
,\;k_{2}=q^{4}\lambda_{+}^{-1}$. Notice that $\varepsilon^{ABC}$ stands for
the three-dimensional $q$-deformed epsilon tensor \cite{qliealg, LWW97} (see
also App.\thinspace\ref{AppA}). Applying the replacements $\hat{R}%
\rightarrow\hat{R}^{-1}$ and $k_{1}\leftrightarrow k_{2}$ to the formula for
the product of three Pauli matrices yields the same identity.

The trace formulae now become%
\begin{align}
\mbox{Tr}_{q}\,\sigma^{A}\sigma^{B}=  &  \;-q^{-2}g^{AB},\nonumber\\
\mbox{Tr}_{q}\,\sigma^{A}\sigma^{B}\sigma^{C}=  &  \;-q^{5}\lambda_{+}%
^{-1/2}\varepsilon^{ABC},\nonumber\\
\mbox{Tr}_{q}\,\sigma^{A}\sigma^{B}\sigma^{C}\sigma^{D}=  &  \;k_{1}%
\,g^{AB}g^{CD}+k_{2}\,g^{AD}g^{BC}\nonumber\\
&  -k_{1}\,g^{AC^{\prime}}\hat{R}^{BC}{}_{C^{\prime}B^{\prime}}g^{B^{\prime}%
D},
\end{align}
with $k_{1}=-q^{6}\lambda_{+}^{-1},$ $k_{2}=-q^{2}\lambda_{+}^{-1}$. The last
relation can alternatively be written down in a form obtained by the
interchanges $\hat{R}\rightarrow\hat{R}^{-1}$ and $k_{1}\leftrightarrow k_{2}$.

There are some Fierz identities concerning the three-dimensional Pauli
matrices:
\eqa
g_{AB}(\sigma^A)_{\alpha\beta}(\sigma^B)_{\gamma\delta}&=&
[[3]]_{q^2}\lambda_+^{-2}\varepsilon_{\alpha\delta^{\prime\prime}}
\hat{R}^{\gamma^{\prime}\beta^{\prime}}{}_{\beta\gamma}
\hat{R}^{\delta^{\prime}\beta^{\prime\prime}}{}_{\beta^{\prime}\delta}
\hat{R}^{\delta^{\prime\prime}\gamma^{\prime\prime}}{}_{\gamma^{\prime}\delta^{\prime}}
\varepsilon_{\gamma^{\prime\prime}\beta^{\prime\prime}}\nn
&-&\lambda_+^{-1}g_{AB}(\sigma^A)_{\alpha\delta^{\prime\prime}}
\hat{R}^{\gamma^{\prime}\beta^{\prime}}{}_{\beta\gamma}
\hat{R}^{\delta^{\prime}\beta^{\prime\prime}}{}_{\beta^{\prime}\delta}
\hat{R}^{\delta^{\prime\prime}\gamma^{\prime\prime}}{}_{\gamma^{\prime}\delta^{\prime}}
(\sigma^B)_{\gamma^{\prime\prime}\beta^{\prime\prime}},\nn
\varepsilon_{\alpha\beta}(\sigma^A)_{\gamma\delta}&=&q^{-4}\lambda_+^{-1}\varepsilon_{\alpha\delta^{\prime\prime}}
\hat{R}^{\gamma^{\prime}\beta^{\prime}}{}_{\beta\gamma}
\hat{R}^{\delta^{\prime}\beta^{\prime\prime}}{}_{\beta^{\prime}\delta}
\hat{R}^{\delta^{\prime\prime}\gamma^{\prime\prime}}{}_{\gamma^{\prime}\delta^{\prime}}
(\sigma^A)_{\gamma^{\prime\prime}\beta^{\prime\prime}}\nn
&+&\lambda_+^{-1}(\sigma^A)_{\alpha\delta^{\prime\prime}}
\hat{R}^{\gamma^{\prime}\beta^{\prime}}{}_{\beta\gamma}
\hat{R}^{\delta^{\prime}\beta^{\prime\prime}}{}_{\beta^{\prime}\delta}
\hat{R}^{\delta^{\prime\prime}\gamma^{\prime\prime}}{}_{\gamma^{\prime}\delta^{\prime}}
\varepsilon_{\gamma^{\prime\prime}\beta^{\prime\prime}}\nn
&+&q^{-1}\lambda_+^{-1/2}\varepsilon^{ABC}g_{CC^{\prime}}g_{BB^{\prime}}
(\sigma^{C^{\prime}})_{\alpha\delta^{\prime\prime}}\nn
&\times&\hat{R}^{\gamma^{\prime}\beta^{\prime}}{}_{\beta\gamma}
\hat{R}^{\delta^{\prime}\beta^{\prime\prime}}{}_{\beta^{\prime}\delta}
\hat{R}^{\delta^{\prime\prime}\gamma^{\prime\prime}}{}_{\gamma^{\prime}\delta^{\prime}}
(\sigma^{B^{\prime}})_{\gamma^{\prime\prime}\beta^{\prime\prime}},\nn
(\sigma^A)_{\alpha\beta}\,\varepsilon_{\gamma\delta}&=&\lambda_+^{-1}\varepsilon_{\alpha\delta^{\prime\prime}}
\hat{R}^{\gamma^{\prime}\beta^{\prime}}{}_{\beta\gamma}
\hat{R}^{\delta^{\prime}\beta^{\prime\prime}}{}_{\beta^{\prime}\delta}
\hat{R}^{\delta^{\prime\prime}\gamma^{\prime\prime}}{}_{\gamma^{\prime}\delta^{\prime}}
(\sigma^A)_{\gamma^{\prime\prime}\beta^{\prime\prime}}\nn
&+&\lambda_+^{-1}(\sigma^A)_{\alpha\delta^{\prime\prime}}
\hat{R}^{\gamma^{\prime}\beta^{\prime}}{}_{\beta\gamma}
\hat{R}^{\delta^{\prime}\beta^{\prime\prime}}{}_{\beta^{\prime}\delta}
\hat{R}^{\delta^{\prime\prime}\gamma^{\prime\prime}}{}_{\gamma^{\prime}\delta^{\prime}}
\varepsilon_{\gamma^{\prime\prime}\beta^{\prime\prime}}\nn
&-&q\lambda_+^{-1/2}\varepsilon^{ABC}g_{CC^{\prime}}g_{BB^{\prime}}
(\sigma^{C^{\prime}})_{\alpha\delta^{\prime\prime}}\nn
&\times&\hat{R}^{\gamma^{\prime}\beta^{\prime}}{}_{\beta\gamma}
\hat{R}^{\delta^{\prime}\beta^{\prime\prime}}{}_{\beta^{\prime}\delta}
\hat{R}^{\delta^{\prime\prime}\gamma^{\prime\prime}}{}_{\gamma^{\prime}\delta^{\prime}}
(\sigma^{B^{\prime}})_{\gamma^{\prime\prime}\beta^{\prime\prime}}.
\ena

Lastly, we give an identity that transforms the three-dimensional quantum
metric from a vectorial basis into a spinorial one (see also the discussion in
Sec.\thinspace\ref{connectchapt}):%
\begin{equation}
g_{AB}(\sigma^{A})_{\alpha\beta}(\sigma^{B})_{\gamma\delta}=-q^{2}\lambda
_{+}^{-1}\varepsilon_{\alpha\beta}\varepsilon_{\gamma\delta}-q^{2}%
\varepsilon_{\alpha\delta}\varepsilon_{\beta\gamma}.
\end{equation}

\subsubsection{The spin matrices $\sigma^{\mu\nu}$ and $\bar{\sigma}^{\mu\nu}%
$}

\textbf{Definition: }In complete analogy to the classical situation we define
the two-dimensional spin matrices $\sigma^{\mu\nu}$ of $q$-Minkowski
space$\ $by%
\begin{equation}
(\sigma^{\mu\nu})_{\alpha}{}^{\beta}\equiv(P_{A})^{\mu\nu}{}_{\kappa\lambda
}(\sigma^{\kappa}\bar{\sigma}^{\lambda})_{\alpha}{}^{\beta}%
.\label{sigmamnmink1}%
\end{equation}
Matrix multiplication is understood in the sense of (\ref{matrixmultmink1})
and $P_{A}$ is the antisymmetric projector of $q$-deformed Minkowski space.
Clearly, we also have 'conjugated' as well as 'inverse' versions of the
two-dimensional spin matrices:
\begin{align}
(\bar{\sigma}^{\mu\nu})_{\dot{\alpha}}{}^{\dot{\beta}} &  \equiv(P_{A}%
)^{\mu\nu}{}_{\kappa\lambda}(\bar{\sigma}^{\kappa}{\sigma}^{\lambda}%
)_{\dot{\alpha}}{}^{\dot{\beta}},\nonumber\\
(\sigma_{\mu\nu}^{-1})_{\alpha}{}^{\beta} &  \equiv(P_{A})^{\kappa\lambda}%
{}_{\mu\nu}(\sigma_{\kappa}^{-1}\bar{\sigma}_{\lambda}^{-1})_{\alpha}{}%
^{\beta},\nonumber\\
(\bar{\sigma}_{\mu\nu}^{-1})_{\dot{\alpha}}{}^{\dot{\beta}} &  \equiv
(P_{A})^{\kappa\lambda}{}_{\mu\nu}(\bar{\sigma}_{\kappa}^{-1}{\sigma}%
_{\lambda}^{-1})_{\dot{\alpha}}{}^{\dot{\beta}}.\label{sigmamnmink2}%
\end{align}
Raising and lowering spinor indices of the spin matrices $\sigma^{\mu\nu}$ and
$\bar{\sigma}^{\mu\nu}\ $is determined by the rules in (\ref{indcon1}).
However, if we consider 'inverse' spin matrices the behavior of their spinor
indices follows from (\ref{covco}) and (\ref{manincon1}).

We write down the matrices in (\ref{sigmamnmink1}) that are not equal to zero:%
\begin{align}
\sigma^{+3}  &  =-q\lambda_{+}^{-1/2}\left(
\begin{array}
[c]{cc}%
0 & 0\\
q^{1/2} & 0
\end{array}
\right)  , & \sigma^{+0}  &  =-q\lambda_{+}^{-1/2}\left(
\begin{array}
[c]{cc}%
0 & 0\\
q^{1/2} & 0
\end{array}
\right)  ,\nonumber\\[0.03in]
\sigma^{+-}  &  =\lambda_{+}^{-1}\left(
\begin{array}
[c]{cc}%
-q & 0\\
0 & q^{-1}%
\end{array}
\right)  , & \sigma^{3+}  &  =\lambda_{+}^{-1/2}\left(
\begin{array}
[c]{cc}%
0 & 0\\
q^{-1/2} & 0
\end{array}
\right)  ,\nonumber\\[0.03in]
\sigma^{33}  &  =q^{-1}\lambda\lambda_{+}^{-1}\left(
\begin{array}
[c]{cc}%
-q & 0\\
0 & q^{-1}%
\end{array}
\right)  , & \sigma^{3-}  &  =q^{1/2}\lambda_{+}^{-1/2}\left(
\begin{array}
[c]{cc}%
0 & 1\\
0 & 0
\end{array}
\right)  ,\nonumber\\[0.03in]
\sigma^{30}  &  =q\lambda_{+}^{-1}\left(
\begin{array}
[c]{cc}%
-q & 0\\
0 & q^{-1}%
\end{array}
\right)  , & \sigma^{0+}  &  =q^{-1/2}\lambda_{+}^{-1/2}\left(
\begin{array}
[c]{cc}%
0 & 0\\
1 & 0
\end{array}
\right)  ,\nonumber\\[0.03in]
\sigma^{03}  &  =-q^{-1}\lambda_{+}^{-1}\left(
\begin{array}
[c]{cc}%
-q & 0\\
0 & q^{-1}%
\end{array}
\right)  , & \sigma^{0-}  &  =-q^{-3/2}\lambda_{+}^{-1/2}\left(
\begin{array}
[c]{cc}%
0 & 1\\
0 & 0
\end{array}
\right)  ,\nonumber\\[0.03in]
\sigma^{-+}  &  =\lambda_{+}^{-1}\left(
\begin{array}
[c]{cc}%
q & 0\\
0 & q^{-1}%
\end{array}
\right)  , & \sigma^{-3}  &  =-q^{-3/2}\lambda_{+}^{-1/2}\left(
\begin{array}
[c]{cc}%
0 & 1\\
0 & 0
\end{array}
\right)  ,\nonumber\\[0.03in]
\sigma^{-0}  &  =q^{1/2}\lambda_{+}^{-1/2}\left(
\begin{array}
[c]{cc}%
0 & 1\\
0 & 0
\end{array}
\right)  . &  &
\end{align}
The explicit form of the other variants of spin matrices are obtained most
easily by means of the correspondences
{\large
   \begin{equation*}
  \xymatrix{
        \left(\sigma^{\mu\nu}\right)_{\alpha{\beta}}\ar@{<->}[rr]^{i)}\ar@{<->}[dd]\ar@{<->}[ddrr]_>>>>>>>{v)}&&
         \left(\bar{\sigma}^{\mu\nu}\right)_{\dot{\alpha}\dot\beta}\ar@{<->}[dd]^{ii)}\ar@{<-}'[dl]^>>>>>>>>{vi)}
          \\ &\ar@{->}[dl]&\\
         \left({\sigma}^{-1}_{\mu\nu}\right)^{{\alpha}\beta}\ar@{<->}[rr]_{iv)}\ar@{<->}[uu]^{iii)} &&
         \left(\bar{\sigma}_{\mu\nu}^{-1}\right)^{\dot\alpha\dot{\beta}}}
     \end{equation*}
}
with
\begin{align}
i)\qquad\bar{\sigma}^{AB}  &  =\sigma^{AB}, & ii)\qquad\bar{\sigma}^{AB}  &
=-\bar{\sigma}_{AB}^{-1},\nonumber\\
\bar{\sigma}^{A0}  &  =-q^{-2}\sigma^{A0}, & \bar{\sigma}^{A0}  &
=-q^{-2}\bar{\sigma}_{A0}^{-1},\nonumber\\
\bar{\sigma}^{0A}  &  =-q^{2}\sigma^{0A}, & \bar{\sigma}^{0A}  &  =-q^{2}%
\bar{\sigma}_{0A}^{-1},\nonumber\\[0.08in]
iii)\qquad{\sigma}^{AB}  &  =-\sigma_{AB}^{-1}, & iv)\qquad{\sigma}_{AB}^{-1}
&  =\bar{\sigma}_{AB}^{-1},\nonumber\\
{\sigma}^{A0}  &  =-q^{-2}\sigma_{A0}^{-1}, & {\sigma}_{A0}^{-1}  &
=-q^{-2}\bar{\sigma}_{A0}^{-1},\nonumber\\
{\sigma}^{0A}  &  =-q^{2}\sigma_{0A}^{-1}, & {\sigma}_{0A}^{-1}  &
=-q^{2}\bar{\sigma}_{0A}^{-1},\nonumber\\[0.08in]
v)\qquad{\sigma}^{AB}  &  =-\bar{\sigma}_{AB}^{-1}, & vi)\qquad\bar{\sigma
}^{AB}  &  =-\sigma_{AB}^{-1},\nonumber\\
{\sigma}^{A0}  &  =\bar{\sigma}_{A0}^{-1}, & \bar{\sigma}^{A0}  &
=\sigma_{A0}^{-1},\nonumber\\
{\sigma}^{0A}  &  =\bar{\sigma}_{0A}^{-1}, & \bar{\sigma}^{0A}  &
=\sigma_{0A}^{-1},
\end{align}
where $A,$ $B\in{\{}+,3,-{\}}$.

\textbf{Fundamental properties: }Next, we list some fundamental features of
the spin matrices. First of all, we have
\begin{align}
(P_{A})^{\mu\nu}{}_{\mu^{\prime}\nu^{\prime}}\sigma^{\mu^{\prime}\nu^{\prime
}}  &  =\sigma^{\mu\nu},\quad(P_{A})^{\mu\nu}{}_{\mu^{\prime}\nu^{\prime}}%
\bar{\sigma}^{\mu^{\prime}\nu^{\prime}}=\bar{\sigma}^{\mu\nu},\nonumber\\
(P_{A})^{\mu^{\prime}\nu^{\prime}}{}_{\mu\nu}\sigma_{\mu^{\prime}\nu^{\prime}%
}^{-1}  &  =\sigma_{\mu\nu}^{-1},\quad(P_{A})^{\mu^{\prime}\nu^{\prime}}%
{}_{\mu\nu}\bar{\sigma}_{\mu\nu}^{-1}=\bar{\sigma}_{\mu\nu}^{-1},
\end{align}
which is a consequence of the definitions in (\ref{sigmamnmink1}) and
(\ref{sigmamnmink2}) together with idempotency of projectors. In view of
$P_{A}=P_{+}+P_{-}$ we more concretely have
\begin{align}
(P_{+})^{\mu\nu}{}_{\rho\lambda}\,\bar{\sigma}^{\rho\lambda}  &  =\bar{\sigma
}^{\mu\nu},\quad(P_{-})^{\mu\nu}{}_{\rho\lambda}\,\bar{\sigma}^{\rho\lambda
}=0,\nonumber\\
(P_{-})^{\mu\nu}{}_{\rho\lambda}\,\sigma^{\rho\lambda}  &  =\sigma^{\mu\nu
},\quad(P_{+})^{\mu\nu}{}_{\rho\lambda}\,\sigma^{\rho\lambda}=0,
\label{P+-SpiMat1}%
\end{align}
and%
\begin{align}
(P_{+})^{\rho\lambda}{}_{\mu\nu}\,\bar{\sigma}_{\rho\lambda}^{-1}  &
=\bar{\sigma}_{\mu\nu}^{-1},\quad(P_{-})^{\rho\lambda}{}_{\mu\nu}\,\bar
{\sigma}_{\rho\lambda}^{-1}=0,\nonumber\\
(P_{-})^{\rho\lambda}{}_{\mu\nu}\,\sigma_{\rho\lambda}^{-1}  &  =\bar{\sigma
}_{\mu\nu}^{-1},\quad(P_{+})^{\rho\lambda}{}_{\mu\nu}\,\sigma_{\rho\lambda
}^{-1}=0. \label{P+-SpiMat2}%
\end{align}

The spin matrices are symmetric in their spinorial indices:%
\begin{align}
(\sigma^{\mu\nu})_{\alpha^{\prime}\beta^{\prime}}S^{\alpha^{\prime}%
\beta^{\prime}}{}_{\alpha\beta} &  =(\sigma^{\mu\nu})_{\alpha\beta}, &
(\sigma_{\mu\nu}^{-1})^{\alpha^{\prime}\beta^{\prime}}S^{\alpha\beta}%
{}_{\alpha^{\prime}\beta^{\prime}} &  =(\sigma_{\mu\nu}^{-1})^{\alpha\beta
},\nonumber\\
(\bar{\sigma}^{\mu\nu})_{\dot{\alpha}^{\prime}\dot{\beta}^{\prime}}%
S^{\dot{\alpha}^{\prime}\dot{\beta}^{\prime}}{}_{\dot{\alpha}\dot{\beta}} &
=(\bar{\sigma}^{\mu\nu})_{\dot{\alpha}\dot{\beta}}, & (\bar{\sigma}_{\mu\nu
}^{-1})^{\dot{\alpha}\dot{\beta}}S^{\dot{\alpha}^{\prime}\dot{\beta}^{\prime}%
}{}_{\dot{\alpha}\dot{\beta}} &  =(\bar{\sigma}_{\mu\nu}^{-1})^{\dot{\alpha
}\dot{\beta}}.
\end{align}
Combining this result with
\begin{equation}
\varepsilon_{\alpha^{\prime}\beta^{\prime}}\,S^{\alpha^{\prime}\beta^{\prime}%
}{}_{\alpha\beta}=S^{\alpha\beta}{}_{\alpha^{\prime}\beta^{\prime}}%
\varepsilon^{\alpha^{\prime}\beta^{\prime}}=0,
\end{equation}
we find that%
\begin{align}
(\sigma^{\mu\nu})_{\alpha\beta}\varepsilon^{\alpha\beta} &  =0,\quad
(\sigma_{\mu\nu}^{-1})^{\alpha\beta}\varepsilon_{\alpha\beta}=0,\nonumber\\
(\bar{\sigma}^{\mu\nu})_{\dot{\alpha}\dot{\beta}}\varepsilon^{\dot{\alpha}%
\dot{\beta}} &  =0,\quad(\bar{\sigma}_{\mu\nu}^{-1})^{\dot{\alpha}\dot{\beta}%
}\varepsilon_{\dot{\alpha}\dot{\beta}}=0.
\end{align}
Using the quantum trace the last relations can alternatively be written as
\begin{equation}
\mbox{Tr}_{q}\,\sigma^{\mu\nu}=0,\quad\mbox{Tr}_{q}\,\sigma_{\mu\nu}%
^{-1}=0,\quad\mbox{Tr}_{q}\,\bar{\sigma}^{\mu\nu}=0,\quad\mbox{Tr}_{q}%
\,\bar{\sigma}_{\mu\nu}^{-1}=0.
\end{equation}

The spin matrices are associated with spin-1 representations of $U_{q}%
(su(2)).$ The symmetrizer $S$ of $U_{q}(su(2))$ maps two spinors onto a space
carrying a spin-1 representation. On this footing we have
\begin{align}
(\sigma_{\mu\nu}^{-1})^{\alpha^{\prime}\beta^{\prime}}(\sigma^{\mu\nu
})_{\alpha\beta}  &  =\lambda_{+}S^{\alpha^{\prime}\beta^{\prime}}{}%
_{\alpha\beta},\quad(\bar{\sigma}_{\mu\nu}^{-1})^{\dot{\alpha}^{\prime}%
\dot{\beta}^{\prime}}(\bar{\sigma}^{\mu\nu})_{\dot{\alpha}\dot{\beta}}%
=\lambda_{+}S^{\dot{\alpha}^{\prime}\dot{\beta}^{\prime}}{}_{\dot{\alpha}%
\dot{\beta}},\nonumber\\
(\sigma^{\mu\nu})_{\alpha\beta}(\sigma_{\mu\nu}^{-1})^{\alpha^{\prime}%
\beta^{\prime}}  &  =\lambda_{+}S^{\alpha^{\prime}\beta^{\prime}}{}%
_{\alpha\beta},\quad(\bar{\sigma}^{\mu\nu})_{\dot{\alpha}\dot{\beta}}%
(\bar{\sigma}_{\mu\nu}^{-1})^{\dot{\alpha}^{\prime}\dot{\beta}^{\prime}%
}=\lambda_{+}S^{\dot{\alpha}^{\prime}\dot{\beta}^{\prime}}{}_{\dot{\alpha}%
\dot{\beta}},
\end{align}
and%
\begin{align}
(\sigma^{\mu\nu})_{\alpha\beta}(\sigma_{\rho\lambda}^{-1})^{\alpha\beta}  &
=\lambda_{+}(P_{-})^{\mu\nu}{}_{\rho\lambda},\quad(\sigma_{\mu\nu}%
^{-1})^{\alpha\beta}(\sigma^{\rho\lambda})_{\alpha\beta}=\lambda_{+}%
(P_{-})^{\rho\lambda}{}_{\mu\nu},\nonumber\\
(\bar{\sigma}^{\mu\nu})_{\dot{\alpha}\dot{\beta}}(\bar{\sigma}_{{\rho}%
{\lambda}}^{-1})^{\dot{\alpha}\dot{\beta}}  &  =\lambda_{+}(P_{+})^{\mu\nu}%
{}_{\rho\lambda},\quad(\bar{\sigma}_{\mu\nu}^{-1})^{\dot{\alpha}\dot{\beta}%
}(\bar{\sigma}^{\rho\lambda})_{\dot{\alpha}\dot{\beta}}=\lambda_{+}%
(P_{+})^{\rho\lambda}{}_{\mu\nu}.
\end{align}

A short look at (\ref{P+-SpiMat1}) and (\ref{P+-SpiMat2}) tells us that the
conjugated spin matrices refer to the projector $P_{+}$ and the unconjugated
ones to the projector $P_{-}$. In this respect, the identities
\begin{equation}
P_{+}P_{-}=P_{-}P_{+}=0,
\end{equation}
imply\textbf{ }that contracting the Lorentz indices of a conjugated spin
matrix with those of an unconjugated one gives zero:
\begin{align}
(\sigma_{\mu\nu}^{-1})^{\gamma\delta}(\bar{\sigma}^{\mu\nu})_{\dot{\alpha}%
\dot{\beta}} &  =0,\quad(\bar{\sigma}_{\mu\nu}^{-1})^{\dot{\gamma}\dot{\delta
}}({\sigma}^{\mu\nu})_{{\alpha}{\beta}}=0,\nonumber\\
(\sigma^{\mu\nu})_{\alpha\beta}(\bar{\sigma}_{\mu\nu}^{-1})^{\dot{\gamma}%
\dot{\delta}} &  =0,\quad(\bar{\sigma}^{\mu\nu})_{\dot{\alpha}\dot{\beta}%
}({\sigma}_{\mu\nu}^{-1})^{{\gamma}{\delta}}=0.\label{ContSpinMat}%
\end{align}

In analogy to the Euclidean case the antisymmetric tensor of $q$-deformed
Minkowski space can be expressed by the projectors $P_{+}$ and $P_{-}$:%
\begin{equation}
\varepsilon^{\mu\nu}{}_{\rho\lambda}=\eta_{\rho^{\prime}\rho}\eta
_{\lambda^{\prime}\lambda}\varepsilon^{\mu\nu\rho^{\prime}\lambda^{\prime}%
}=q^{-2}\lambda_{+}((P_{+})^{\mu\nu}{}_{\rho\lambda}-(P_{-})^{\mu\nu}{}%
_{\rho\lambda}).
\end{equation}
From this formula together with\ the identities in (\ref{P+-SpiMat1}) it
follows, at once, that%
\begin{align}
\varepsilon^{\mu\nu}{}_{\rho\lambda}\sigma^{\rho\lambda}  &  =-q^{-2}%
\lambda_{+}((P_{+})^{\mu\nu}{}_{\rho\lambda}-(P_{-})^{\mu\nu}{}_{\rho\lambda
})\sigma^{\rho\lambda}=q^{-2}\lambda_{+}\sigma^{\rho\lambda},\nonumber\\
\varepsilon^{\mu\nu}{}_{\rho\lambda}\bar{\sigma}^{\rho\lambda}  &
=-q^{-2}\lambda_{+}((P_{+})^{\mu\nu}{}_{\rho\lambda}-(P_{-})^{\mu\nu}{}%
_{\rho\lambda})\bar{\sigma}^{\rho\lambda}=-q^{-2}\lambda_{+}\bar{\sigma}%
^{\rho\lambda}.
\end{align}

There is a remarkable relationship between spin matrices and their 'inverse'
counterparts. We found that%
\begin{equation}
(\sigma^{\mu\nu})_{\alpha\beta}=(\sigma^{-1})_{\beta\alpha}^{\mu\nu}%
,\quad(\bar{\sigma}^{\mu\nu})_{\dot{\alpha}\dot{\beta}}=(\bar{\sigma}%
^{-1})_{\dot{\beta}\dot{\alpha}}^{\mu\nu},
\end{equation}
and%
\begin{equation}
(\sigma^{\mu\nu})_{\alpha\beta}=(\sigma^{-1})_{\alpha\beta}^{\nu\mu}%
,\quad(\bar{\sigma}^{\mu\nu})_{\dot{\alpha}\dot{\beta}}=(\bar{\sigma}%
^{-1})_{\dot{\alpha}\dot{\beta}}^{\nu\mu},
\end{equation}
These identities can be checked by making use of
\begin{equation}
(\sigma^{\mu})_{\dot{\beta}\alpha}=-(\bar{\sigma}^{-1})_{\alpha\dot{\beta}%
}^{\mu},\quad(\bar{\sigma}^{\mu})_{\beta\dot{\alpha}}=-(\sigma^{-1}%
)_{\dot{\alpha}\beta}^{\mu}.
\end{equation}

\textbf{Formulae concerning spin matrices: }Next, we would like to present
$q$-analogs of some well-known relations concerning spin and Pauli matrices
(see, for example, Refs. \cite{Wessbagger}, \cite{Core}, and \cite{Grimm}). We
start with
\begin{align}
\sigma^{\mu\nu} &  =\lambda_{+}^{-1}\eta^{\mu\nu}+\sigma^{\mu}\bar{\sigma
}^{\nu},\nonumber\\
\bar{\sigma}^{\mu\nu} &  =\lambda_{+}^{-1}\eta^{\mu\nu}+\bar{\sigma}^{\mu
}{\sigma}^{\nu},\label{simgamunumink0}\\[0.1in]
\sigma^{\mu\nu}\sigma^{\lambda} &  =-\frac{1}{2}\lambda_{+}(P_{A})^{\mu\nu}%
{}_{\mu^{\prime}\nu^{\prime}}\eta^{\nu^{\prime}\lambda}\sigma^{\mu^{\prime}%
}-\frac{q^{2}}{2}\varepsilon^{\mu\nu\lambda\rho}\sigma_{\rho},\nonumber\\
\bar{\sigma}^{\mu\nu}\bar{\sigma}^{\lambda} &  =-\frac{1}{2}\lambda_{+}%
(P_{A})^{\mu\nu}{}_{\mu^{\prime}\nu^{\prime}}\eta^{\nu^{\prime}\lambda}%
\bar{\sigma}^{\mu^{\prime}}+\frac{q^{2}}{2}\varepsilon^{\mu\nu\lambda\rho}%
\bar{\sigma}_{\rho},\label{simgamunumink1}\\[0.1in]
\bar{\sigma}^{\mu}\sigma^{\nu\lambda} &  =-\frac{1}{2}\lambda_{+}(P_{A}%
)^{\nu\lambda}{}_{\nu^{\prime}\lambda^{\prime}}\eta^{\mu\nu^{\prime}}%
\bar{\sigma}^{\lambda^{\prime}}-\frac{q^{2}}{2}\varepsilon^{\mu\nu\lambda\rho
}\bar{\sigma}_{\rho},\nonumber\\
{\sigma}^{\mu}\bar{\sigma}^{\nu\lambda} &  =-\frac{1}{2}\lambda_{+}%
(P_{A})^{\nu\lambda}{}_{\nu^{\prime}\lambda^{\prime}}\eta^{\mu\nu^{\prime}%
}\bar{\sigma}^{\lambda^{\prime}}+\frac{q^{2}}{2}\varepsilon^{\mu\nu\lambda
\rho}\bar{\sigma}_{\rho}.\label{simgamunumink}%
\end{align}
Finally, the product of two spin matrices decomposes as
\begin{align}
{\sigma}^{\mu\nu}{\sigma}^{\kappa\lambda}= &  \;\frac{1}{2}(P_{A})^{\mu\nu}%
{}_{\mu^{\prime}\nu^{\prime}}(P_{A})^{\kappa\lambda}{}_{\kappa^{\prime}%
\lambda^{\prime}}\eta^{\mu^{\prime}\lambda^{\prime}}\eta^{\nu^{\prime}%
\kappa^{\prime}}+\frac{q^{2}}{2}\lambda_{+}^{-1}\varepsilon^{\mu\nu
\kappa\lambda}\nonumber\\
&  \,-\;\lambda_{+}(P_{A})^{\mu\nu}{}_{\mu^{\prime}\nu^{\prime}}%
(P_{A})^{\kappa\lambda}{}_{\kappa^{\prime}\lambda^{\prime}}\eta^{\nu^{\prime
}\kappa^{\prime}}\eta^{\mu^{\prime}\lambda^{\prime}}\sigma^{\mu^{\prime
}\lambda^{\prime}},\nonumber\\
\bar{\sigma}^{\mu\nu}\bar{\sigma}^{\kappa\lambda}= &  \;\frac{1}{2}%
(P_{A})^{\mu\nu}{}_{\mu^{\prime}\nu^{\prime}}(P_{A})^{\kappa\lambda}{}%
_{\kappa^{\prime}\lambda^{\prime}}\eta^{\mu^{\prime}\lambda^{\prime}}\eta
^{\nu^{\prime}\kappa^{\prime}}-\frac{q^{2}}{2}\lambda_{+}^{-1}\varepsilon
^{\mu\nu\kappa\lambda}\nonumber\\
&  \,-\;\lambda_{+}(P_{A})^{\mu\nu}{}_{\mu^{\prime}\nu^{\prime}}%
(P_{A})^{\kappa\lambda}{}_{\kappa^{\prime}\lambda^{\prime}}\eta^{\nu^{\prime
}\kappa^{\prime}}\eta^{\mu^{\prime}\lambda^{\prime}}\bar{\sigma}^{\mu^{\prime
}\lambda^{\prime}}.
\end{align}
Taking the quantum trace of the last two relations we obtain%
\begin{align}
\mbox{Tr}_{q}(\sigma^{\mu\nu}\sigma^{\kappa\lambda})= &  \;\frac{1}{2}%
\lambda_{+}(P_{A})^{\mu\nu}{}_{\mu^{\prime}\nu^{\prime}}(P_{A})^{\kappa
\lambda}{}_{\kappa^{\prime}\lambda^{\prime}}\eta^{\mu^{\prime}\lambda^{\prime
}}\eta^{\nu^{\prime}\kappa^{\prime}}-\frac{q^{2}}{2}\varepsilon^{\mu\nu
\kappa\lambda},\nonumber\\
\mbox{Tr}_{q}(\bar{\sigma}^{\mu\nu}\bar{\sigma}^{\kappa\lambda})= &
\;\frac{1}{2}\lambda_{+}(P_{A})^{\mu\nu}{}_{\mu^{\prime}\nu^{\prime}}%
(P_{A})^{\kappa\lambda}{}_{\kappa^{\prime}\lambda^{\prime}}\eta^{\mu^{\prime
}\lambda^{\prime}}\eta^{\nu^{\prime}\kappa^{\prime}}+\frac{q^{2}}%
{2}\varepsilon^{\mu\nu\kappa\lambda}.\label{sigmamunumink2}%
\end{align}

The formulae for 'inverse' matrices follow from (\ref{simgamunumink0}%
)-(\ref{sigmamunumink2}) by applying the replacements in (\ref{sigmainvrules})
together with
\begin{equation}
\sigma^{\mu\nu}\rightarrow\sigma_{\mu\nu}^{-1},\qquad\bar{\sigma}^{\mu\nu
}\rightarrow\bar{\sigma}_{\mu\nu}^{-1},\qquad(P_{A})^{\mu\nu}{}_{\mu^{\prime
}\nu^{\prime}}\rightarrow(P_{A})^{\mu^{\prime}\nu^{\prime}}{}_{\mu\nu}.
\end{equation}
Once again, the above formulae do not depend on whether we work with bosonic
or fermionic Pauli matrices. As already remarked in the previous section this
is due to our thoroughly chosen conventions.

Contracting Lorentz indices with the quantum metric one can get from
(\ref{simgamunumink0}) and (\ref{simgamunumink}) that
\begin{align}
(\sigma^{\mu\nu})_{\alpha\alpha^{\prime}}(\sigma^{\rho})_{\beta\dot{\beta}%
}\,\eta_{\nu\rho}  &  =k_{1}\,\varepsilon_{\alpha^{\prime}\beta}(\sigma^{\mu
})_{\alpha\dot{\beta}}+k_{2}\,\varepsilon_{\alpha\alpha^{\prime}}(\sigma^{\mu
})_{\beta\dot{\beta}},\nonumber\\
(\sigma^{\nu})_{\alpha\dot{\alpha}}(\bar{\sigma}^{\rho\mu})_{\dot{\beta}%
\dot{\beta}^{\prime}}\,\eta_{\nu\rho}  &  =-k_{1}\,\varepsilon_{\alpha
^{\prime}\beta}(\sigma^{\mu})_{\alpha\dot{\beta}}-k_{2}\,\varepsilon
_{\alpha\alpha^{\prime}}(\sigma^{\mu})_{\beta\dot{\beta}},\\[0.08in]
(\bar{\sigma}^{\mu\nu})_{\dot{\alpha}\dot{\alpha}^{\prime}}(\bar{\sigma}%
^{\rho})_{\dot{\beta}\beta}\,\eta_{\nu\rho}  &  =-k_{1}\,\varepsilon
_{\dot{\alpha}^{\prime}\dot{\beta}}(\bar{\sigma}^{\mu})_{\dot{\alpha}\beta
}-k_{2}\,\varepsilon_{\dot{\alpha}\dot{\alpha}^{\prime}}(\bar{\sigma}^{\mu
})_{\dot{\beta}\beta},\nonumber\\
(\bar{\sigma}^{\nu})_{\dot{\alpha}\alpha}(\sigma^{\rho\mu})_{\beta^{\prime
}\beta}\,\eta_{\nu\rho}  &  =k_{1}\,\varepsilon_{\dot{\alpha}^{\prime}%
\dot{\beta}}(\bar{\sigma}^{\mu})_{\dot{\alpha}\beta}+k_{2}\,\varepsilon
_{\dot{\alpha}\dot{\alpha}^{\prime}}(\bar{\sigma}^{\mu})_{\dot{\beta}\beta},
\end{align}
where $k_{1}=-1,$ $k_{2}=-\lambda_{+}^{-1},$ $k_{3}=1,$ $k_{4}=\lambda
_{+}^{-1}$. There are also versions with 'inverse' matrices, which we simply
get by the replacements%
\begin{gather}
\eta_{\mu\nu}\rightarrow\eta^{\mu\nu},\qquad\varepsilon_{\alpha\beta
}\rightarrow-\varepsilon^{\alpha\beta},\nonumber\\
(\sigma^{\mu})_{\alpha\dot{\alpha}}\rightarrow(\sigma_{\mu}^{-1})^{\alpha
\dot{\alpha}},\qquad(\bar{\sigma}^{\mu})_{\alpha\dot{\alpha}}\rightarrow
(\bar{\sigma}_{\mu}^{-1})^{\alpha\dot{\alpha}},\nonumber\\
(\sigma^{\mu\nu})_{\alpha\alpha^{\prime}}\rightarrow(\sigma_{\mu\nu}%
^{-1})^{\alpha\alpha^{\prime}},\qquad(\bar{\sigma}^{\mu\nu})_{\dot{\alpha}%
\dot{\alpha}^{\prime}}\rightarrow(\bar{\sigma}_{\mu\nu}^{-1})^{\dot{\alpha
}\dot{\alpha}^{\prime}}.
\end{gather}

Now, we come to some very useful decomposition formulae involving spin
matrices. First of all, we have
\begin{align}
(\sigma^{\mu})_{\alpha\dot{\alpha}}(\sigma^{\nu})_{\beta\dot{\beta}}= &
\;k_{1}\,\varepsilon_{\alpha\beta^{\prime}}(\hat{R}^{-1})^{\beta^{\prime}%
\dot{\alpha}^{\prime}}{}_{\dot{\alpha}\beta}\varepsilon_{\dot{\alpha}^{\prime
}\dot{\beta}}\eta^{\mu\nu}\nonumber\\
&  +k_{2}\,\varepsilon_{\alpha\beta^{\prime}}(\hat{R}^{-1})^{\beta^{\prime
}\dot{\alpha}^{\prime}}{}_{\dot{\alpha}\beta}(\bar{\sigma}^{\mu\nu}%
)_{\dot{\alpha}^{\prime}\dot{\beta}}\nonumber\\
&  +k_{3}\,(\sigma^{\mu\nu})_{\alpha\beta^{\prime}}(\hat{R}^{-1}%
)^{\beta^{\prime}\dot{\alpha}^{\prime}}{}_{\dot{\alpha}\beta}\varepsilon
_{\dot{\alpha}^{\prime}\dot{\beta}}\nonumber\\
&  +k_{4}\,(\sigma^{\mu\rho})_{\alpha\beta^{\prime}}(\hat{R}^{-1}%
)^{\beta^{\prime}\dot{\alpha}^{\prime}}{}_{\dot{\alpha}\beta}\eta_{\rho\kappa
}(\bar{\sigma}^{\kappa\nu})_{\dot{\alpha}^{\prime}\dot{\beta}}%
,\label{fierzsigmaminkneu1}%
\end{align}
where $\hat{R}$ and $\hat{R}^{-1}$ respectively denote the $\hat{R}$-matrix of
$U_{q}(su(2))$ and its inverse. The values of the coefficients\ are
$k_{1}=q^{-2}\lambda_{+}^{-2},$ $k_{2}=q^{2}\lambda_{+}^{-1},$ $k_{3}%
=-q^{2}\lambda_{+}^{-1},$ $k_{4}=q^{2}$. Performing the interchanges%
\begin{equation}
\sigma\leftrightarrow\bar{\sigma},\qquad\hat{R}\leftrightarrow\hat{R}%
^{-1},\label{IntFierz}%
\end{equation}
we get a second decomposition formula from (\ref{fierzsigmaminkneu1}). In this
case the coefficients take on the values $k_{1}=-q^{-2}\lambda_{+}^{-2},$
$k_{2}=-q^{-2}\lambda_{+}^{-1},$ $k_{3}=q^{-2}\lambda_{+}^{-1},$ $k_{4}%
=q^{-2}$.

We can also find a variant of (\ref{fierzsigmaminkneu1}) with one Pauli matrix
being replaced by a conjugated one:
\begin{align}
(\sigma^{\mu})_{\alpha\dot{\alpha}}(\bar{\sigma}^{\nu})_{\dot{\beta}\beta}= &
\;k_{1}\,\varepsilon_{\alpha\beta}\varepsilon_{\dot{\alpha}\dot{\beta}}%
\eta^{\mu\nu}+k_{2}\,(\sigma^{\mu\nu})_{\alpha\beta}\varepsilon_{\dot{\alpha
}\dot{\beta}}\nonumber\\
&  +k_{3}\,\varepsilon_{\alpha\beta^{\prime}}(\hat{R}^{-1})^{\beta^{\prime
}\dot{\alpha}^{\prime}}{}_{\dot{\alpha}\beta^{\prime\prime}}(\bar{\sigma}%
^{\mu\nu})_{^{\prime}\alpha^{\prime}\dot{\beta}^{\prime}}(\hat{R}^{-1}%
)^{\beta^{\prime\prime}\dot{\beta}^{\prime}}{}_{\dot{\beta}\beta}\nonumber\\
&  +k_{4}\,(\sigma^{\mu\rho})_{\alpha\beta^{\prime}}(\hat{R}^{-1}%
)^{\beta^{\prime}\dot{\alpha}^{\prime}}{}_{\dot{\alpha}\beta^{\prime\prime}%
}(\bar{\sigma}^{\kappa\nu})_{\dot{\alpha}^{\prime}\dot{\beta}^{\prime}}%
(\hat{R}^{-1})^{\beta^{\prime\prime}\dot{\beta}^{\prime}}{}_{\dot{\beta}\beta
}\eta^{\rho\kappa},\label{fierzsigmaminkneu2}%
\end{align}
where $k_{1}=-\lambda_{+}^{-2},$ $k_{2}=-\lambda_{+}^{-1},$ $k_{3}%
=q\lambda_{+}^{-1},$ $k_{4}=q.$ Modifying the expressions in
(\ref{fierzsigmaminkneu1}) according to (\ref{IntFierz}) leads us to another
relation with $k_{1}=-\lambda_{+}^{-2},$ $k_{2}=\lambda_{+}^{-1},$
$k_{3}=-q^{-1}\lambda_{+}^{-1},$ $k_{4}=q^{-1}$.

The above decomposition formulae can also be written in terms of 'inverse'
$\sigma$-matrices. In this manner, we have to perform the following
substitutions to (\ref{fierzsigmaminkneu1}), (\ref{sigmamunumink2}), and their
variants obtained by (\ref{IntFierz}):%
\begin{gather}
(\sigma^{\mu})_{\alpha\dot{\beta}}\leftrightarrow(\bar{\sigma}_{\mu}%
^{-1})^{\dot{\alpha}\beta},\qquad(\bar{\sigma}^{\mu})^{\dot{\alpha}\beta
}\leftrightarrow(\sigma_{\mu}^{-1})_{\alpha\dot{\beta}},\nonumber\\
(\sigma^{\mu\nu})_{\alpha\beta}\leftrightarrow(\bar{\sigma}_{\mu\nu}%
^{-1})_{\dot{\alpha}\dot{\beta}},\qquad(\bar{\sigma}^{\mu\nu})_{\dot{\alpha
}\dot{\beta}}\leftrightarrow(\sigma_{\mu\nu}^{-1})^{\alpha\beta},\nonumber\\
\eta^{\mu\nu}\leftrightarrow\eta_{\mu\nu},\qquad\varepsilon_{\alpha\beta
}\leftrightarrow-\varepsilon^{\alpha\beta},\nonumber\\
\hat{R}^{\alpha\beta}{}_{\alpha^{\prime}\beta^{\prime}}\leftrightarrow\hat
{R}^{\alpha^{\prime}\beta^{\prime}}{}_{\alpha\beta},\qquad(\hat{R}%
^{-1})^{\alpha\beta}{}_{\alpha^{\prime}\beta^{\prime}}\leftrightarrow(\hat
{R}^{-1})^{\alpha^{\prime}\beta^{\prime}}{}_{\alpha\beta}.\label{SubInv}%
\end{gather}

\subsubsection{Additional relations for $\sigma^{AB}$}

Again, we can define spin matrices also for the three-dimensional
Euclidean space. For this, we use the Pauli matrices as described in
Sec. \thinspace\ref{eu3reladd} as well as the antisymmetric projector $(P_A)^{AB}{}_{CD}$
for the three-dimensional Euclidean space.
We define
\eqa
\sigma^{AB}\equiv(P_A)^{AB}{}_{CD}\,\sigma^{C}\sigma^{D},
\ena
with $(AB)\in {\{}+3,+-,3-,3+,-+,-3,++,--,33{\}}$.
All remarks 
concerning conjugation, taking the 'inverse' versions or
raising indices and defining matrix multiplication remain 
valid as described for the Minkowski case and in
Sec. \thinspace\ref{eu3reladd}.

Comparing the resulting matrices with their corresponding Minkowski
counterparts, we get
\eqa
\sigma^{AB}_{Eu3}=-q^2\sigma^{AB}_{Mink}.
\ena
This results from the different normalization factors of the
used projectors.  

The completeness and orthonormality relations are
\eqa
(\sigma_{AB}^{-1})^{\alpha\beta}(\sigma^{AB})_{\gamma\delta}
&=&q^{-2}(q^4+1)\lambda_+^{-1}S^{\alpha\beta}{}_{\gamma\delta},\nn
(\sigma^{AB})_{\alpha\beta}(\sigma_{CD}^{-1})^{\alpha\beta}
&=&q^{-2}(q^4+1)\lambda_+^{-1}(P_A)^{AB}{}_{CD}.
\ena

At last, we are able to express the Pauli matrices and the spin
matrices via the identifications
\eqa
(\sigma^{AB})_{\alpha\beta}&=&-q^{-3}\lambda_+^{1/2}\varepsilon^{ABC}g_{CD}(\sigma^D)_{\alpha\beta},\nn
(\sigma^{D})_{\alpha\beta}&=&-q^3(q^4+1)\lambda_+^{-1/2}\varepsilon_{BAC}g^{DC}(\sigma^{AB})_{\alpha\beta}.
\ena

\section{Converting spinorial and vectorial objects with the help of Pauli
matrices \label{connectchapt}}

After having developed the formalism of $\sigma$-matrices, we are well
prepared to show how spinorial objects can be converted into vectorial ones
and vice versa. Due to its importance we give a detailed description of the subject.

\textbf{The metric:} Let us start with conversion formulae for the quantum
metric. Taking the quantum metric of $q$-deformed Minkowski space as it was
given in (\ref{asdf2}) and (\ref{asdf3}) we have to use the following formulae
to get objects with Lorentz indices:
\begin{equation}
\eta_{\mu\nu}=(\bar{\sigma}_{\mu}^{-1})^{\bar{\imath}j}\,C_{(\bar{\imath
}j)(\bar{k}l)}\,(\bar{\sigma}_{\nu}^{-1})^{\bar{k}l},\qquad\eta^{\mu\nu}%
=(\bar{\sigma}^{\mu})_{\bar{\imath}j}\,C^{(\bar{\imath}j)(\bar{k}l)}%
\,(\bar{\sigma}^{\nu})_{\bar{k}l}.\label{kopp1}%
\end{equation}
The reverse formulae are
\begin{equation}
C_{(\bar{\imath}j)(\bar{k}l)}=(\bar{\sigma}^{\mu})_{\bar{\imath}j}\,\eta
_{\mu\nu}\,(\bar{\sigma}^{\nu})_{\bar{k}l},\qquad C^{(\bar{\imath}j)(\bar
{k}l)}=(\bar{\sigma}_{\mu}^{-1})^{\bar{\imath}j}\,\eta^{\mu\nu}\,(\bar{\sigma
}_{\nu}^{-1})^{\bar{k}l}.\label{kopp2}%
\end{equation}
Analogous formulae hold if we start from
\begin{equation}
C_{(i\bar{j})(k\bar{l})}=-q^{2}\varepsilon_{ii^{\prime}}(\hat{R}%
^{-1})^{i^{\prime}\bar{l}^{\prime}}{}_{{\bar{j}}k}\varepsilon_{{\bar{l}%
}^{\prime}\bar{l}},\qquad C^{(i\bar{j})(k\bar{l})}=-q^{-2}\varepsilon
^{ii^{\prime}}\hat{R}^{{\bar{j}}k}{}_{i^{\prime}\bar{l}^{\prime}}%
\varepsilon^{{\bar{l}}^{\prime}\bar{l}},
\end{equation}
and use Pauli matrices without a bar.

Notice that formulae (\ref{kopp1}) and (\ref{kopp2}) carry over to the
Euclidean case if we substitute Pauli matrices with a bar by those with a
tilde. In the case of the Euclidean space, however, the two types of Pauli
matrices can be identified (see also Sec.\thinspace\ref{koppkap}), so it does
not matter which type of Pauli matrices we work with.

The above reasonings show a pattern that is valid for general conversion
processes: the Pauli matrices $(\sigma^{\mu})_{i\bar{j}}$ and $(\bar{\sigma
}^{\mu})_{\bar{\imath}j}$ convert a lower Lorentz index $\mu$ into two lower
spinor indices, i.e. $(i\bar{j})$ and $(\bar{\imath}j)$, respectively
[cf.\thinspace(\ref{kopp2})], or they convert two upper spinor indices
$(i\bar{j})$ or $(\bar{\imath}j)$ into an upper Lorentz index $\mu$
[cf.\thinspace(\ref{kopp1})]. The 'inverse' matrices behave the other way round.

\textbf{The $\hat{R}$-matrix and its projectors:} From the explanations so
far\ it should be clear how to convert the $\hat{R}$-matrix and its projectors
from a spinor basis into a vectorial one and vice verse. So we only give an
example:
\begin{align}
\hat{R}^{\mu\nu}{}_{\rho\lambda}  &  =(\sigma^{\mu})_{i\bar{j}}(\sigma^{\nu
})_{k\bar{l}}\hat{R}^{(i\bar{j})(k\bar{l})}{}_{(i^{\prime}\bar{j}^{\prime
})(k^{\prime}\bar{l}^{\prime})}(\sigma_{\rho}^{-1})_{i^{\prime}\bar{j}%
^{\prime}}(\sigma_{\lambda}^{-1})_{k^{\prime}\bar{l}^{\prime}},\nonumber\\
\hat{R}^{(i\bar{j})(k\bar{l})}{}_{(i^{\prime}\bar{j}^{\prime})(k^{\prime}%
\bar{l}^{\prime})}  &  =(\sigma_{\mu}^{-1})^{i\bar{j}}(\sigma_{\nu}%
^{-1})^{k\bar{l}}\hat{R}^{\mu\nu}{}_{\rho\lambda}(\sigma^{\rho})_{i^{\prime
}\bar{j}^{\prime}}(\sigma^{\lambda})_{k^{\prime}\bar{l}^{\prime}}.
\end{align}
The conversion formulae for the projectors are of the same form, i.e. in the
above formulae one is allowed to replace the $\hat{R}$-matrix with the
projector under consideration. The formulae apply to the
four-dimensional Euclidean space as well. 

\textbf{The epsilon tensor:} The formulae are
\begin{align}
\varepsilon^{\mu\nu\rho\lambda} &  =q^{-2}(\sigma^{\mu})_{i\bar{\imath}%
}(\sigma^{\nu})_{j\bar{j}}\varepsilon^{(i\bar{\imath})(j\bar{j})(k\bar
{k})(l\bar{l})}(\sigma^{\rho})_{k\bar{k}}(\sigma^{\lambda})_{l\bar{l}%
},\nonumber\\
\varepsilon^{(i\bar{\imath})(j\bar{j})(k\bar{k})(l\bar{l})} &  =q^{2}%
(\sigma_{\mu}^{-1})^{i\bar{\imath}}(\sigma_{\nu}^{-1})^{j\bar{j}}%
\varepsilon^{\mu\nu\rho\lambda}(\sigma_{\rho}^{-1})^{k\bar{k}}(\sigma
_{\lambda}^{-1})^{l\bar{l}},
\end{align}
and similar ones with conjugated Pauli matrices. Notice that the normalization
factors were introduced to be consistent with the epsilon tensor from our
former paper \cite{qliealg}. Its non-vanishing components are listed in
App.\thinspace\ref{AppA}. In the case of the four-dimensional
Euclidean space the same formulae hold, but with setting the
prefactors equal to one.

\textbf{Formulae for converting special tensors:} First of all, let us recall
that tensors with two spinor indices can be decomposed into an antisymmetric
and a symmetric part, i.e.
\begin{equation}
T^{\alpha\beta}=T^{[\alpha\beta]}+T^{(\alpha\beta)}=-\lambda_{+}%
^{-1}\varepsilon^{\alpha\beta}T+T^{(\alpha\beta)},
\end{equation}
with
\begin{equation}
T=\varepsilon_{\alpha^{\prime}\beta^{\prime}}T^{\alpha^{\prime}\beta^{\prime}%
},\quad T^{(\alpha\beta)}=S^{\alpha\beta}{}_{\alpha^{\prime}\beta^{\prime}%
}T^{\alpha^{\prime}\beta^{\prime}}.
\end{equation}
These relations are a direct consequence of (\ref{prodecman}) and
(\ref{projmaneps}).

Next, we consider a tensor with two dotted and two undotted spinor indices.
Its decomposition into irreducible constituents reads as
\begin{equation}
X^{\alpha\beta\dot{\alpha}\dot{\beta}}=\lambda_{+}^{-2}\varepsilon
^{\alpha\beta}\varepsilon^{\dot{\alpha}\dot{\beta}}X+\lambda_{+}%
^{-1}\varepsilon^{\alpha\beta}X^{(\dot{\alpha}\dot{\beta})}+\lambda_{+}%
^{-1}\varepsilon^{\dot{\alpha}\dot{\beta}}X^{(\alpha\beta)}+X^{(\alpha
\beta)(\dot{\alpha}\dot{\beta})}, \label{ZerTen}%
\end{equation}
with
\begin{align}
X  &  =\varepsilon_{\alpha^{\prime}\beta^{\prime}}\varepsilon_{\dot{\alpha
}^{\prime}\dot{\beta}^{\prime}}X^{\alpha^{\prime}\beta^{\prime}\dot{\alpha
}^{\prime}\dot{\beta}^{\prime}},\nonumber\\
X^{(\alpha\beta)}  &  =\varepsilon_{\dot{\alpha}^{\prime}\dot{\beta}^{\prime}%
}S^{\alpha\beta}{}_{\alpha^{\prime}\beta^{\prime}}X^{\alpha^{\prime}%
\beta^{\prime}\dot{\alpha}^{\prime}\dot{\beta}^{\prime}},\nonumber\\
X^{(\dot{\alpha}\dot{\beta})}  &  =\varepsilon_{\alpha^{\prime}\beta^{\prime}%
}S^{\dot{\alpha}\dot{\beta}}{}_{\dot{\alpha}^{\prime}\dot{\beta}^{\prime}%
}X^{\alpha^{\prime}\beta^{\prime}\dot{\alpha}^{\prime}\dot{\beta}^{\prime}%
},\nonumber\\
X^{(\alpha\beta)(\dot{\alpha}\dot{\beta})}  &  =S^{\alpha\beta}{}%
_{\alpha^{\prime}\beta^{\prime}}S^{\dot{\alpha}\dot{\beta}}{}_{\dot{\alpha
}^{\prime}\dot{\beta}^{\prime}}X^{\alpha^{\prime}\beta^{\prime}\dot{\alpha
}^{\prime}\dot{\beta}^{\prime}}. \label{TensZer}%
\end{align}
Notice that the tensors in (\ref{TensZer}) correspond to the representations
$(0,0)$, $(1,0)$, $(0,1),$ and $(1,1).$

For a tensor transforming according to the representation $(1,0)\oplus(0,1)$
(like the field-strength tensor in electrodynamics) we have
\begin{equation}
F^{\mu\nu}=f_{(1,0)}^{\mu\nu}+f_{(0,1)}^{\mu\nu},\label{effmunu}
\end{equation}
with%
\begin{equation}
f_{(1,0)}^{\mu\nu}=\bar k\,(\bar{\sigma}^{\mu\nu})_{\dot{\alpha}\dot{\beta}}%
F^{\dot{\alpha}\dot{\beta}},\qquad f_{(0,1)}^{\mu\nu}=k \,(\sigma^{\mu\nu
})_{\alpha\beta}F^{\alpha\beta}. \label{242}
\end{equation}
The inverse spin matrices project out the constituents referring to the
representations $(1,0)$ and $(0,1)$ [cf.\thinspace(\ref{ContSpinMat})]:
\begin{equation}
F^{\dot{\alpha}\dot{\beta}}=\bar k_1(\bar{\sigma}_{\mu\nu}^{-1}%
)^{\dot{\alpha}\dot{\beta}}F^{\mu\nu},\qquad F^{{\alpha}{\beta}}=k_1({\sigma}_{\mu\nu}^{-1})^{{\alpha}{\beta}}F^{\mu\nu}.
\label{243} \end{equation}
The relations in (\ref{242}) are consistent with those in (\ref{243})
iff the constants satisfy the condition $k k_1=\bar k\bar
k_1=\lambda_+^{-1}$. These reasonings
hold for both four-dimensional spaces.

Realizing that%
\begin{align}
(\sigma^{\mu\nu})_{\alpha{\beta}} &  =S^{{\alpha}^{\prime}{\beta}^{\prime}}%
{}_{\alpha\beta}(\sigma^{\mu})_{\alpha^{\prime}\dot{\alpha}^{\prime\prime}%
}\varepsilon^{\dot{\alpha}^{\prime\prime}\dot{\beta}^{\prime\prime}}%
(\bar{\sigma}^{\nu})_{\dot{\beta}^{\prime\prime}\beta^{\prime}},\nonumber\\
(\bar{\sigma}^{\mu\nu})_{\dot{\alpha}\dot{\beta}} &  =-S^{\dot{\alpha}%
^{\prime}\dot{\beta}^{\prime}}{}_{\dot{\alpha}\dot{\beta}}(\bar{\sigma}^{\mu
})_{\dot{\alpha}^{\prime}{\alpha^{\prime\prime}}}\varepsilon^{{\alpha}%
^{\prime\prime}{\beta}^{\prime\prime}}({\sigma}^{\nu})_{{\beta}^{\prime\prime
}\dot{\beta}^{\prime}},\label{realize}
\end{align}
one can also show that
\begin{align}
f_{(1,0)}^{\mu\nu} &  =-\bar k (\bar{\sigma}^{\mu})_{\dot{\alpha}\alpha^{\prime}%
}\varepsilon^{\alpha^{\prime}\alpha^{\prime\prime}}(\sigma^{\nu}%
)_{\alpha^{\prime\prime}\dot{\beta}}F^{\dot{\alpha}\dot{\beta}},\nonumber\\
f_{(0,1)}^{\mu\nu} &  =k(\sigma^{\mu})_{\alpha\dot{\alpha}^{\prime}}%
\varepsilon^{\dot{\alpha}^{\prime}\dot{\alpha}^{\prime\prime}}(\bar{\sigma
}^{\nu})_{\dot{\alpha}^{\prime\prime}\beta}F^{\alpha\beta}.
\end{align}
The inversions of these identities are
\begin{align}
F^{\dot{\alpha}\dot{\beta}} &  =-\bar k_1(\bar{\sigma}_{\mu}^{-1})^{\dot{\alpha}%
\beta^{\prime}}\varepsilon_{\beta^{\prime}\beta^{\prime\prime}}(\sigma_{\nu
}^{-1})^{\beta^{\prime\prime}\dot{\beta}}f_{(1,0)}^{\mu\nu},\nonumber\\
F^{\alpha\beta} &  =k_1(\sigma_{\mu}^{-1})^{\alpha\dot{\beta}^{\prime}%
}\varepsilon_{\dot{\beta}^{\prime}\dot{\beta}^{\prime\prime}}(\bar{\sigma
}_{\nu}^{-1})^{\dot{\beta}^{\prime\prime}\beta}f_{(0,1)}^{\mu\nu}.\label{effmunu2}
\end{align}
The identities (\ref{effmunu}-\ref{effmunu2}) hold for the
four-dimensional Euclidean space, too, but with neglecting the
minus sign in (\ref{realize}-\ref{effmunu2}).

It is rather instructive to examine the irreducible constituents of an
arbitrary tensor with two contravariant Lorentz indices. To this end we use
the Fierz identity (\ref{fierzsigmaminkneu1}). The $k_{1}$-part of
(\ref{fierzsigmaminkneu1}) projects onto a one-dimensional space. It carries
the representation $(0,0)$ and refers to the quantum metric. A tensor
$T^{\mu\nu}$ that transforms like a scalar satisfies $T^{\mu\nu}=(P_{0}%
)^{\mu\nu}{}_{\mu^{\prime}\nu^{\prime}}T^{\mu^{\prime}\nu^{\prime}}$. With the
$k_{1}$-part of (\ref{fierzsigmaminkneu1}) we find for its spinorial
equivalent:
\begin{equation}
T^{\alpha\dot{\alpha}\beta\dot{\beta}}=k_{1}\varepsilon^{\alpha\beta^{\prime}%
}\hat{R}^{\dot{\alpha}\beta}{}_{\beta^{\prime}\dot{\alpha}^{\prime}%
}\varepsilon^{\dot{\alpha}^{\prime}\dot{\beta}}T,\quad T\equiv\eta_{\mu\nu
}T^{\mu\nu}.\label{TMatUm}%
\end{equation}
The field-strength tensor being subject to $F^{\mu\nu}=(P_{A})^{\mu\nu}{}%
_{\mu^{\prime}\nu^{\prime}}F^{\mu^{\prime}\nu^{\prime}}$ possesses as spinor
equivalent
\begin{equation}
F^{\alpha\dot{\alpha}\beta\dot{\beta}}=\,k_{2}\varepsilon^{\alpha\beta
^{\prime}}\hat{R}^{\dot{\alpha}\beta}{}_{\beta^{\prime}\dot{\alpha}^{\prime}%
}F^{\dot{\alpha}^{\prime}\dot{\beta}}+k_{3}F^{\alpha\beta^{\prime}}\hat
{R}^{\dot{\alpha}\beta}{}_{\beta^{\prime}\dot{\alpha}^{\prime}}\varepsilon
^{\dot{\alpha}^{\prime}\dot{\beta}},
\end{equation}
where the two summands again correspond to the representations $(1,0)$ and
$(0,1).$ Finally, we consider a symmetric tensor with $S^{\mu\nu}=(P_{S}%
)^{\mu\nu}{}_{\mu^{\prime}\nu^{\prime}}S^{\mu^{\prime}\nu^{\prime}}$. Its
spinor equivalent becomes%
\begin{equation}
S^{\alpha\dot{\alpha}\beta\dot{\beta}}=k_{4}(\sigma_{\mu\rho}^{-1}%
)^{\alpha\beta^{\prime}}(\hat{R}^{-1})^{\dot{\alpha}\beta}{}_{\beta^{\prime
}\dot{\alpha}^{\prime}}(\bar{\sigma}_{\kappa\nu}^{-1})^{\dot{\alpha}^{\prime
}\dot{\beta}}\eta^{\rho\kappa}S^{\mu\nu}.\label{SMatUm}%
\end{equation}

The expressions in (\ref{TMatUm})-(\ref{SMatUm}) hold for $q$-deformed Minkowski
space. If we deal with four-dimensional $q$-deformed Euclidean space we have to
start our considerations from the Fierz identity (\ref{fierzsigmaeu4neu}) and
drop all $\hat{R}$-matrices from our results. More concretely, we simply apply
the following substitutions to the results in (\ref{TMatUm})-(\ref{SMatUm}):%
\begin{equation}
\hat{R}^{\dot{\alpha}\beta}{}_{\beta^{\prime}\dot{\alpha}^{\prime}}%
\rightarrow\delta_{\dot{\alpha}^{\prime}}^{\dot{\alpha}}\delta_{\beta^{\prime
}}^{\beta},\qquad(\hat{R}^{-1})^{\dot{\alpha}\beta}{}_{\beta^{\prime}%
\dot{\alpha}^{\prime}}\rightarrow\delta_{\dot{\alpha}^{\prime}}^{\dot{\alpha}%
}\delta_{\beta^{\prime}}^{\beta}.\label{RTwist}%
\end{equation}

We conclude this section with an expression for the kinetic term of
electrodynamics. This requires to know the identities
\begin{align}
\eta_{\mu\mu^{\prime}}\eta_{\nu\nu^{\prime}}(\sigma^{\mu\nu})_{\alpha\beta
}(\sigma^{\mu^{\prime}\nu^{\prime}})_{\alpha^{\prime}\beta^{\prime}} &
=-q\varepsilon_{\alpha\beta^{\prime}}\varepsilon_{\beta\alpha^{\prime}}%
-q^{-1}\varepsilon_{\alpha\alpha^{\prime\prime}}\hat{R}^{\alpha^{\prime\prime
}\beta^{\prime\prime}}{}_{\beta\alpha^{\prime}}\varepsilon_{\beta
^{\prime\prime}\beta^{\prime}},\nonumber\\
&  =-q^{-1}\varepsilon_{\alpha\beta^{\prime}}\varepsilon_{\beta\alpha^{\prime
}}-q\varepsilon_{\alpha\alpha^{\prime\prime}}(\hat{R}^{-1})^{\alpha
^{\prime\prime}\beta^{\prime\prime}}{}_{\beta\alpha^{\prime}}\varepsilon
_{\beta^{\prime\prime}\beta^{\prime}}.\label{UmRech}%
\end{align}
Notice that these equalities remain valid if we substitute the spin matrices
on the left-hand side by their 'conjugated' counterparts. Moreover, if we
apply the substitutions (\ref{SubInv}) to the expressions in (\ref{UmRech}) we
obtain the corresponding identities in terms of inverse spin
matrices. Notice that Eq. (\ref{UmRech}) also holds for the
four-dimensional Euclidean space without any changes.

Finally, (\ref{UmRech}) enables us to express the kinetic term of
electrodynamics in a spinorial basis:%
\begin{align}
F^{\mu\nu}F_{\mu\nu}= &  \;F^{\mu\nu}F^{\mu^{\prime}\nu^{\prime}}\eta_{\mu
\mu^{\prime}}\eta_{\nu\nu^{\prime}}\nonumber\\
= &  \;((\bar{\sigma}^{\mu\nu})_{\dot{\alpha}\dot{\beta}}F^{\dot{\alpha}%
\dot{\beta}}+(\sigma^{\mu\nu})_{\alpha\beta}F^{\alpha\beta})\nonumber\\
&  \times((\bar{\sigma}_{\mu\nu}^{-1})^{\dot{\alpha}^{\prime}\dot{\beta
}^{\prime}}F_{\dot{\alpha}^{\prime}\dot{\beta}^{\prime}}+(\sigma_{\mu\nu}%
^{-1})^{\alpha^{\prime}\beta^{\prime}}F_{\alpha^{\prime}\beta^{\prime}%
})\nonumber\\
= &  \;-2q(\varepsilon_{\dot{\alpha}\dot{\beta}^{\prime}}\varepsilon
_{\dot{\beta}\dot{\alpha}^{\prime}}F^{\dot{\alpha}\dot{\beta}}F^{\dot{\alpha
}^{\prime}\dot{\beta}^{\prime}}+\varepsilon_{\alpha\beta^{\prime}}%
\varepsilon_{\beta\alpha^{\prime}}F^{\alpha\beta}F^{\alpha^{\prime}%
\beta^{\prime}}).\label{kineticterm}%
\end{align}

\section{Fierz identities for four-dimensional $q$-defor{\-}med
spaces\label{Fierz}}

Fierz identities are rules that enable us to reorder spinors in a product of
two bilinear forms. There are plenty of them, but we concentrate on
$q$-analogs of the Fierz identities given in Ref. \cite{bailin}. The main
purpose within this section is to show that it is possible to write down
$q$-deformed analogs of the most important Fierz identities. Once again, our
reasonings refer to the four-dimensional $q$-deformed Euclidean space and the
$q$-deformed Minkowski space.

Calculating $q$-deformed Fierz identities enforces us to fix commutation
relations between different sets of spinor components. Let us say a few words
about this matter. Each Fierz identity comprises four types of spinors which
we shall denote by $\theta^{i},$ $\phi^{i},$ $\eta^{i},$ and $\chi^{i}$. The
bilinear forms the Fierz identities are made of contain one conjugated and one
unconjugated spinor. Thus, we additionally have to take account of the
conjugated spinors $\bar{\theta}^{i},$ $\bar{\phi}^{i},$ $\bar{\eta}^{i},$ and
$\bar{\chi}^{i}$. (In the Euclidean case the role of the conjugated spinors is
played by those with a tilde.)

Now, we come to a consistent choice of commutation relations between the
different spinors. At first, components of the same spinor have to satisfy the
usual quantum plane relations. If we deal with\ symmetrized spinors we have,
for example,%
\begin{equation}
\bar{\eta}^{1}\bar{\eta}^{2}=q\,\bar{\eta}^{2}\bar{\eta}^{1},\qquad{\theta
}^{1}{\theta}^{2}=q\,{\theta}^{2}{\theta}^{1}.
\end{equation}
In the case of antisymmetrized spinors, however, the commutation relations
instead become%
\begin{equation}
\bar{\eta}^{1}\bar{\eta}^{2}=-q^{-1}\,\bar{\eta}^{2}\bar{\eta}^{1}%
,\qquad{\theta}^{1}{\theta}^{2}=-q^{-1}\,{\theta}^{2}{\theta}^{1}.
\end{equation}
We see that these relations are the same for conjugated and unconjugated spinors.

Next, we have to specify the commutation relations between components of
different spinors. We first discuss the situation for $q$-Lorentz symmetry.
For the relations between conjugated spinor components and unconjugated ones
we choose the braiding relation (\ref{braidqplane1}), if we deal with
symmetrized spinors. In this manner we have, for example,%
\begin{equation}
\bar{\theta}^{\dot{\alpha}}\theta^{\beta}=q\hat{R}_{\beta^{\prime}\dot{\alpha
}^{\prime}}^{\dot{\alpha}\beta}\theta^{\beta^{\prime}}\bar{\theta}%
^{\dot{\alpha}^{\prime}}, \label{BraidFierzAnf}%
\end{equation}
It is reasonable to assume that this braiding extends to all combinations of
conjugated and unconjugated spinors, i.e., for example,%
\begin{equation}
\bar{\theta}^{\dot{\alpha}}\eta^{\beta}=q\hat{R}_{\beta^{\prime}\dot{\alpha
}^{\prime}}^{\dot{\alpha}\beta}\eta^{\beta^{\prime}}\bar{\theta}^{\dot{\alpha
}^{\prime}}.
\end{equation}

It remains to determine the relations between two different spinors that are
both either unconjugated or conjugated. We have four options to fix these
relations. They can be represented as
\begin{align}
\mbox{R1}:\theta^{\alpha}\eta^{\beta} &  =q\hat{R}_{\beta^{\prime}%
\alpha^{\prime}}^{\alpha\beta}\eta^{\beta^{\prime}}\theta^{\alpha^{\prime}%
},\quad\bar{\theta}^{\dot{\alpha}}\bar{\eta}^{\dot{\beta}}=q\hat{R}%
_{\dot{\beta}^{\prime}\dot{\alpha}^{\prime}}^{\dot{\alpha}\dot{\beta}}%
\bar{\eta}^{\dot{\beta}^{\prime}}\bar{\theta}^{\dot{\alpha}^{\prime}%
};\nonumber\\
\mbox{R2}:\theta^{\alpha}\eta^{\beta} &  =q\hat{R}_{\beta^{\prime}%
\alpha^{\prime}}^{\alpha\beta}\eta^{\beta^{\prime}}\theta^{\alpha^{\prime}%
},\quad\bar{\theta}^{\dot{\alpha}}\bar{\eta}^{\dot{\beta}}=q^{-1}(\hat{R}%
^{-1})_{\dot{\beta}^{\prime}\dot{\alpha}^{\prime}}^{\dot{\alpha}\dot{\beta}%
}\bar{\eta}^{\dot{\beta}^{\prime}}\bar{\theta}^{\dot{\alpha}^{\prime}%
};\nonumber\\
\mbox{R3}:\theta^{\alpha}\eta^{\beta} &  =q^{-1}(\hat{R}^{-1})_{\beta^{\prime
}\alpha^{\prime}}^{\alpha\beta}\eta^{\beta^{\prime}}\theta^{\alpha^{\prime}%
},\quad\bar{\theta}^{\dot{\alpha}}\bar{\eta}^{\dot{\beta}}=q\hat{R}%
_{\dot{\beta}^{\prime}\dot{\alpha}^{\prime}}^{\dot{\alpha}\dot{\beta}}%
\bar{\eta}^{\dot{\beta}^{\prime}}\bar{\theta}^{\dot{\alpha}^{\prime}%
};\nonumber\\
\mbox{R4}:\theta^{\alpha}\eta^{\beta} &  =q^{-1}(\hat{R}^{-1})_{\beta^{\prime
}\alpha^{\prime}}^{\alpha\beta}\eta^{\beta^{\prime}}\theta^{\alpha^{\prime}%
},\quad\bar{\theta}^{\dot{\alpha}}\bar{\eta}^{\dot{\beta}}=q^{-1}(\hat{R}%
^{-1})_{\dot{\beta}^{\prime}\dot{\alpha}^{\prime}}^{\dot{\alpha}\dot{\beta}%
}\bar{\eta}^{\dot{\beta}^{\prime}}\bar{\theta}^{\dot{\alpha}^{\prime}%
}.\label{braidopt}%
\end{align}

If we\ consider antisymmetrized spinors the braiding between conjugated and
unconjugated spinors should be governed by (\ref{braidqplane2}), i.e.%
\begin{equation}
\bar{\theta}^{\dot{\alpha}}\theta^{\beta}=-q^{-1}\hat{R}_{\beta^{\prime}%
\dot{\alpha}^{\prime}}^{\dot{\alpha}\beta}\theta^{\beta^{\prime}}\bar{\theta
}^{\dot{\alpha}^{\prime}},\qquad\bar{\theta}^{\dot{\alpha}}\eta^{\beta
}=-q^{-1}\hat{R}_{\beta^{\prime}\dot{\alpha}^{\prime}}^{\dot{\alpha}\beta}%
\eta^{\beta^{\prime}}\bar{\theta}^{\dot{\alpha}},
\end{equation}
and the four options in (\ref{braidopt}) modify to
\begin{align}
\mbox{R1}:\theta^{\alpha}\eta^{\beta} &  =-q^{-1}\hat{R}_{\beta^{\prime}%
\alpha^{\prime}}^{\alpha\beta}\eta^{\beta^{\prime}}\theta^{\alpha^{\prime}%
},\quad\bar{\theta}^{\dot{\alpha}}\bar{\eta}^{\dot{\beta}}=-q^{-1}\hat
{R}_{\dot{\beta}^{\prime}\dot{\alpha}^{\prime}}^{\dot{\alpha}\dot{\beta}}%
\bar{\eta}^{\dot{\beta}^{\prime}}\bar{\theta}^{\dot{\alpha}^{\prime}%
};\nonumber\\
\mbox{R2}:\theta^{\alpha}\eta^{\beta} &  =-q^{-1}\hat{R}_{\beta^{\prime}%
\alpha^{\prime}}^{\alpha\beta}\eta^{\beta^{\prime}}\theta^{\alpha^{\prime}%
},\quad\bar{\theta}^{\dot{\alpha}}\bar{\eta}^{\dot{\beta}}=-q(\hat{R}%
^{-1})_{\dot{\beta}^{\prime}\dot{\alpha}^{\prime}}^{\dot{\alpha}\dot{\beta}%
}\bar{\eta}^{\dot{\beta}^{\prime}}\bar{\theta}^{\dot{\alpha}^{\prime}%
};\nonumber\\
\mbox{R3}:\theta^{\alpha}\eta^{\beta} &  =-q(\hat{R}^{-1})_{\beta^{\prime
}\alpha^{\prime}}^{\alpha\beta}\eta^{\beta^{\prime}}\theta^{\alpha^{\prime}%
},\quad\bar{\theta}^{\dot{\alpha}}\bar{\eta}^{\dot{\beta}}=-q^{-1}\hat
{R}_{\dot{\beta}^{\prime}\dot{\alpha}^{\prime}}^{\dot{\alpha}\dot{\beta}}%
\bar{\eta}^{\dot{\beta}^{\prime}}\bar{\theta}^{\dot{\alpha}^{\prime}%
};\nonumber\\
\mbox{R4}:\theta^{\alpha}\eta^{\beta} &  =-q(\hat{R}^{-1})_{\beta^{\prime
}\alpha^{\prime}}^{\alpha\beta}\eta^{\beta^{\prime}}\theta^{\alpha^{\prime}%
},\quad\bar{\theta}^{\dot{\alpha}}\bar{\eta}^{\dot{\beta}}=-q(\hat{R}%
^{-1})_{\dot{\beta}^{\prime}\dot{\alpha}^{\prime}}^{\dot{\alpha}\dot{\beta}%
}\bar{\eta}^{\dot{\beta}^{\prime}}\bar{\theta}^{\dot{\alpha}^{\prime}%
}.\label{BraidFierzEnd}%
\end{align}

To sum up, different spinors are braided if we lay down $q$-deformed Lorentz
symmetry as space-time symmetry. In the case of $U_{q}(so(4))$, however,
things simplify drastically, since the different spinors are no longer
braided. Thus, all $\hat{R}$-matrices and $q$'s in (\ref{BraidFierzAnf}%
)-(\ref{BraidFierzEnd}) have to vanish for that case [cf.\thinspace
Eq.\thinspace(\ref{RTwist})].

We wish to proceed in this section as follows. First, we present the
$q$-analog to a Fierz rearrangement rule of Ref. \cite{bailin}. It is written
down in its most general form and depends on several undetermined
coefficients. These formulae hold for both $q$-deformed Minkowski space and
four-dimensional $q$-deformed Euclidean space, if we insert the $\hat{R}%
$-matrix and the projectors of the quantum space under consideration. After
each Fierz identity we give different specifications of its coefficients.
These specifications correspond to the following choices: Do we have bosonic
or fermionic spinors, conjugated or unconjugated matrices, and which of the
four different possibilities in (\ref{braidopt}) is used? In this way we get
for both four-dimensional quantum spaces and each Fierz identity sixteen
possible specifications of the corresponding coefficients. A further
degeneracy comes from the fact that in some formulae we can replace the
$\hat{R}$-matrix (of the quantum space under consideration) by their inverse.
We ignore this possibility for lack of space.

Before starting with our listing we collect the definitions of the different
bilinear forms in spinor space:
\begin{align}
\theta\phi &  \equiv\varepsilon_{\alpha\beta}\;\theta^{\alpha}\phi^{\beta
},\nonumber\\
\bar{\theta}\bar{\phi} &  \equiv\varepsilon_{\dot{\alpha}\dot{\beta}}%
\;\bar{\theta}^{\dot{\alpha}}\bar{\phi}^{\dot{\beta}},\nonumber\\[0.06in]
\theta\sigma^{\mu}\bar{\phi} &  \equiv\theta^{\alpha}(\sigma^{\mu}%
)_{\alpha\dot{\beta}}\bar{\phi}^{\dot{\beta}},\nonumber\\
\bar{\theta}\bar{\sigma}^{\mu}\phi &  \equiv\bar{\theta}^{\dot{\alpha}}%
(\bar{\sigma}^{\mu})_{\dot{\alpha}\beta}\phi^{\beta},\nonumber\\[0.06in]
\theta\sigma^{\mu\nu}\phi &  \equiv\theta^{\alpha}(\sigma^{\mu\nu}%
)_{\alpha\beta}{\phi}^{\beta},\nonumber\\
\bar{\theta}\bar{\sigma}^{\mu\nu}\bar{\phi} &  \equiv\bar{\theta}^{\dot
{\alpha}}(\bar{\sigma}^{\mu\nu})_{\dot{\alpha}\dot{\beta}}{\bar{\phi}}%
^{\dot{\beta}}.\label{deffierz}%
\end{align}
Notice that in the Euclidean case the objects with bar have to be replaced by
those with tilde. It should also be mentioned that the Pauli matrices have
been defined in such a way that%
\begin{equation}
\theta\sigma^{\mu}\bar{\phi}\equiv\pm\bar{\phi}\bar{\sigma}^{\mu}\theta,
\end{equation}
where the minus sign holds for antisymmetrized spinors. Moreover, if we assume
that%
\begin{equation}
\theta^{\alpha}\phi^{\beta}=k\hat{R}_{\beta^{\prime}\alpha^{\prime}}%
^{\alpha\beta}\phi^{\beta^{\prime}}\theta^{\alpha^{\prime}},\quad\bar{\theta
}^{\dot{\alpha}}\bar{\phi}^{\dot{\beta}}=k\hat{R}_{\dot{\beta}^{\prime}%
\dot{\alpha}^{\prime}}^{\dot{\alpha}\dot{\beta}}\bar{\phi}^{\dot{\beta
}^{\prime}}\bar{\theta}^{\dot{\alpha}^{\prime}},\label{BraidRel}%
\end{equation}
where $k$ denotes a real constant, we have%
\begin{gather}
\theta\phi=-kq^{-1}\phi\theta,\quad\bar{\theta}\bar{\phi}=-kq^{-1}\bar{\phi
}\bar{\theta},\nonumber\\
\theta\sigma^{\mu\nu}\phi=kq\,\phi\sigma^{\mu\nu}\theta,\quad\bar{\theta}%
\bar{\sigma}^{\mu\nu}\bar{\phi}=kq\,\bar{\phi}\bar{\sigma}^{\mu\nu}\bar
{\theta}.\label{VerBil}%
\end{gather}
To prove these identities we first\ contract the braiding relations in
(\ref{BraidRel}) with the spinor metric or the spin matrices and then make use
of%
\begin{equation}
\varepsilon_{\alpha\beta}\hat{R}_{\beta^{\prime}\alpha^{\prime}}^{\alpha\beta
}=-q^{-1}\varepsilon_{\beta^{\prime}\alpha^{\prime}},\quad\sigma_{\alpha\beta
}^{\mu\nu}R_{\beta^{\prime}\alpha^{\prime}}^{\alpha\beta}=q\sigma
_{\beta^{\prime}\alpha^{\prime}}^{\mu\nu},\quad\bar{\sigma}_{\dot{\alpha}%
\dot{\beta}}^{\mu\nu}R_{\dot{\beta}^{\prime}\dot{\alpha}^{\prime}}%
^{\dot{\alpha}\dot{\beta}}=q\bar{\sigma}_{\dot{\beta}^{\prime}\dot{\alpha
}^{\prime}}^{\mu\nu}.
\end{equation}
Finally, let us note that starting in (\ref{BraidRel}) from braiding relations
with $\hat{R}^{-1}$ would yield expressions with $q$ and $q^{-1}$ being interchanged.

Now, we begin with the simplest Fierz identity whose undeformed counterpart is
called '\textbf{identity (A.4)}' in Ref. \cite{bailin}. It takes the general
form
\begin{equation}
(\theta\phi)(\bar{\chi}\bar{\eta})=k_{1}\;\eta_{\mu\nu}(\theta\sigma_{\mu}%
\bar{\eta})(\bar{\chi}\bar{\sigma}^{\nu}\phi).\label{Fierz1}%
\end{equation}
The values for the coefficient $k_{1}$ are specified in the following manner:

\begin{center}%
\begin{tabular}
[c]{r||c|c|c|c}\hline
Euclid & R1 & R2 & R3 & R4\\\hline
bos-uncon & $1$ & $1$ & $1$ & $1$\\\hline
bos-con & $1$ & $1$ & $1$ & $1$\\\hline
ferm-uncon & $-q^{-2}$ & $-q^{2}$ & $-q^{-2}$ & $-q^{2}$\\\hline
ferm-con & $-q^{-2}$ & $-q^{-2}$ & $-q^{2}$ & $-q^{2}$\\\hline
\end{tabular}
\newline\newline\newline%
\begin{tabular}
[c]{r||c|c|c|c}\hline
Mink & R1 & R2 & R3 & R4\\\hline
bos-uncon & $-q^{-3}$ & $-q^{-3}$ & $-q^{-3}$ & $-q^{-3}$\\\hline
bos-con & $-q^{3}$ & $-q^{3}$ & $-q^{3}$ & $-q^{3}$\\\hline
ferm-uncon & $q^{-1}$ & $q^{3}$ & $q^{3}$ & $q^{-1}$\\\hline
ferm-con & $q$ & $q$ & $q^{-3}$ & $q^{-3}$\\\hline
\end{tabular}
\newline
\end{center}

\noindent The terminus 'uncon' refers to the expression as it stands in
(\ref{Fierz1}), whereas the terminus 'con' is understood to mean the
following replacements applied to the Fierz identity:%
\begin{gather}
\theta\leftrightarrow\bar{\theta},\quad\phi\leftrightarrow\bar{\phi}%
,\quad{\chi}\leftrightarrow{\bar{\chi},\quad\eta}\leftrightarrow{\bar{\eta}%
}\nonumber\\
\sigma^{\mu\nu}\leftrightarrow\bar{\sigma}^{\mu\nu},\quad\sigma^{\nu
}\leftrightarrow\bar{\sigma}^{\nu}.
\end{gather}

\noindent\textbf{Identity (A.3)} in Ref. \cite{bailin}:
\begin{equation}
(\theta\phi)({\chi}{\eta})=k_{1}\;(\theta\eta)(\chi\phi)+\;k_{2}\;\eta
_{\nu\lambda}\eta_{\mu\kappa}(\theta\sigma^{\mu\nu}\eta)(\chi\sigma
^{\lambda\kappa}\phi).
\end{equation}
Table for $k_{1}$:

\begin{center}%
\begin{tabular}
[c]{r||c|c|c|c}\hline
Euclid & R1 & R2 & R3 & R4\\\hline
bos-uncon & $q^{-3}\lambda_{+}^{-1}$ & $q^{-3}\lambda_{+}^{-1}$ &
$q^{3}\lambda_{+}^{-1}$ & $q^{3}\lambda_{+}^{-1}$\\\hline
bos-con & $q^{-3}\lambda_{+}^{-1}$ & $q^{3}\lambda_{+}^{-1}$ & $q^{-3}%
\lambda_{+}^{-1}$ & $q^{3}\lambda_{+}^{-1}$\\\hline
ferm-uncon & $-q^{-1}\lambda_{+}^{-1}$ & $-q^{-1}\lambda_{+}^{-1}$ &
$-q\lambda_{+}^{-1}$ & $-q\lambda_{+}^{-1}$\\\hline
ferm-con & $-q^{-1}\lambda_{+}^{-1}$ & $-q\lambda_{+}^{-1}$ & $-q^{-1}%
\lambda_{+}^{-1}$ & $-q\lambda_{+}^{-1}$\\\hline
\end{tabular}
\newline\newline\newline%
\begin{tabular}
[c]{r||c|c|c|c}\hline
Mink & R1 & R2 & R3 & R4\\\hline
bos-uncon & $q^{-3}\lambda_{+}^{-1}$ & $q^{-3}\lambda_{+}^{-1}$ &
$q^{3}\lambda_{+}^{-1}$ & $q^{3}\lambda_{+}^{-1}$\\\hline
bos-con & $q^{3}\lambda_{+}^{-1}$ & $q^{-3}\lambda_{+}^{-1}$ & $q^{-3}%
\lambda_{+}^{-1}$ & $q^{3}\lambda_{+}^{-1}$\\\hline
ferm-uncon & $-q^{3}\lambda_{+}^{-1}$ & $-q^{3}\lambda_{+}^{-1}$ &
$-q^{-3}\lambda_{+}^{-1}$ & $-q^{-3}\lambda_{+}^{-1}$\\\hline
ferm-con & $-q^{3}\lambda_{+}^{-1}$ & $-q^{-3}\lambda_{+}^{-1}$ &
$-q^{3}\lambda_{+}^{-1}$ & $-q^{-3}\lambda_{+}^{-1}$\\\hline
\end{tabular}
\newline

\end{center}

\noindent Table for $k_{2}$:

\begin{center}%
\begin{tabular}
[c]{r||c|c|c|c}\hline
Euclid & R1 & R2 & R3 & R4\\\hline
bos-uncon & $q^{-3}\lambda_{+}^{-1}$ & $q^{-3}\lambda_{+}^{-1}$ &
$q^{3}\lambda_{+}^{-1}$ & $q^{3}\lambda_{+}^{-1}$\\\hline
bos-con & $q^{-3}\lambda_{+}^{-1}$ & $q^{3}\lambda_{+}^{-1}$ & $q^{-3}%
\lambda_{+}^{-1}$ & $q^{3}\lambda_{+}^{-1}$\\\hline
ferm-uncon & $-q^{-1}\lambda_{+}^{-1}$ & $-q^{-1}\lambda_{+}^{-1}$ &
$-q\lambda_{+}^{-1}$ & $-q\lambda_{+}^{-1}$\\\hline
ferm-con & $-q^{-1}\lambda_{+}^{-1}$ & $-q\lambda_{+}^{-1}$ & $-q^{-1}%
\lambda_{+}^{-1}$ & $-q\lambda_{+}^{-1}$\\\hline
\end{tabular}
\newline\newline\newline%
\begin{tabular}
[c]{r||c|c|c|c}\hline
Mink & R1 & R2 & R3 & R4\\\hline
bos-uncon & $q^{-3}\lambda_{+}^{-1}$ & $q^{-3}\lambda_{+}^{-1}$ &
$q^{3}\lambda_{+}^{-1}$ & $q^{3}\lambda_{+}^{-1}$\\\hline
bos-con & $q^{3}\lambda_{+}^{-1}$ & $q^{-3}\lambda_{+}^{-1}$ & $q^{-3}%
\lambda_{+}^{-1}$ & $q^{3}\lambda_{+}^{-1}$\\\hline
ferm-uncon & $-q^{3}\lambda_{+}^{-1}$ & $-q^{-3}\lambda_{+}^{-1}$ &
$-q^{3}\lambda_{+}^{-1}$ & $-q^{-3}\lambda_{+}^{-1}$\\\hline
ferm-con & $-q^{3}\lambda_{+}^{-1}$ & $-q^{_{3}}\lambda_{+}^{-1}$ &
$-q^{3}\lambda_{+}^{-1}$ & $-q^{_{3}}\lambda_{+}^{-1}$\\\hline
\end{tabular}
\newline
\end{center}

\noindent\textbf{Identity (A.5)} in Ref. \cite{bailin}:
\begin{equation}
(\theta\phi)(\chi\sigma^{\mu}\bar{\eta})=k_{1}\;(\theta\sigma^{\mu}\bar{\eta
})(\chi\phi)+k_{2}\;\eta_{\rho\nu}(\theta\sigma^{\rho}\bar{\eta})(\chi
\sigma^{\nu\mu}\phi).
\end{equation}

\noindent Table for $k_{1}$:

\begin{center}%
\begin{tabular}
[c]{r||c|c|c|c}\hline
Euclid & R1 & R2 & R3 & R4\\\hline
bos-uncon & $\lambda_{+}^{-1}$ & $\lambda_{+}^{-1}$ & $\lambda_{+}^{-1}$ &
$\lambda_{+}^{-1}$\\\hline
bos-con & $\lambda_{+}^{-1}$ & $\lambda_{+}^{-1}$ & $\lambda_{+}^{-1}$ &
$\lambda_{+}^{-1}$\\\hline
ferm-uncon & $-q^{2}\lambda_{+}^{-1}$ & $-q^{2}\lambda_{+}^{-1}$ &
$-q^{-2}\lambda_{+}^{-1}$ & $-q^{-2}\lambda_{+}^{-1}$\\\hline
ferm-con & $-q^{2}\lambda_{+}^{-1}$ & $-q^{-2}\lambda_{+}^{-1}$ &
$-q^{2}\lambda_{+}^{-1}$ & $-q^{-2}\lambda_{+}^{-1}$\\\hline
\end{tabular}
\newline\newline\newline%
\begin{tabular}
[c]{r||c|c|c|c}\hline
Mink & R1 & R2 & R3 & R4\\\hline
bos-uncon & $q^{-3}\lambda_{+}^{-1}$ & $q^{-3}\lambda_{+}^{-1}$ &
$q^{-3}\lambda_{+}^{-1}$ & $q^{-3}\lambda_{+}^{-1}$\\\hline
bos-con & $q^{3}\lambda_{+}^{-1}$ & $q^{3}\lambda_{+}^{-1}$ & $q^{3}%
\lambda_{+}^{-1}$ & $q^{3}\lambda_{+}^{-1}$\\\hline
ferm-uncon & $-q^{3}\lambda_{+}^{-1}$ & $-q^{3}\lambda_{+}^{-1}$ &
$-q^{-1}\lambda_{+}^{-1}$ & $-q^{-1}\lambda_{+}^{-1}$\\\hline
ferm-con & $-q^{-3}\lambda_{+}^{-1}$ & $-q^{3}\lambda_{+}^{-1}$ &
$-q^{3}\lambda_{+}^{-1}$ & $-q^{-3}\lambda_{+}^{-1}$\\\hline
\end{tabular}
\newline
\end{center}

\noindent Table for $k_{2}$:

\begin{center}%
\begin{tabular}
[c]{r||c|c|c|c}\hline
Euclid & R1 & R2 & R3 & R4\\\hline
bos-uncon & $-q^{-2}$ & $-q^{-2}$ & $-q^{2}$ & $-q^{2}$\\\hline
bos-con & $-q^{-2}$ & $-q^{2}$ & $-q^{-2}$ & $-q^{2}$\\\hline
ferm-uncon & $1$ & $1$ & $1$ & $1$\\\hline
ferm-con & $1$ & $1$ & $1$ & $1$\\\hline
\end{tabular}
\newline\newline\newline%
\begin{tabular}
[c]{r||c|c|c|c}\hline
Mink & R1 & R2 & R3 & R4\\\hline
bos-uncon & $-q^{-5}$ & $-q^{-5}$ & $-q^{-1}$ & $-q^{-1}$\\\hline
bos-con & $q^{5}$ & $q$ & $q^{5}$ & $q$\\\hline
ferm-uncon & $q$ & $q$ & $q$ & $q$\\\hline
ferm-con & $-q^{-1}$ & $-q^{-1}$ & $-q^{-1}$ & $-q^{-1}$\\\hline
\end{tabular}
\newline
\end{center}

\noindent\textbf{Identity (A.6)} in Ref. \cite{bailin}:
\begin{equation}
(\theta\phi)(\bar{\chi}\bar{\sigma}^{\mu}{\eta})=k_{1}\;(\theta\eta)(\bar
{\chi}\bar{\sigma}^{\mu}\phi)+k_{2}\;\eta_{\nu\rho}(\theta\sigma^{\mu\nu}%
\eta)(\bar{\chi}\bar{\sigma}^{\rho}\phi).
\end{equation}

\noindent Table for $k_{1}$:

\begin{center}%
\begin{tabular}
[c]{r||c|c|c|c}\hline
Euclid & R1 & R2 & R3 & R4\\\hline
bos-uncon & $q^{-3}\lambda_{+}^{-1}$ & $q^{-3}\lambda_{+}^{-1}$ &
$q^{3}\lambda_{+}^{-1}$ & $q^{3}\lambda_{+}^{-1}$\\\hline
bos-con & $q^{-3}\lambda_{+}^{-1}$ & $q^{3}\lambda_{+}^{-1}$ & $q^{-3}%
\lambda_{+}^{-1}$ & $q^{3}\lambda_{+}^{-1}$\\\hline
ferm-uncon & $-q^{-1}\lambda_{+}^{-1}$ & $-q^{-1}\lambda_{+}^{-1}$ &
$-q\lambda_{+}^{-1}$ & $-q\lambda_{+}^{-1}$\\\hline
ferm-con & $-q^{-1}\lambda_{+}^{-1}$ & $-q\lambda_{+}^{-1}$ & $-q^{-1}%
\lambda_{+}^{-1}$ & $-q\lambda_{+}^{-1}$\\\hline
\end{tabular}
\newline\newline\newline%
\begin{tabular}
[c]{r||c|c|c|c}\hline
Mink & R1 & R2 & R3 & R4\\\hline
bos-uncon & $q^{-3}\lambda_{+}^{-1}$ & $q^{-3}\lambda_{+}^{-1}$ &
$q^{3}\lambda_{+}^{-1}$ & $q^{3}\lambda_{+}^{-1}$\\\hline
bos-con & $q^{3}\lambda_{+}^{-1}$ & $q^{-3}\lambda_{+}^{-1}$ & $q^{-3}%
\lambda_{+}^{-1}$ & $q^{3}\lambda_{+}^{-1}$\\\hline
ferm-uncon & $-q^{-1}\lambda_{+}^{-1}$ & $-q^{-1}\lambda_{+}^{-1}$ &
$-q\lambda_{+}^{-1}$ & $-q\lambda_{+}^{-1}$\\\hline
ferm-con & $-q\lambda_{+}^{-1}$ & $-q\lambda_{+}^{-1}$ & $-q\lambda_{+}^{-1}$
& $-q\lambda_{+}^{-1}$\\\hline
\end{tabular}
\newline
\end{center}

\noindent Table for $k_{2}$:

\begin{center}%
\begin{tabular}
[c]{r||c|c|c|c}\hline
Euclid & R1 & R2 & R3 & R4\\\hline
bos-uncon & $-q^{-1}$ & $-q^{-1}$ & $-q$ & $-q$\\\hline
bos-con & $-q^{-1}$ & $-q$ & $-q^{-1}$ & $-q$\\\hline
ferm-uncon & $q$ & $q$ & $q^{-1}$ & $q^{-1}$\\\hline
ferm-con & $q$ & $q^{-1}$ & $q$ & $q^{-1}$\\\hline
\end{tabular}
\newline\newline\newline%
\begin{tabular}
[c]{r||c|c|c|c}\hline
Mink & R1 & R2 & R3 & R4\\\hline
bos-uncon & $-q^{-1}$ & $-q^{-1}$ & $-q$ & $-q$\\\hline
bos-con & $q$ & $q^{-1}$ & $q^{-1}$ & $q$\\\hline
ferm-uncon & $q$ & $q$ & $q^{-1}$ & $q^{-1}$\\\hline
ferm-con & $-q^{-1}$ & $-q$ & $-q$ & $-q^{-1}$\\\hline
\end{tabular}
\newline
\end{center}

\noindent\textbf{Identity (A.7)} in Ref. \cite{bailin}:
\begin{align}
(\theta\sigma^{\mu}\bar{\phi})(\chi\sigma^{\nu}\bar{\eta})= &  \;k_{1}%
\;(\theta\sigma^{\mu}\bar{\eta})(\chi\sigma^{\nu}\bar{\phi})+\;k_{2}\;\hat
{R}^{\mu\nu}{}_{\mu^{\prime}\nu^{\prime}}(\theta\sigma^{\mu^{\prime}}\bar
{\eta})(\chi\sigma^{\nu^{\prime}}\bar{\phi})\nonumber\\
&  +\;k_{3}\;\eta^{\mu\nu}\eta_{\rho\lambda}(\theta\sigma^{\rho}\bar{\eta
})(\chi\sigma^{\lambda}\bar{\phi})\nonumber\\
&  +\;k_{4}\;\varepsilon^{\mu\nu\kappa\lambda}\eta_{\lambda\rho}\eta
_{\kappa\beta}(\theta\sigma^{\rho}\bar{\eta})(\chi\sigma^{\beta}\bar{\phi}).
\end{align}

\noindent Table for $k_{1}$:

\begin{center}%
\begin{tabular}
[c]{r||c|c|c|c}\hline
Euclid & R1 & R2 & R3 & R4\\\hline
bos-uncon & $q^{-2}-1/2$ & $q^{2}/2$ & $q^{-2}-1/2$ & $q^{2}/2$\\\hline
bos-con & $q^{-2}-1/2$ & $q^{-2}-1/2$ & $q^{2}/2$ & $q^{2}/2$\\\hline
ferm-uncon & $q^{2}/2-1$ & $-1/2$ & $q^{2}/2-1$ & $-1/2$\\\hline
ferm-con & $q^{2}/2-1$ & $q^{2}/2-1$ & $-1/2$ & $-1/2$\\\hline
\end{tabular}
\newline\newline\newline%
\begin{tabular}
[c]{r||c|c|c|c}\hline
Mink & R1 & R2 & R3 & R4\\\hline
bos-uncon & $q^{2}-1/2$ & $q^{-2}/2$ & $q^{2}-1/2$ & $q^{-2}/2$\\\hline
bos-con & $q^{2}-1/2$ & $q^{-2}/2$ & $q^{2}-1/2$ & $q^{2}-1/2$\\\hline
ferm-uncon & $-(2-q^{-2})/2$ & $-1/2$ & $-1/2$ & $-(2-q^{-2})/2$\\\hline
ferm-con & $-1/2$ & $-(2-q^{2})/2$ & $-(2-q^{-2})/2$ & $-(2-q^{-2})/2$\\\hline
\end{tabular}
\newline
\end{center}

\noindent Table for $k_{2}$:

\begin{center}%
\begin{tabular}
[c]{r||c|c|c|c}\hline
Euclid & R1 & R2 & R3 & R4\\\hline
bos-uncon & $q^{-1}/2$ & $q/2$ & $q^{-1}/2$ & $q/2$\\\hline
bos-con & $q^{-1}/2$ & $q^{-1}/2$ & $q/2$ & $q/2$\\\hline
ferm-uncon & $-q/2$ & $-q^{-1}/2$ & $-q/2$ & $-q^{-1}/2$\\\hline
ferm-con & $-q/2$ & $-q/2$ & $-q^{-1}/2$ & $-q^{-1}/2$\\\hline
\end{tabular}
\newline\newline\newline%
\begin{tabular}
[c]{r||c|c|c|c}\hline
Mink & R1 & R2 & R3 & R4\\\hline
bos-uncon & $q^{2}/2$ & $1/2$ & $q^{2}/2$ & $1/2$\\\hline
bos-con & $q^{2}/2$ & $1/2$ & $q^{2}/2$ & $q^{2}/2$\\\hline
ferm-uncon & $-1/2$ & $q^{2}/2$ & $q^{2}/2$ & $-1/2$\\\hline
ferm-con & $-q^{2}/2$ & $-1/2$ & $-1/2$ & $-1/2$\\\hline
\end{tabular}
\newline
\end{center}

\noindent Table for $k_{3}$:

\begin{center}%
\begin{tabular}
[c]{r||c|c|c|c}\hline
Euclid & R1 & R2 & R3 & R4\\\hline
bos-uncon & $-q^{-2}/2$ & $-1/2$ & $-q^{-2}/2$ & $-1/2$\\\hline
bos-con & $-q^{-2}/2$ & $-q^{-2}/2$ & $-1/2$ & $-1/2$\\\hline
ferm-uncon & $1/2$ & $q^{-2}/2$ & $1/2$ & $q^{-2}/2$\\\hline
ferm-con & $1/2$ & $1/2$ & $q^{-2}/2$ & $q^{-2}/2$\\\hline
\end{tabular}
\newline\newline\newline%
\begin{tabular}
[c]{r||c|c|c|c}\hline
Mink & R1 & R2 & R3 & R4\\\hline
bos-uncon & $q^{2}/2$ & $-1/2$ & $q^{2}/2$ & $-1/2$\\\hline
bos-con & $q^{-2}/2$ & $-1/2$ & $-q^{2}/2$ & $-q^{2}/q$\\\hline
ferm-uncon & $1/2$ & $q^{2}/2$ & $q^{2}/2$ & $1/2$\\\hline
ferm-con & $q^{2}/2$ & $1/2$ & $1/2$ & $1/2$\\\hline
\end{tabular}
\newline
\end{center}

\noindent Table for $k_{4}$:

\begin{center}%
\begin{tabular}
[c]{r||c|c|c|c}\hline
Euclid & R1 & R2 & R3 & R4\\\hline
bos-uncon & $q^{-3}/2$ & $q^{-1}/2$ & $q^{-3}/2$ & $q^{-1}/2$\\\hline
bos-con & $-q^{-3}/2$ & $-q^{-3}/2$ & $-q^{-1}/2$ & $-q^{-1}/2$\\\hline
ferm-uncon & $-q^{-1}/2$ & $-q^{-3}/2$ & $-q^{-1}/2$ & $-q^{-3}/2$\\\hline
ferm-con & $q^{-1}/2$ & $q^{-1}/2$ & $q^{-3}/2$ & $q^{-3/2}$\\\hline
\end{tabular}
\newline\newline\newline%
\begin{tabular}
[c]{r||c|c|c|c}\hline
Mink & R1 & R2 & R3 & R4\\\hline
bos-uncon & $-q^{3}/2$ & $-q/2$ & $-q^{3}/2$ & $-q/2$\\\hline
bos-con & $q/2$ & $q/2$ & $q^{3}/2$ & $q^{3}/2$\\\hline
ferm-uncon & $q/2$ & $q^{3}/2$ & $q^{3}/2$ & $q/2$\\\hline
ferm-con & $-q^{3}/2$ & $-q^{3}/2$ & $-q/2$ & $-q/2$\\\hline
\end{tabular}
\newline
\end{center}

\noindent\textbf{Identity (A.8)} in Ref. \cite{bailin}:
\begin{align}
(\theta\sigma^{\mu}\bar{\phi})(\bar{\chi}\bar{\sigma}^{\nu}{\eta})= &
\;k_{1}\;\eta^{\mu\nu}(\theta\eta)(\bar{\chi}\bar{\phi})+\;k_{2}%
\;(\theta\sigma^{\mu\nu}\eta)(\bar{\chi}\bar{\phi})\nonumber\\
&  +\;k_{3}\;(\theta\eta)(\bar{\chi}\bar{\sigma}^{\mu\nu}\bar{\phi
})\nonumber\\
&  +\;k_{4}\;\hat{R}^{\mu\nu}{}_{\mu^{\prime}\nu^{\prime}}\eta_{\lambda\rho
}(\theta\sigma^{\mu^{\prime}\lambda}\eta)(\bar{\chi}\bar\sigma^{\rho\nu^{\prime}%
}\bar{\phi}).
\end{align}

\noindent Table for $k_{1}$:

\begin{center}%
\begin{tabular}
[c]{r||c|c|c|c}\hline
Euclid & R1 & R2 & R3 & R4\\\hline
bos-uncon & $\lambda_{+}^{-2}$ & $\lambda_{+}^{-2}$ & $\lambda_{+}^{-2}$ &
$\lambda_{+}^{-2}$\\\hline
bos-con & $\lambda_{+}^{-2}$ & $\lambda_{+}^{-2}$ & $\lambda_{+}^{-2}$ &
$\lambda_{+}^{-2}$\\\hline
ferm-uncon & $-q^{2}\lambda_{+}^{-2}$ & $-q^{-2}\lambda_{+}^{-2}$ &
$-q^{2}\lambda_{+}^{-2}$ & $-q^{-2}\lambda_{+}^{-2}$\\\hline
ferm-con & $-q^{2}\lambda_{+}^{-2}$ & $-q^{2}\lambda_{+}^{-2}$ &
$-q^{-2}\lambda_{+}^{-2}$ & $-q^{-2}\lambda_{+}^{-2}$\\\hline
\end{tabular}
\newline\newline\newline%
\begin{tabular}
[c]{r||c|c|c|c}\hline
Mink & R1 & R2 & R3 & R4\\\hline
bos-uncon & $-q^{3}\lambda_{+}^{-2}$ & $-q^{3}\lambda_{+}^{-2}$ &
$-q^{3}\lambda_{+}^{-2}$ & $-q^{3}\lambda_{+}^{-2}$\\\hline
bos-con & $-q^{-3}\lambda_{+}^{-2}$ & $-q^{-3}\lambda_{+}^{-2}$ &
$-q^{-3}\lambda_{+}^{-2}$ & $-q^{-3}\lambda_{+}^{-2}$\\\hline
ferm-uncon & $q^{-3}\lambda_{+}^{-2}$ & $q\lambda_{+}^{-1}$ & $q\lambda
_{+}^{-1}$ & $q^{-3}\lambda_{+}^{-2}$\\\hline
ferm-con & $q^{3}\lambda_{+}^{-2}$ & $q^{3}\lambda_{+}^{-2}$ & $q^{-1}%
\lambda_{+}^{-2}$ & $q^{-1}\lambda_{+}^{-2}$\\\hline
\end{tabular}
\newline
\end{center}

\noindent Table for $k_{2}$:

\begin{center}%
\begin{tabular}
[c]{r||c|c|c|c}\hline
Euclid & R1 & R2 & R3 & R4\\\hline
bos-uncon & $\lambda_{+}^{-1}$ & $\lambda_{+}^{-1}$ & $\lambda_{+}^{-1}$ &
$\lambda_{+}^{-1}$\\\hline
bos-con & $\lambda_{+}^{-1}$ & $\lambda_{+}^{-1}$ & $\lambda_{+}^{-1}$ &
$\lambda_{+}^{-1}$\\\hline
ferm-uncon & $-q^{2}\lambda_{+}^{-1}$ & $-q^{-2}\lambda_{+}^{-1}$ &
$-q^{2}\lambda_{+}^{-1}$ & $-q^{-2}\lambda_{+}^{-1}$\\\hline
ferm-con & $-q^{2}\lambda_{+}^{-1}$ & $-q^{2}\lambda_{+}^{-1}$ &
$-q^{-2}\lambda_{+}^{-1}$ & $-q^{-2}\lambda_{+}^{-1}$\\\hline
\end{tabular}
\newline\newline\newline%
\begin{tabular}
[c]{r||c|c|c|c}\hline
Mink & R1 & R2 & R3 & R4\\\hline
bos-uncon & $-q^{3}\lambda_{+}^{-1}$ & $-q^{3}\lambda_{+}^{-1}$ &
$-q^{3}\lambda_{+}^{-1}$ & $-q^{3}\lambda_{+}^{-1}$\\\hline
bos-con & $q^{-3}\lambda_{+}^{-1}$ & $q^{-3}\lambda_{+}^{-1}$ & $q^{-3}%
\lambda_{+}^{-1}$ & $q^{-3}\lambda_{+}^{-1}$\\\hline
ferm-uncon & $q^{3}\lambda_{+}^{-1}$ & $q\lambda_{+}^{-1}$ & $q\lambda
_{+}^{-1}$ & $q^{3}\lambda_{+}^{-1}$\\\hline
ferm-con & $-q^{3}\lambda_{+}^{-1}$ & $-q^{3}\lambda_{+}^{-1}$ &
$-q^{-1}\lambda_{+}^{-1}$ & $-q^{-1}\lambda_{+}^{-1}$\\\hline
\end{tabular}
\newline
\end{center}

\noindent Table for $k_{3}$:

\begin{center}%
\begin{tabular}
[c]{r||c|c|c|c}\hline
Euclid & R1 & R2 & R3 & R4\\\hline
bos-uncon & $-q^{-2}\lambda_{+}^{-1}$ & $-q^{2}\lambda_{+}^{-1}$ &
$-q^{-2}\lambda_{+}^{-1}$ & $-q^{2}\lambda_{+}^{-1}$\\\hline
bos-con & $-q^{-2}\lambda_{+}^{-1}$ & $-q^{-2}\lambda_{+}^{-1}$ &
$-q^{2}\lambda_{+}^{-1}$ & $-q^{2}\lambda_{+}^{-1}$\\\hline
ferm-uncon & $\lambda_{+}^{-1}$ & $\lambda_{+}^{-1}$ & $\lambda_{+}^{-1}$ &
$\lambda_{+}^{-1}$\\\hline
ferm-con & $\lambda_{+}^{-1}$ & $\lambda_{+}^{-1}$ & $\lambda_{+}^{-1}$ &
$\lambda_{+}^{-1}$\\\hline
\end{tabular}
\newline\newline\newline%
\begin{tabular}
[c]{r||c|c|c|c}\hline
Mink & R1 & R2 & R3 & R4\\\hline
bos-uncon & $-q^{5}\lambda_{+}^{-1}$ & $-q\lambda_{+}^{-1}$ & $-q\lambda
_{+}^{-1}$ & $-q^{5}\lambda_{+}^{-1}$\\\hline
bos-con & $q^{-5}\lambda_{+}^{-1}$ & $q^{-1}\lambda_{+}^{-1}$ & $q^{-1}%
\lambda_{+}^{-1}$ & $q^{-5}\lambda_{+}^{-1}$\\\hline
ferm-uncon & $q^{-1}\lambda_{+}^{-1}$ & $q^{-1}\lambda_{+}^{-1}$ &
$q^{-1}\lambda_{+}^{-1}$ & $q^{-1}\lambda_{+}^{-1}$\\\hline
ferm-con & $-q\lambda_{+}^{-1}$ & $-q\lambda_{+}^{-1}$ & $-q\lambda_{+}^{-1}$
& $-q\lambda_{+}^{-1}$\\\hline
\end{tabular}
\newline
\end{center}

\noindent Table for $k_{4}:$

\begin{center}%
\begin{tabular}
[c]{r||c|c|c|c}\hline
Euclid & R1 & R2 & R3 & R4\\\hline
bos-uncon & $-q^{-3}$ & $-q$ & $-q^{-3}$ & $-q$\\\hline
bos-con & $-q^{-3}$ & $-q^{-3}$ & $-q$ & $-q$\\\hline
ferm-uncon & $q^{-1}$ & $q^{-1}$ & $q^{-1}$ & $q^{-1}$\\\hline
ferm-con & $q^{-1}$ & $q^{-1}$ & $q^{-1}$ & $q^{-1}$\\\hline
\end{tabular}
\newline\newline\newline%
\begin{tabular}
[c]{r||c|c|c|c}\hline
Mink & R1 & R2 & R3 & R4\\\hline
bos-uncon & $-q^{7}$ & $-q^{3}$ & $-q^{3}$ & $-q^{7}$\\\hline
bos-con & $-q^{-3}$ & $-q$ & $-q$ & $-q^{-3}$\\\hline
ferm-uncon & $q$ & $q$ & $q$ & $q$\\\hline
ferm-con & $q^{3}$ & $q^{3}$ & $q^{3}$ & $q^{3}$\\\hline
\end{tabular}
\newline
\end{center}

\noindent\textbf{Identity (A.9)} in Ref. \cite{bailin}:
\begin{align}
(\theta\phi)(\chi\sigma^{\mu\nu}\eta)  &  =k_{1}\;(\theta\eta)(\chi\sigma
^{\mu\nu}\phi)+\;k_{2}\;(\theta\sigma^{\mu\nu}\eta)(\chi\phi)\nonumber\\
&  +\;k_{3}\;\eta_{\lambda\rho}(\theta\sigma^{\mu\lambda}\eta)(\chi\sigma
^{\rho\nu}\phi)\nonumber\\
&  +\;k_{4}\;\hat{R}^{\mu\nu}{}_{\mu^{\prime}\nu^{\prime}}\eta_{\lambda\rho
}(\theta\sigma^{\mu^{\prime}\lambda}\eta)(\chi\sigma^{\rho\nu^{\prime}}\phi).
\end{align}

\noindent Table for $k_{1}$:

\begin{center}%
\begin{tabular}
[c]{r||c|c|c|c}\hline
Euclid & R1 & R2 & R3 & R4\\\hline
bos-uncon & $q^{-3}\lambda_{+}^{-1}$ & $q^{-3}\lambda_{+}^{-1}$ &
$q^{3}\lambda_{+}^{-1}$ & $q^{3}\lambda_{+}^{-1}$\\\hline
bos-con & $q^{-3}\lambda_{+}^{-1}$ & $q^{3}\lambda_{+}^{-1}$ & $q^{-3}%
\lambda_{+}^{-1}$ & $q^{3}\lambda_{+}^{-1}$\\\hline
ferm-uncon & $-q^{-1}\lambda_{+}^{-1}$ & $-q^{-1}\lambda_{+}^{-1}$ &
$-q\lambda_{+}^{-1}$ & $-q\lambda_{+}^{-1}$\\\hline
ferm-con & $-q^{-1}\lambda_{+}^{-1}$ & $-q\lambda_{+}^{-1}$ & $-q^{-1}%
\lambda_{+}^{-1}$ & $-q\lambda_{+}^{-1}$\\\hline
\end{tabular}
\newline\newline\newline%
\begin{tabular}
[c]{r||c|c|c|c}\hline
Mink & R1 & R2 & R3 & R4\\\hline
bos-uncon & $q^{-7}\lambda_{+}^{-1}$ & $q^{-7}\lambda_{+}^{-1}$ &
$q^{7}\lambda_{+}^{-1}$ & $q^{7}\lambda_{+}^{-1}$\\\hline
bos-con & $q^{7}\lambda_{+}^{-1}$ & $q^{-7}\lambda_{+}^{-1}$ & $q^{7}%
\lambda_{+}^{-1}$ & $q^{-7}\lambda_{+}^{-1}$\\\hline
ferm-uncon & $-q^{-1}\lambda_{+}^{-1}$ & $-q^{-1}\lambda_{+}^{-1}$ &
$q^{-1}\lambda_{+}^{-1}$ & $q^{-1}\lambda_{+}^{-1}$\\\hline
ferm-con & $-q\lambda_{+}^{-1}$ & $-q^{-1}\lambda_{+}^{-1}$ & $-q\lambda
_{+}^{-1}$ & $-q^{-1}\lambda_{+}^{-1}$\\\hline
\end{tabular}
\newline
\end{center}

\noindent Table for $k_{2}$:

\begin{center}%
\begin{tabular}
[c]{r||c|c|c|c}\hline
Euclid & R1 & R2 & R3 & R4\\\hline
bos-uncon & $q\lambda_{+}^{-1}$ & $q\lambda_{+}^{-1}$ & $q^{-1}\lambda
_{+}^{-1}$ & $q^{-1}\lambda_{+}^{-1}$\\\hline
bos-con & $q\lambda_{+}^{-1}$ & $q^{-1}\lambda_{+}^{-1}$ & $q\lambda_{+}^{-1}$
& $q^{-1}\lambda_{+}^{-1}$\\\hline
ferm-uncon & $-q^{3}\lambda_{+}^{-1}$ & $-q^{3}\lambda_{+}^{-1}$ &
$-q^{-3}\lambda_{+}^{-1}$ & $-q^{-3}\lambda_{+}^{-1}$\\\hline
ferm-con & $-q^{3}\lambda_{+}^{-1}$ & $-q^{-3}\lambda_{+}^{-1}$ &
$-q^{3}\lambda_{+}^{-1}$ & $-q^{-3}\lambda_{+}^{-1}$\\\hline
\end{tabular}
\newline\newline\newline%
\begin{tabular}
[c]{r||c|c|c|c}\hline
Mink & R1 & R2 & R3 & R4\\\hline
bos-uncon & $q^{-3}\lambda_{+}^{-1}$ & $q^{-3}\lambda_{+}^{-1}$ &
$q^{3}\lambda_{+}^{-1}$ & $q^{3}\lambda_{+}^{-1}$\\\hline
bos-con & $q^{3}\lambda_{+}^{-1}$ & $q^{-3}\lambda_{+}^{-1}$ & $q^{3}%
\lambda_{+}^{-1}$ & $q^{-3}\lambda_{+}^{-1}$\\\hline
ferm-uncon & $-q^{3}\lambda_{+}^{-1}$ & $-q^{3}\lambda_{+}^{-1}$ &
$-q^{-3}\lambda_{+}^{-1}$ & $-q^{-3}\lambda_{+}^{-1}$\\\hline
ferm-con & $-q^{-3}\lambda_{+}^{-1}$ & $-q^{3}\lambda_{+}^{-1}$ &
$-q^{-3}\lambda_{+}^{-1}$ & $-q^{3}\lambda_{+}^{-1}$\\\hline
\end{tabular}
\newline
\end{center}

\noindent Table for $k_{3}$:

\begin{center}%
\begin{tabular}
[c]{r||c|c|c|c}\hline
Euclid & R1 & R2 & R3 & R4\\\hline
bos-uncon & $-q^{-1}\lambda_{+}^{-1}$ & $-q^{-1}\lambda_{+}^{-1}$ &
$-q\lambda_{+}^{-1}$ & $-q\lambda_{+}^{-1}$\\\hline
bos-con & $-q^{-2}\lambda_{+}^{-1}$ & $-\lambda_{+}^{-1}$ & $-q^{-2}%
\lambda_{+}^{-1}$ & $-\lambda_{+}^{-1}$\\\hline
ferm-uncon & $\lambda_{+}^{-1}$ & $\lambda_{+}^{-1}$ & $q^{-2}\lambda_{+}%
^{-1}$ & $q^{-2}\lambda_{+}^{-1}$\\\hline
ferm-con & $\lambda_{+}^{-1}$ & $q^{-2}\lambda_{+}^{-1}$ & $\lambda_{+}^{-1}$
& $q^{-2}\lambda_{+}^{-1}$\\\hline
\end{tabular}
\newline\newline\newline%
\begin{tabular}
[c]{r||c|c|c|c}\hline
Mink & R1 & R2 & R3 & R4\\\hline
bos-uncon & $-q^{-4}\lambda_{+}^{-1}$ & $-q^{-4}\lambda_{+}^{-1}$ &
$-q^{6}\lambda_{+}^{-1}$ & $-q^{6}\lambda_{+}^{-1}$\\\hline
bos-con & $q^{6}\lambda_{+}^{-1}$ & $q^{-4}\lambda_{+}^{-1}$ & $q^{6}%
\lambda_{+}^{-1}$ & $q^{-4}\lambda_{+}^{-1}$\\\hline
ferm-uncon & $q^{2}\lambda_{+}^{-1}$ & $q^{2}\lambda_{+}^{-1}$ & $\lambda
_{+}^{-1}$ & $\lambda_{+}^{-1}$\\\hline
ferm-con & $-\lambda_{+}^{-1}$ & $-q^{2}\lambda_{+}^{-1}$ & $-\lambda_{+}%
^{-1}$ & $-q^{2}\lambda_{+}^{-1}$\\\hline
\end{tabular}
\newline
\end{center}

\noindent Table for $k_{4}$:

\begin{center}%
\begin{tabular}
[c]{r||c|c|c|c}\hline
Euclid & R1 & R2 & R3 & R4\\\hline
bos-uncon & $q\lambda_{+}^{-1}$ & $q\lambda_{+}^{-1}$ & $q^{3}\lambda_{+}%
^{-1}$ & $q^{3}\lambda_{+}^{-1}$\\\hline
bos-con & $q\lambda_{+}^{-1}$ & $q^{3}\lambda_{+}^{-1}$ & $q\lambda_{+}^{-1}$
& $q^{3}\lambda_{+}^{-1}$\\\hline
ferm-uncon & $-q^{3}\lambda_{+}^{-1}$ & $-q^{3}\lambda_{+}^{-1}$ &
$-q\lambda_{+}^{-1}$ & $-q\lambda_{+}^{-1}$\\\hline
ferm-con & $-q^{3}\lambda_{+}^{-1}$ & $-q\lambda_{+}^{-1}$ & $-q^{3}%
\lambda_{+}^{-1}$ & $-q\lambda_{+}^{-1}$\\\hline
\end{tabular}
\newline\newline\newline%
\begin{tabular}
[c]{r||c|c|c|c}\hline
Mink & R1 & R2 & R3 & R4\\\hline
bos-uncon & $q^{-6}\lambda_{+}^{-1}$ & $q^{-6}\lambda_{+}^{-1}$ &
$q^{4}\lambda_{+}^{-1}$ & $q^{4}\lambda_{+}^{-1}$\\\hline
bos-con & $-q^{4}\lambda_{+}^{-1}$ & $-q^{-6}\lambda_{+}^{-1}$ &
$-q^{4}\lambda_{+}^{-1}$ & $-q^{-6}\lambda_{+}^{-1}$\\\hline
ferm-uncon & $-\lambda_{+}^{-1}$ & $-\lambda_{+}^{-1}$ & $-q^{-2}\lambda
_{+}^{-1}$ & $-q^{-2}\lambda_{+}^{-1}$\\\hline
ferm-con & $q^{-2}\lambda_{+}^{-1}$ & $\lambda_{+}^{-1}$ & $\lambda_{+}^{-1}$
& $q^{-2}\lambda_{+}^{-1}$\\\hline
\end{tabular}
\newline
\end{center}

\noindent\textbf{Identity (A.10)} in Ref. \cite{bailin}:
\begin{align}
(\theta\phi)(\bar{\chi}\bar{\sigma}^{\mu\nu}\bar{\eta})= &  \;k_{1}%
\;(P_{A})^{\mu\nu}{}_{\mu^{\prime}\nu^{\prime}}(\theta\sigma^{\mu^{\prime}%
}\bar{\eta})(\bar{\chi}\bar{\sigma}^{\nu^{\prime}}\phi)\nonumber\\
&  +\;k_{2}\;\varepsilon^{\mu\nu\kappa\lambda}\eta_{\lambda\rho}\eta
_{\kappa\rho^{\prime}}(\theta\sigma^{\rho}\bar{\eta})(\bar{\chi}\bar{\sigma
}^{\rho^{\prime}}\phi).
\end{align}

\noindent Table for $k_{1}$:

\begin{center}%
\begin{tabular}
[c]{r||c|c|c|c}\hline
Euclid & R1 & R2 & R3 & R4\\\hline
bos-uncon & $-q^{2}\lambda_{+}/2$ & $-q^{-2}\lambda_{+}/2$ & $-q^{2}%
\lambda_{+}/2$ & $-q^{-2}\lambda_{+}/2$\\\hline
bos-con & $-q^{2}\lambda_{+}/2$ & $-q^{2}\lambda_{+}/2$ & $-q^{-2}\lambda
_{+}/2$ & $-q^{-2}\lambda_{+}/2$\\\hline
ferm-uncon & $\lambda_{+}/2$ & $\lambda_{+}/2$ & $\lambda_{+}/2$ &
$\lambda_{+}/2$\\\hline
ferm-con & $\lambda_{+}/2$ & $\lambda_{+}/2$ & $\lambda_{+}/2$ & $\lambda
_{+}/2$\\\hline
\end{tabular}
\newline\newline\newline%
\begin{tabular}
[c]{r||c|c|c|c}\hline
Mink & R1 & R2 & R3 & R4\\\hline
bos-uncon & $q^{-1}\lambda_{+}/2$ & $-q^{-5}\lambda_{+}/2$ & $q^{-1}%
\lambda_{+}/2$ & $-q^{-5}\lambda_{+}/2$\\\hline
bos-con & $-q^{-1}\lambda_{+}/2$ & $-q^{-5}\lambda_{+}/2$ & $-q^{-1}%
\lambda_{+}/2$ & $-q^{-5}\lambda_{+}/2$\\\hline
ferm-uncon & $q/2\lambda_{+}$ & $q/2\lambda_{+}$ & $q/2\lambda_{+}$ &
$q/2\lambda_{+}$\\\hline
ferm-con & $-q^{-1}\lambda_{+}^{-1}/2$ & $-q^{-1}\lambda_{+}^{-1}/2$ &
$-q^{-1}\lambda_{+}^{-1}/2$ & $-q^{-1}\lambda_{+}^{-1}/2$\\\hline
\end{tabular}
\newline
\end{center}

\noindent Table for $k_{2}:$

\begin{center}%
\begin{tabular}
[c]{r||c|c|c|c}\hline
Euclid & R1 & R2 & R3 & R4\\\hline
bos-uncon & $-1/2$ & $-q^{-4}/2$ & $-1/2$ & $-q^{-4}/2$\\\hline
bos-con & $1/2$ & $1/2$ & $q^{-4}/2$ & $q^{-4}/2$\\\hline
ferm-uncon & $q^{-2}/2$ & $q^{-2}/2$ & $q^{-2}/2$ & $q^{-2}/2$\\\hline
ferm-con & $-q^{-2}/2$ & $-q^{-2}/2$ & $-q^{-2}/2$ & $-q^{-2}/2$\\\hline
\end{tabular}
\newline\newline\newline%
\begin{tabular}
[c]{r||c|c|c|c}\hline
Mink & R1 & R2 & R3 & R4\\\hline
bos-uncon & $q/2$ & $q^{-3}/2$ & $q/2$ & $q^{-3}/2$\\\hline
bos-con & $q/2$ & $q^{-3}/2$ & $q/2$ & $q^{-3}/2$\\\hline
ferm-uncon & $-q^{3}/2$ & $-q^{3}/2$ & $-q^{3}/2$ & $-q^{3}/2$\\\hline
ferm-con & $-q/2$ & $-q/2$ & $-q/2$ & $-q/2$\\\hline
\end{tabular}
\newline
\end{center}

\noindent\textbf{Identity (A.11)} in Ref. \cite{bailin}:
\begin{align}
(\theta\sigma^{\mu\nu}\phi)(\chi\sigma^{\lambda}\bar{\eta})= &  \;k_{1}%
\;(P_{A})^{\mu\nu}{}_{\mu^{\prime}\nu^{\prime}}\eta^{\nu^{\prime}\lambda
}(\theta\sigma^{\mu^{\prime}}\bar{\eta})(\chi\phi)\nonumber\\
&  +\;k_{2}\;(P_{A})^{\mu\nu}{}_{\mu^{\prime}\nu^{\prime}}(\theta\sigma
^{\mu^{\prime}}\bar{\eta})(\chi\sigma^{\nu^{\prime}\lambda}\phi)\nonumber\\
&  +\;k_{3}\;\varepsilon^{\mu\nu\lambda\rho}\eta_{\rho\gamma}(\theta
\sigma^{\gamma}\bar{\eta})(\chi\phi)\nonumber\\
&  +\;k_{4}\;\varepsilon^{\mu\nu\kappa\rho}\eta_{\rho\alpha}\eta
_{\kappa\gamma}(\theta\sigma^{\alpha}\bar{\eta})(\chi\sigma^{\gamma\lambda
}\phi).
\end{align}

\noindent Table for $k_{1}$:

\begin{center}%
\begin{tabular}
[c]{r||c|c|c|c}\hline
Euclid & R1 & R2 & R3 & R4\\\hline
bos-uncon & $1/2$ & $1/2$ & $1/2$ & $1/2$\\\hline
bos-con & $1/2$ & $1/2$ & $1/2$ & $1/2$\\\hline
ferm-uncon & $-q^{2}/2$ & $-q^{2}/2$ & $-q^{-2}/2$ & $-q^{-2}/2$\\\hline
ferm-con & $-q^{2}/2$ & $-q^{-2}/2$ & $-q^{2}/2$ & $-q^{-2}/2$\\\hline
\end{tabular}
\newline\newline\newline%
\begin{tabular}
[c]{r||c|c|c|c}\hline
Mink & R1 & R2 & R3 & R4\\\hline
bos-uncon & $q^{-3}/2$ & $q^{-3}/2$ & $q^{-3}/2$ & $q^{-3}/2$\\\hline
bos-con & $-q^{3}/2$ & $-q^{3}/2$ & $-q^{3}/2$ & $-q^{3}/2$\\\hline
ferm-uncon & $-q^{3}/2$ & $-q^{3}/2$ & $-q^{-1}/2$ & $-q^{-1}/2$\\\hline
ferm-con & $q^{3}/2$ & $q/2$ & $q/2$ & $q^{3}/2$\\\hline
\end{tabular}
\newline
\end{center}

\noindent Table for $k_{2}$:

\begin{center}%
\begin{tabular}
[c]{r||c|c|c|c}\hline
Euclid & R1 & R2 & R3 & R4\\\hline
bos-uncon & $-q^{-2}\lambda_{+}/2$ & $-q^{-2}\lambda_{+}/2$ & $-q^{2}%
\lambda_{+}/2$ & $-q^{2}\lambda_{+}/2$\\\hline
bos-con & $-q^{-2}\lambda_{+}/2$ & $-q^{2}\lambda_{+}/2$ & $-q^{-2}\lambda
_{+}/2$ & $-q^{2}\lambda_{+}/2$\\\hline
ferm-uncon & $\lambda_{+}/2$ & $\lambda_{+}/2$ & $\lambda_{+}/2$ &
$\lambda_{+}/2$\\\hline
ferm-con & $\lambda_{+}/2$ & $\lambda_{+}/2$ & $\lambda_{+}/2$ & $\lambda
_{+}/2$\\\hline
\end{tabular}
\newline\newline\newline%
\begin{tabular}
[c]{r||c|c|c|c}\hline
Mink & R1 & R2 & R3 & R4\\\hline
bos-uncon & $-q^{-5}\lambda_{+}/2$ & $-q^{-5}\lambda_{+}/2$ & $q^{-1}%
\lambda_{+}/2$ & $q^{-1}\lambda_{+}/2$\\\hline
bos-con & $-q^{5}\lambda_{+}/2$ & $-q\lambda_{+}/2$ & $-q\lambda_{+}/2$ &
$-q^{5}\lambda_{+}/2$\\\hline
ferm-uncon & $q\lambda_{+}/2$ & $q\lambda_{+}/2$ & $q\lambda_{+}/2$ &
$q\lambda_{+}/2$\\\hline
ferm-con & $q^{-1}\lambda_{+}/2$ & $q^{-1}\lambda_{+}/2$ & $q^{-1}\lambda
_{+}/2$ & $q^{-1}\lambda_{+}/2$\\\hline
\end{tabular}
\newline
\end{center}

\noindent Table for $k_{3}$:

\begin{center}%
\begin{tabular}
[c]{r||c|c|c|c}\hline
Euclid & R1 & R2 & R3 & R4\\\hline
bos-uncon & $-q^{-2}\lambda_{+}^{-1}/2$ & $-q^{-2}\lambda_{+}^{-1}/2$ &
$-q^{-2}\lambda_{+}^{-1}/2$ & $-q^{-2}\lambda_{+}^{-1}/2$\\\hline
bos-con & $q^{-2}\lambda_{+}^{-1}/2$ & $q^{-2}\lambda_{+}^{-1}/2$ &
$q^{-2}\lambda_{+}^{-1}/2$ & $q^{-2}\lambda_{+}^{-1}/2$\\\hline
ferm-uncon & $\lambda_{+}^{-1}/2$ & $\lambda_{+}^{-1}/2$ & $q^{-4}\lambda
_{+}^{-1}/2$ & $q^{-4}\lambda_{+}^{-1}/2$\\\hline
ferm-con & $-\lambda_{+}^{-1}/2$ & $-q^{-4}\lambda_{+}^{-1}/2$ & $-\lambda
_{+}^{-1}/2$ & $-q^{-4}\lambda_{+}^{-1}/2$\\\hline
\end{tabular}
\newline\newline\newline%
\begin{tabular}
[c]{r||c|c|c|c}\hline
Mink & R1 & R2 & R3 & R4\\\hline
bos-uncon & $q^{-1}\lambda_{+}^{-1}/2$ & $q^{-1}\lambda_{+}^{-1}/2$ &
$q^{-1}\lambda_{+}^{-1}/2$ & $q^{-1}\lambda_{+}^{-1}/2$\\\hline
bos-con & $q^{5}\lambda_{+}^{-1}/2$ & $q^{5}\lambda_{+}^{-1}/2$ &
$q^{5}\lambda_{+}^{-1}/2$ & $q^{5}\lambda_{+}^{-1}/2$\\\hline
ferm-uncon & $-q^{5}\lambda_{+}^{-1}/2$ & $q^{5}\lambda_{+}^{-1}/2$ &
$-q\lambda_{+}^{-1}/2$ & $-q\lambda_{+}^{-1}/2$\\\hline
ferm-con & $-q^{-1}\lambda_{+}^{-1}/2$ & $q^{3}\lambda_{+}^{-1}/2$ &
$q^{3}\lambda_{+}^{-1}/2$ & $-q^{-1}\lambda_{+}^{-1}/2$\\\hline
\end{tabular}
\newline
\end{center}

\noindent Table for $k_{4}$:

\begin{center}%
\begin{tabular}
[c]{r||c|c|c|c}\hline
Euclid & R1 & R2 & R3 & R4\\\hline
bos-uncon & $q^{-4}/2$ & $q^{-4}/2$ & $1/2$ & $1/2$\\\hline
bos-con & $-q^{-4}/2$ & $-1/2$ & $-q^{-4}/2$ & $-1/2$\\\hline
ferm-uncon & $-q^{-2}/2$ & $-q^{-2}/2$ & $-q^{-2}/2$ & $-q^{-2}/2$\\\hline
ferm-con & $q^{-2}/2$ & $q^{-2}/2$ & $q^{-2}/2$ & $q^{-2}/2$\\\hline
\end{tabular}
\newline\newline\newline%
\begin{tabular}
[c]{r||c|c|c|c}\hline
Mink & R1 & R2 & R3 & R4\\\hline
bos-uncon & $-q^{-3}/2$ & $-q^{-3}/2$ & $-q/2$ & $-q/2$\\\hline
bos-con & $q^{7}/2$ & $q^{3}/2$ & $q^{3}/2$ & $q^{7}/2$\\\hline
ferm-uncon & $q^{3}/2$ & $q^{3}/2$ & $q^{3}/2$ & $q^{3}/2$\\\hline
ferm-con & $-q/2$ & $-q/2$ & $-q/2$ & $-q/2$\\\hline
\end{tabular}
\newline
\end{center}

\noindent\textbf{Identity (A.12)} in Ref. \cite{bailin}:
\begin{align}
(\theta\sigma^{\mu\nu}\phi)(\bar{\chi}\bar{\sigma}^{\lambda}{\eta})= &
\;k_{1}\;(P_{A})^{\mu\nu}{}_{\mu^{\prime}\nu^{\prime}}\eta^{\nu^{\prime
}\lambda}(\theta\eta)(\bar{\chi}\bar{\sigma}^{\mu^{\prime}}\phi)\nonumber\\
&  +\;k_{2}\;(P_{A})^{\mu\nu}{}_{\mu^{\prime}\nu^{\prime}}\hat{R}^{\nu
^{\prime}\lambda}{}_{\nu^{\prime\prime}\lambda^{\prime\prime}}(\theta
\sigma^{\mu^{\prime}\nu^{\prime\prime}}\eta)(\bar{\chi}\bar{\sigma}%
^{\lambda^{\prime\prime}}\phi)\nonumber\\
&  +\;k_{3}\;\varepsilon^{\mu\nu\lambda\rho}\eta_{\rho\rho^{\prime}}%
(\theta\eta)(\bar{\chi}\bar{\sigma}^{\rho^{\prime}}\phi)\nonumber\\
&  +\;k_{4}\;\varepsilon^{\mu\nu\rho\kappa}\hat{R}^{\rho^{\prime}\lambda}%
{}_{\rho^{\prime\prime}\lambda^{\prime}}\eta_{\kappa\kappa^{\prime}}\eta
_{\rho\rho^{\prime}}(\theta\sigma^{\kappa^{\prime}\rho^{\prime\prime}}%
\eta)(\bar{\chi}\bar{\sigma}^{\lambda^{\prime}}\phi).
\end{align}

\noindent Table for $k_{1}$:

\begin{center}%
\begin{tabular}
[c]{r||c|c|c|c}\hline
Euclid & R1 & R2 & R3 & R4\\\hline
bos-uncon & / & / & $-q/2$ & $-q/2$\\\hline
bos-con & / & $-q/2$ & / & $-q/2$\\\hline
ferm-uncon & / & / & $q^{-1}/2$ & $q^{-1}/2$\\\hline
ferm-con & / & $q/2$ & / & $q/2$\\\hline
\end{tabular}
\newline\newline\newline%
\begin{tabular}
[c]{r||c|c|c|c}\hline
Mink & R1 & R2 & R3 & R4\\\hline
bos-uncon & $-q^{-1}/2$ & $-q^{-1}/2$ & / & /\\\hline
bos-con & $q^{-1}/2$ & $q^{-1}/2$ & / & /\\\hline
ferm-uncon & $q/2$ & $q/2$ & / & /\\\hline
ferm-con & / & $-q/2$ & $-q/2$ & /\\\hline
\end{tabular}
\newline
\end{center}

\noindent Table for $k_{2}$:

\begin{center}%
\begin{tabular}
[c]{r||c|c|c|c}\hline
Euclid & R1 & R2 & R3 & R4\\\hline
bos-uncon & / & / & $q^{-1}\lambda_{+}^{-1}/2$ & $q^{-1}\lambda_{+}^{-1}%
/2$\\\hline
bos-con & / & $-q^{-1}\lambda_{+}^{-1}/2$ & / & $-q^{-1}\lambda_{+}^{-1}%
/2$\\\hline
ferm-uncon & / & / & $-q^{-3}\lambda_{+}^{-1}/2$ & $-q^{-3}\lambda_{+}^{-1}%
/2$\\\hline
ferm-con & / & $q^{-1}\lambda_{+}^{-1}/2$ & / & $q^{-1}\lambda_{+}^{-1}%
/2$\\\hline
\end{tabular}
\newline\newline\newline%
\begin{tabular}
[c]{r||c|c|c|c}\hline
Mink & R1 & R2 & R3 & R4\\\hline
bos-uncon & $-q\lambda_{+}^{-1}/2$ & $-q\lambda_{+}^{-1}/2$ & / & /\\\hline
bos-con & $-q\lambda_{+}^{-1}/2$ & $-q\lambda_{+}^{-1}/2$ & / & /\\\hline
ferm-uncon & $q^{3}\lambda_{+}^{-1}/2$ & $q^{3}\lambda_{+}^{-1}/2$ & / &
/\\\hline
ferm-con & / & $q^{3}\lambda_{+}^{-1}/2$ & $q^{3}\lambda_{+}^{-1}/2$ &
/\\\hline
\end{tabular}
\newline
\end{center}

\noindent Table for $k_{3}$:

\begin{center}%
\begin{tabular}
[c]{r||c|c|c|c}\hline
Euclid & R1 & R2 & R3 & R4\\\hline
bos-uncon & / & / & $\lambda_{+}/2$ & $\lambda_{+}/2$\\\hline
bos-con & / & $\lambda_{+}/2$ & / & $\lambda_{+}/2$\\\hline
ferm-uncon & / & / & $-q^{-2}\lambda_{+}/2$ & $-q^{-2}\lambda_{+}/2$\\\hline
ferm-con & / & $\lambda_{+}/2$ & / & $\lambda_{+}/2$\\\hline
\end{tabular}
\newline\newline\newline%
\begin{tabular}
[c]{r||c|c|c|c}\hline
Mink & R1 & R2 & R3 & R4\\\hline
bos-uncon & $-q\lambda_{+}/2$ & $-q\lambda_{+}/2$ & / & /\\\hline
bos-con & $q\lambda_{+}/2$ & $q\lambda_{+}/2$ & / & /\\\hline
ferm-uncon & $-q^{3}\lambda_{+}/2$ & $-q^{3}\lambda_{+}/2$ & / & /\\\hline
ferm-con & / & $q^{3}\lambda_{+}/2$ & $q^{3}\lambda_{+}/2$ & /\\\hline
\end{tabular}
\newline
\end{center}

\noindent Table for $k_{4}$:

\begin{center}%
\begin{tabular}
[c]{r||c|c|c|c}\hline
Euclid & R1 & R2 & R3 & R4\\\hline
bos-uncon & / & / & $-q^{-2}/2$ & $-q^{-2}/2$\\\hline
bos-con & / & $q^{-2}/2$ & / & $q^{-2}/2$\\\hline
ferm-uncon & / & / & $q^{-4}/2$ & $q^{-4}/2$\\\hline
ferm-con & / & $q^{-2}/2$ & / & $q^{-2}/2$\\\hline
\end{tabular}
\newline\newline\newline%
\begin{tabular}
[c]{r||c|c|c|c}\hline
Mink & R1 & R2 & R3 & R4\\\hline
bos-uncon & $q^{3}/2$ & $q^{3}/2$ & / & /\\\hline
bos-con & $q^{3}/2$ & $q^{3}/2$ & / & /\\\hline
ferm-uncon & $-q^{5}/2$ & $-q^{5}/2$ & / & /\\\hline
ferm-con & / & $q^{5}/2$ & $q^{5}/2$ & /\\\hline
\end{tabular}
\newline
\end{center}

\noindent\textbf{Identity (A.13)} in Ref. \cite{bailin}:
\begin{align}
(\theta\sigma^{\mu\nu}\phi)(\bar{\chi}\bar{\sigma}^{\kappa\lambda}\bar{\eta})=
&  \;k_{1}\;(P_{A})^{\mu\nu}{}_{\mu^{\prime}\nu^{\prime}}(P_{A})^{\kappa
\lambda}{}_{\kappa^{\prime}\lambda^{\prime}}\hat{R}^{\nu^{\prime}%
\kappa^{\prime}}{}_{\kappa^{\prime\prime}\nu^{\prime\prime}}\eta
^{\mu^{\prime}\kappa^{\prime\prime}}\eta^{\nu^{\prime\prime}\lambda^{\prime}%
}\eta_{\rho\rho^{\prime}}(\theta\sigma^{\rho}\bar{\eta})(\bar{\chi}%
\sigma^{\rho^{\prime}}\phi)\nonumber\\
&  +\;k_{2}\;(P_{A})^{\mu\nu}{}_{\nu^{\prime\prime}\kappa^{\prime\prime}%
}(P_{A})^{\kappa\lambda}{}_{\kappa^{\prime}\lambda^{\prime}}\hat{R}%
^{\nu^{\prime\prime}\lambda^{\prime}}{}_{\lambda^{\prime\prime}\nu
^{\prime\prime\prime}}\eta^{\kappa^{\prime\prime}\kappa^{\prime}}(\theta
\sigma^{\lambda^{\prime\prime}}\bar{\eta})(\bar{\chi}\sigma^{\nu^{\prime
\prime\prime}}\phi)\nonumber\\
&  +\;k_{3}\;(P_{A})^{\mu\nu}{}_{\mu^{\prime}\nu^{\prime}}(P_{A}%
)^{\kappa\lambda}{}_{\kappa^{\prime}\lambda^{\prime}}\hat{R}^{\nu^{\prime
}\kappa^{\prime}}{}_{\kappa^{\prime\prime}\nu^{\prime\prime}}\eta^{\mu
^{\prime}\kappa^{\prime\prime}}(\theta\sigma^{\nu^{\prime\prime}}\bar{\eta
})(\bar{\chi}\bar{\sigma}^{\lambda^{\prime}}\phi)\nonumber\\
&  +\;k_{4}\;\varepsilon^{\mu\nu\kappa^{\prime}\rho}(P_{A})^{\kappa\lambda}%
{}_{\kappa^{\prime}\lambda^{\prime}}\eta_{\rho\rho^{\prime}}(\theta
\sigma^{\rho^{\prime}}\bar{\eta})(\bar{\chi}\bar{\sigma}^{\lambda^{\prime}%
}\phi)\nonumber\\
&  +\;k_{5}\;\varepsilon^{\nu^{\prime}\kappa\lambda\rho}(P_{A})^{\mu\nu}%
{}_{\mu^{\prime}\nu^{\prime}}\eta_{\rho\rho^{\prime}}(\theta\sigma
^{\mu^{\prime}}\bar{\eta})(\bar{\chi}\bar{\sigma}^{\rho^{\prime}}\phi).
\end{align}

\noindent Table for $k_{1}$:

\begin{center}%
\begin{tabular}
[c]{r||c|c|c|c}\hline
Euclid & R1 & R2 & R3 & R4\\\hline
bos-uncon & $-q^{-1}/2$ & $-q^{-5}/2$ & $-q^{-1}/2$ & $-q^{-5}/2$\\\hline
bos-con & $-q^{-1}/2$ & $-q^{-1}/2$ & $-q^{-5}/2$ & $-q^{-5}/2$\\\hline
ferm-uncon & $q^{-3}/2$ & $q^{-3}/2$ & $q^{-3}/2$ & $q^{-3}/2$\\\hline
ferm-con & $q^{-3}/2$ & $q^{-3}/2$ & $q^{-3}/2$ & $q^{-3}/2$\\\hline
\end{tabular}
\newline\newline\newline%
\begin{tabular}
[c]{r||c|c|c|c}\hline
Mink & R1 & R2 & R3 & R4\\\hline
bos-uncon & $-q^{-1}/2$ & $-q^{3}/2$ & $-q^{3}/2$ & $-q^{3}/2$$-q^{-1}%
/2$\\\hline
bos-con & $-q^{5}/2$ & $-q^{5}/2$ & $-q^{5}/2$ & $-q^{5}/2$\\\hline
ferm-uncon & $q^{9}/2$ & $q^{5}/2$ & $q^{5}/2$ & $q^{9}/2$\\\hline
ferm-con & $q^{-1}/2$ & $q^{-1}/2$ & $q^{3}/2$ & $q^{3}/2$\\\hline
\end{tabular}
\newline
\end{center}

\noindent Table for $k_{2}$:

\begin{center}%
\begin{tabular}
[c]{r||c|c|c|c}\hline
Euclid & R1 & R2 & R3 & R4\\\hline
bos-uncon & $-q^{2}\lambda_{+}/2$ & $-q^{-2}\lambda_{+}/2$ & $-q^{2}%
\lambda_{+}/2$ & $-q^{-2}\lambda_{+}/2$\\\hline
bos-con & $-q^{2}\lambda_{+}/2$ & $-q^{2}\lambda_{+}/2$ & $-q^{-2}\lambda
_{+}/2$ & $-q^{-2}\lambda_{+}/2$\\\hline
ferm-uncon & $\lambda_{+}/2$ & $\lambda_{+}/2$ & $\lambda_{+}/2$ &
$\lambda_{+}/2$\\\hline
ferm-con & $\lambda_{+}/2$ & $\lambda_{+}/2$ & $\lambda_{+}/2$ & $\lambda
_{+}/2$\\\hline
\end{tabular}
\newline\newline\newline%
\begin{tabular}
[c]{r||c|c|c|c}\hline
Mink & R1 & R2 & R3 & R4\\\hline
bos-uncon & $-q^{4}\lambda_{+}/2$ & $-\lambda_{+}/2$ & $-\lambda_{+}/2$ &
$-q^{4}\lambda_{+}/2$\\\hline
bos-con & $-q^{2}\lambda_{+}/2$ & $-q^{2}\lambda_{+}/2$ & $-q^{2}\lambda
_{+}/2$ & $-q^{2}\lambda_{+}/2$\\\hline
ferm-uncon & $q^{2}\lambda_{+}/2$ & $q^{2}\lambda_{+}/2$ & $q^{2}\lambda
_{+}/2$ & $q^{2}\lambda_{+}/2$\\\hline
ferm-con & $\lambda_{+}/2$ & $\lambda_{+}/2$ & $\lambda_{+}/2$ & $\lambda
_{+}/2$\\\hline
\end{tabular}
\newline
\end{center}

\noindent Table for $k_{3}$:

\begin{center}%
\begin{tabular}
[c]{r||c|c|c|c}\hline
Euclid & R1 & R2 & R3 & R4\\\hline
bos-uncon & $\lambda_{+}/2$ & $q^{-4}\lambda_{+}/2$ & $\lambda_{+}/2$ &
$q^{-4}\lambda_{+}/2$\\\hline
bos-con & $\lambda_{+}/2$ & $\lambda_{+}/2$ & $q^{-4}\lambda_{+}/2$ &
$q^{-4}\lambda_{+}/2$\\\hline
ferm-uncon & $-q^{-2}\lambda_{+}/2$ & $-q^{-2}\lambda_{+}/2$ & $-q^{-2}%
\lambda_{+}/2$ & $-q^{-2}\lambda_{+}/2$\\\hline
ferm-con & $-q^{-2}\lambda_{+}/2$ & $-q^{-2}\lambda_{+}/2$ & $-q^{-2}%
\lambda_{+}/2$ & $-q^{-2}\lambda_{+}/2$\\\hline
\end{tabular}
\newline\newline\newline%
\begin{tabular}
[c]{r||c|c|c|c}\hline
Mink & R1 & R2 & R3 & R4\\\hline
bos-uncon & $q^{_{2}}\lambda_{+}/2$ & $q^{2}\lambda_{+}/2$ & $q^{2}\lambda
_{+}/2$ & $q^{-2}\lambda_{+}/2$\\\hline
bos-con & $q^{4}\lambda_{+}/2$ & $q^{4}\lambda_{+}/2$ & $q^{4}\lambda_{+}/2$ &
$q^{4}\lambda_{+}/2$\\\hline
ferm-uncon & $-q^{4}\lambda_{+}/2$ & $-q^{4}\lambda_{+}/2$ & $-q^{4}%
\lambda_{+}/2$ & $-q^{4}\lambda_{+}/2$\\\hline
ferm-con & $-q^{2}\lambda_{+}/2$ & $-q^{2}\lambda_{+}/2$ & $-q^{2}\lambda
_{+}/2$ & $-q^{2}\lambda_{+}/2$\\\hline
\end{tabular}
\newline
\end{center}

Table for $k_{4}$:

\begin{center}%
\begin{tabular}
[c]{r||c|c|c|c}\hline
Euclid & R1 & R2 & R3 & R4\\\hline
bos-uncon & $\lambda_{+}/4$ & $q^{-4}\lambda_{+}/4$ & $\lambda_{+}/4$ &
$q^{-4}\lambda_{+}/4$\\\hline
bos-con & $-\lambda_{+}/4$ & $-\lambda_{+}/4$ & $-q^{-4}\lambda_{+}/4$ &
$-q^{-4}\lambda_{+}/4$\\\hline
ferm-uncon & $-q^{-2}\lambda_{+}/4$ & $-q^{-2}\lambda_{+}/4$ & $-q^{-2}%
\lambda_{+}/4$ & $-q^{-2}\lambda_{+}/4$\\\hline
ferm-con & $q^{-2}\lambda_{+}/4$ & $q^{-2}\lambda_{+}/4$ & $q^{-2}\lambda
_{+}/4$ & $q^{-2}\lambda_{+}/4$\\\hline
\end{tabular}
\newline\newline\newline%
\begin{tabular}
[c]{r||c|c|c|c}\hline
Mink & R1 & R2 & R3 & R4\\\hline
bos-uncon & $q^{-3}\lambda_{+}/4$ & $-q\lambda_{+}/4$ & $-q\lambda_{+}/4$ &
$q^{-3}\lambda_{+}/4$\\\hline
bos-con & $q^{3}\lambda_{+}/4$ & $q^{3}\lambda_{+}/4$ & $q^{3}\lambda_{+}/4$ &
$q^{3}\lambda_{+}/4$\\\hline
ferm-uncon & $q^{3}\lambda_{+}/4$ & $q^{3}\lambda_{+}/4$ & $q^{3}\lambda
_{+}/4$ & $q^{3}\lambda_{+}/4$\\\hline
ferm-con & $-q\lambda_{+}/4$ & $-q\lambda_{+}/4$ & $-q\lambda_{+}/4$ &
$-q\lambda_{+}/4$\\\hline
\end{tabular}
\newline
\end{center}

\noindent Table for $k_{5}$:

\begin{center}%
\begin{tabular}
[c]{r||c|c|c|c}\hline
Euclid & R1 & R2 & R3 & R4\\\hline
bos-uncon & $\lambda_{+}/4$ & $q^{-4}\lambda_{+}/4$ & $\lambda_{+}/4$ &
$q^{-4}\lambda_{+}/4$\\\hline
bos-con & $-\lambda_{+}/4$ & $-\lambda_{+}/4$ & $-q^{-4}\lambda_{+}/4$ &
$-q^{-4}\lambda_{+}/4$\\\hline
ferm-uncon & $-q^{-2}\lambda_{+}/4$ & $-q^{-2}\lambda_{+}/4$ & $-q^{-2}%
\lambda_{+}/4$ & $-q^{-2}\lambda_{+}/4$\\\hline
ferm-con & $-q\lambda_{+}/4$ & $-q\lambda_{+}/4$ & $-q\lambda_{+}/4$ &
$-q\lambda_{+}/4$\\\hline
\end{tabular}
\newline\newline\newline%
\begin{tabular}
[c]{r||c|c|c|c}\hline
Mink & R1 & R2 & R3 & R4\\\hline
bos-uncon & $q^{-3}\lambda_{+}/4$ & $-q\lambda_{+}/4$ & $-q\lambda_{+}/4$ &
$q^{-3}\lambda_{+}/4$\\\hline
bos-con & $q^{3}\lambda_{+}/4$ & $q^{3}\lambda_{+}/4$ & $q^{3}\lambda_{+}/4$ &
$q^{3}\lambda_{+}/4$\\\hline
ferm-uncon & $q^{3}\lambda_{+}/4$ & $q^{3}\lambda_{+}/4$ & $q^{3}\lambda
_{+}/4$ & $q^{3}\lambda_{+}/4$\\\hline
ferm-con & $-q\lambda_{+}/4$ & $-q\lambda_{+}/4$ & $-q\lambda_{+}/4$ &
$-q\lambda_{+}/4$\\\hline
\end{tabular}
\newline
\end{center}

\noindent\textbf{Acknowledgements}\newline First of all we are very grateful
to Eberhard Zeidler for his invitation to the MPI Leipzig, his special
interest in our work and his financial support. Furthermore we would like to
thank Fabian Bachmaier and Ina Stein for their steady support. Finally, we
thank Dieter L\"{u}st for kind hospitality.

\appendix

\section{$q$-Deformed quantum spaces\label{AppA}}

The aim of this appendix is the following. For the quantum spaces under
consideration we list the defining commutation relations. In addition to this,
we write down the non-vanishing elements of their quantum metric and
$q$-deformed epsilon tensor.

The two-dimensional quantum plane is described in Sec.\thinspace
\ref{qplanekap} in great detail. In the case of three-dimensional $q$-deformed
Euclidean space the commutation relations between its coordinates $X^{A},$
$A\in\{+,3,-\},$ read
\begin{align}
X^{3}X^{\pm} &  =q^{\pm2}X^{\pm}X^{3},\\
X^{-}X^{+} &  =X^{+}X^{-}+\lambda X^{3}X^{3}.\nonumber
\end{align}
The non-vanishing elements of the quantum metric are
\begin{equation}
g^{+-}=-q,\quad g^{33}=1,\quad g^{-+}=-q^{-1}.\label{metriceu3app}%
\end{equation}
As usual, covariant coordinates are introduced by
\begin{equation}
X_{A}=g_{AB}X^{B},
\end{equation}
with $g_{AB}$ being the inverse of $g^{AB}$. The non-vanishing components of
the three-dimensional $q$-deformed epsilon tensor take the form%
\begin{align}
\varepsilon^{-3+} &  =-q^{-4}, & \varepsilon^{3-+} &  =q^{-2},\nonumber\\
\varepsilon^{-+3} &  =q^{-2}, & \varepsilon^{+-3} &  =-q^{-2},\nonumber\\
\varepsilon^{3+-} &  =-q^{-2}, & \varepsilon^{+3-} &  =1,\nonumber\\
\varepsilon^{333} &  =-q^{-2}\lambda. &  & \label{epseu3app}%
\end{align}

Next we come to four-dimensional $q$-deformed Euclidean space. For its
coordinates $X^{i},$ $i=1,\ldots,4,$ we have the relations
\begin{align}
X^{1}X^{j} &  =qX^{j}X^{1},\nonumber\\
X^{j}X^{4} &  =qX^{4}X^{j},\quad j=1,2,\nonumber\\
X^{2}X^{3} &  =X^{3}X^{2},\nonumber\\
X^{4}X^{1} &  =X^{1}X^{4}+\lambda X^{2}X^{3}.\label{koordeu4app}%
\end{align}
The metric has the non-vanishing components
\begin{equation}
g^{14}=q^{-1},\quad g^{23}=g^{32}=1,\quad g^{41}=q.\label{metriceu4app}%
\end{equation}
Its inverse denoted by $g_{ij}$ can again be used to introduce covariant
coordinates, i.e.
\begin{equation}
X_{i}=g_{ij}X^{j}.
\end{equation}
The non-vanishing components of the epsilon tensor of four-dimensional
$q$-deformed Euclidean become (see also Refs. \cite{Maj-Eps, Fio93, Mey96})%
\begin{align}
\varepsilon^{1234} &  =1, & \varepsilon^{1432} &  =-q^{2}, & \varepsilon
^{2413} &  =-q^{2},\nonumber\\
\varepsilon^{2134} &  =-q, & \varepsilon^{4132} &  =q^{2}, & \varepsilon
^{4213} &  =q^{3},\nonumber\\
\varepsilon^{1324} &  =-1, & \varepsilon^{3412} &  =q^{2}, & \varepsilon
^{2341} &  =-q^{2},\nonumber\\
\varepsilon^{3124} &  =q, & \varepsilon^{4312} &  =-q^{3}, & \varepsilon
_{3241} &  =q^{2},\nonumber\\
\varepsilon^{2314} &  =q^{2}, & \varepsilon^{1243} &  =-q, & \varepsilon
^{2431} &  =q^{3},\nonumber\\
\varepsilon^{3214} &  =-q^{2}, & \varepsilon^{2143} &  =q^{2}, &
\varepsilon^{4231} &  =-q^{4},\nonumber\\
\varepsilon^{1342} &  =q, & \varepsilon^{1423} &  =q^{2}, & \varepsilon^{3421}
&  =-q^{3},\nonumber\\
\varepsilon^{3142} &  =-q^{2}, & \varepsilon^{4123} &  =-q^{2}, &
\varepsilon^{4321} &  =q^{4},\label{epseu4app}%
\end{align}
together with the non-classical components%
\begin{equation}
\varepsilon^{3232}=-\varepsilon^{2323}=-q^{2}\lambda.
\end{equation}

Now, we come to $q$-deformed Minkowski space \cite{SWZ91,OSWZ92,Maj91}. Its
coordinates are subjected to the relations
\begin{gather}
X^{\mu}X^{0}=X^{0}X^{\mu},\quad\mu\in{\{}0,+,-,3{\},}\nonumber\\
X^{3}X^{\pm}-q^{\pm2}X^{\pm}X^{3}=-q\lambda X^{0}X^{\pm},\nonumber\\
X^{-}X^{+}-X^{+}X^{-}=\lambda(X^{3}X^{3}-X^{0}X^{3}),\label{koordminkapp}%
\end{gather}
and its metric is given by%
\begin{equation}
\eta^{00}=-1,\quad\eta^{33}=1,\quad\eta^{+-}=-q,\quad\eta^{-+}=-q^{-1}%
.\label{metricminkapp}%
\end{equation}
As usual, the metric can be used to raise and lower indices. The non-vanishing
components of the $q$-deformed epsilon tensor read%
\begin{align}
\varepsilon^{+30-} &  =1, & \varepsilon^{+-03} &  =-q^{-2}, & \varepsilon
^{3-+0} &  =q^{-2},\nonumber\\
\varepsilon^{3+0-} &  =-q^{-2}, & \varepsilon^{-+03} &  =q^{-2}, &
\varepsilon^{-3+0} &  =q^{-4},\nonumber\\
\varepsilon^{+03-} &  =-1, & \varepsilon^{0-+3} &  =q^{-2}, & \varepsilon
^{30-+} &  =-q^{-2},\nonumber\\
\varepsilon^{0+3-} &  =1, & \varepsilon^{-0+3} &  =-q^{-2}, & \varepsilon
^{03-+} &  =q^{-2},\nonumber\\
\varepsilon^{30+-} &  =q^{-2}, & \varepsilon^{+3-0} &  =-1, & \varepsilon
^{3-0+} &  =q^{-2},\nonumber\\
\varepsilon^{03+-} &  =-q^{-2}, & \varepsilon^{3+-0} &  =q^{-2}, &
\varepsilon^{-30+} &  =-q^{-4},\nonumber\\
\varepsilon^{+0-3} &  =q^{-2}, & \varepsilon^{+-30} &  =q^{-2}, &
\varepsilon^{0-3+} &  =-q^{-4},\nonumber\\
\varepsilon^{0+-3} &  =-q^{-2}, & \varepsilon^{-+30} &  =-q^{-2}, &
\varepsilon^{-03+} &  =q^{-4},\label{epsminkapp}%
\end{align}
and%
\begin{align}
\varepsilon^{0-0+} &  =q^{-3}\lambda, & \varepsilon^{-0+0} &  =-q^{-3}%
\lambda,\\
\varepsilon^{0333} &  =-q^{-2}\lambda, & \varepsilon^{3330} &  =q^{-2}%
\lambda,\nonumber\\
\varepsilon^{3033} &  =+q^{-2}\lambda, & \varepsilon^{3030} &  =-q^{-2}%
\lambda,\nonumber\\
\varepsilon^{3303} &  =-q^{-2}\lambda, & \varepsilon^{+0-0} &  =-q^{-1}%
\lambda,\nonumber\\
\varepsilon^{0303} &  =q^{-2}\lambda, & \varepsilon^{0+0-} &  =q^{-1}%
\lambda.\nonumber
\end{align}
Lowering the indices of the epsilon tensor is achieved by the quantum metric:%
\begin{equation}
\varepsilon_{\mu\nu\rho\sigma}=\eta_{\mu\mu^{\prime}}\eta_{\nu\nu^{\prime}%
}\eta_{\rho\rho^{\prime}}\eta_{\sigma\sigma^{\prime}}\varepsilon^{\mu^{\prime
}\nu^{\prime}\rho^{\prime}\sigma^{\prime}}.
\end{equation}

\end{document}